\documentclass[a4paper,UKenglish,cleveref, autoref, thm-restate]{lipics-v2021}
\pdfoutput=1 %uncomment to ensure pdflatex processing (mandatatory e.g. to submit to arXiv)
\hideLIPIcs  %uncomment to remove references to LIPIcs series (logo, DOI, ...), e.g. when preparing a pre-final version to be uploaded to arXiv or another public repository

\usepackage{amsopn}
\usepackage{enumerate}
\usepackage{tikz}
\usetikzlibrary{trees}
\usepackage[utf8]{inputenc}
\usepackage{amsmath}
\usepackage{amssymb,amsfonts}
\usepackage[T1]{fontenc}
\usepackage{bm}
\usepackage{graphicx}
\usepackage{hyperref}
\usepackage{wrapfig}
\usepackage{blindtext}
\usepackage{array}
\usepackage{makecell}
\usepackage{mathtools}
\usepackage{float}
\usepackage{algpseudocode}
\usepackage[linesnumbered,lined,boxed,commentsnumbered,noend]{algorithm2e}
\SetKwFunction{Recursion}{Recursion}
\SetKwFunction{Valid}{Valid}
\SetKwFunction{GenPath}{Gen\_Path}
\usepackage{stmaryrd}
\usepackage{thmtools}
\usepackage{thm-restate}
\usepackage{cleveref}
\newcommand{\classize}[1]{{\ensuremath{\mathsf{#1}}}\xspace}
\newcommand{\totp}{\classize{TotP}}
\newcommand{\fpspace}{\classize{FPSPACE}}
\newcommand{\spspace}{\classize{\#PSPACE}}
\newcommand{\totpspace}{\classize{{TotPSPACE}}}

\newcommand{\sP}{\classize{\#P}}
\newcommand{\sL}{\classize{\#L}}
\newcommand{\spanl}{\classize{SpanL}}
\newcommand{\spanp}{\classize{SpanP}}
\newcommand{\spanpspace}{\classize{SpanPSPACE}}
\newcommand{\spanpspacep}{\classize{SpanPSPACE(poly)}}
\newcommand{\sPE}{\classize{\#PE}}
\newcommand{\PH}{\classize{PH}}
\newcommand{\up}{\classize{UP}}
\newcommand{\cP}{\classize{P}}
\newcommand{\NP}{\classize{NP}}
\newcommand{\MaxSNP}{\classize{MaxSNP}}
\newcommand{\NL}{\classize{NL}}
\newcommand{\PSPACE}{\classize{PSPACE}}
\newcommand{\RP}{\classize{RP}}
\newcommand{\FP}{\classize{FP}}

\newcommand{\FPRAS}{\classize{FPRAS}}
\newcommand{\existsSOclass}{\classize{\exists SO}}
\newcommand{\sigmaone}{\classize{\#\Sigma_1}}
\newcommand{\rsigmatwo}{\classize{\#R\Sigma_2}}
\newcommand{\rpihone}{\classize{\#R\Pi H_1}}

\newcommand{\qsoclass}{\classize{QSO}}
\newcommand{\sigqsofoclass}{\classize{\Sigma QSO(FO)}}
\newcommand{\classlam}{\classize{\Lambda}}
\newcommand{\classC}{\classize{C}}
\newcommand{\sotcclass}{\classize{SO(TC)}}

\newcommand{\totclasslfpu}{\classize{R_{so}^r\Sigma_{so}^r(LFP)}}
\newcommand{\totclasslfpf}{\classize{R_{so}^r\Sigma_{so}^r(FO)}}
\newcommand{\spanclass}{\classize{R_{fo}\Sigma_{fo}(FO)}}
\newcommand{\spanpspaceclass}{\classize{R_{so}\Sigma_{so} (SO)}}
\newcommand{\pspaceclass}{\classize{R_{so}^{r}\Sigma_{so}(SO)}}

\newcommand{\sSAT}{\textsc{\#Sat}}

\newcommand{\snfa}{\#\textsc{NFA}}

\newcommand{\logicize}[1]{{\ensuremath{\mathtt{#1}}}\xspace}
\newcommand{\loglam}{\logicize{\Lambda}}
\newcommand{\fo}{\logicize{FO}}
\newcommand{\folfp}{\logicize{FO(LFP)}}
\newcommand{\fotc}{\logicize{FO(TC)}}
\newcommand{\sotc}{\logicize{SO(TC)}}
\newcommand{\SO}{\logicize{SO}}
\newcommand{\existsSO}{\logicize{\exists SO}}
\newcommand{\qso}{\logicize{QSO}}
\newcommand{\qfo}{\logicize{QFO}}
\newcommand{\sigqso}{\logicize{\Sigma QSO}}
\newcommand{\sigqsofo}{\logicize{\Sigma QSO(FO)}}

\newcommand{\totlogiclfpth}{\logicize{R_{so}^r\Sigma_{so}^r(LFP)}}
\newcommand{\totlogiclfpf}{\logicize{R_{so}^r\Sigma_{so}^r(FO)}}
\newcommand{\spanlogic}{\logicize{R_{fo}\Sigma_{fo}(FO)}}
\newcommand{\sigmasol}{\logicize{\Sigma SO(\underline{\Lambda})}}
\newcommand{\sigmafol}{\logicize{\Sigma FO(\underline{\Lambda})}}
\newcommand{\genlog}{\logicize{R_{L_1}\Sigma_{L_2}(L_3)}}
\newcommand{\lone}{\logicize{L_1}}
\newcommand{\ltwo}{\logicize{L_2}}
\newcommand{\lthree}{\logicize{L_3}}
\newcommand{\folower}{\logicize{fo}}
\newcommand{\solower}{\logicize{so}}
\newcommand{\supr}{\logicize{r}}
\newcommand{\sigmaqsol}{\logicize{\Sigma QSO(\Lambda)}}
\newcommand{\sigmaqfou}{\logicize{\Sigma FO(\underline{FO})}}
\newcommand{\sigmaqsou}{\logicize{\Sigma SO(\underline{SO})}}
\newcommand{\pspacelogic}{\logicize{R_{so}^{r}\Sigma_{so}(SO)}}
\newcommand{\spanpspacelogic}{\logicize{R_{so}\Sigma_{so} (SO)}}
\newcommand{\sigmalfp}{\logicize{\Sigma SO(\underline{LFP})}}
\newcommand{\lfpu}{\logicize{\underline{LFP}}}
\newcommand{\sigmalfpu}{\logicize{\Sigma SO^r(\underline{LFP})}}
\newcommand{\sigmaforu}{\logicize{\Sigma SO^r(\underline{FO})}}
\newcommand{\sigmasofo}{\logicize{\Sigma SO(\underline{FO})}}

\newcommand{\arity}{\mathsf{arity}}

\newcommand{\lfpd}{\ensuremath{\mathrm{\mathbf{lfp}}}}
\newcommand{\lfp}{\ensuremath{\mathrm{lfp}}}

\newcommand{\enc}{\ensuremath{\mathrm{enc}}}
\newcommand{\quantsigma}{\ensuremath{\mathsf{\Sigma}}\xspace}
\newcommand{\quantpi}{\ensuremath{\mathsf{\Pi}}\xspace}

\newcommand{\circl}{\ensuremath{\circ}}
\newcommand{\rel}{\ensuremath{\mathrm{Expl}}}
\newcommand\mydashv{\mathrel{\stackrel{\makebox[0pt]{\mbox{\normalfont \scriptsize a}}}{\dashv}}}
\newcommand\vartextvisiblespace[1][.5em]{%
  \makebox[#1]{%
    \kern.07em
    \vrule height.3ex
    \hrulefill
    \vrule height.3ex
    \kern.07em
  }% <-- don't forget this one!
}
\renewcommand{\gets}{:=}

\title{Counting Computations with Formulae: Logical Characterisations of Counting Complexity Classes} %TODO Please add

\titlerunning{Logical Characterisations of Counting Complexity Classes} %TODO optional, please use if title is longer than one line

\author{Antonis Achilleos}{Department of Computer Science, Reykjavik University, Iceland  \and \url{https://sites.google.com/view/antonisachilleos} }{antonios@ru.is}
{https://orcid.org/0000-0002-1314-333X}{}
\author{Aggeliki Chalki}
{Department of Computer Science, Reykjavik University, Iceland \and \url{https://aggelikichal.github.io/}}{angelikic@ru.is}{https://orcid.org/0000-0001-5378-0467}
{}

\authorrunning{A. Achilleos and A. Chalki} %TODO mandatory. First: Use abbreviated first/middle names. Second (only in severe cases): Use first author plus 'et al.'

\Copyright{Antonis Achilleos and Aggeliki Chalki} %TODO mandatory, please use full first names. LIPIcs license is "CC-BY";  http://creativecommons.org/licenses/by/3.0/

\ccsdesc[500]{Theory of computation~Complexity theory and logic}
\ccsdesc[500]{Theory of computation~Complexity classes}
\keywords{descriptive complexity, quantitative logics, counting problems, \sP} %TODO mandatory; please add comma-separated list of keywords

\category{} %optional, e.g. invited paper

\relatedversion{} %optional, e.g. full version hosted on arXiv, HAL, or other respository/website
\funding{This work has been funded by the projects ``Open Problems in the Equational Logic of Processes (OPEL)'' (grant no.~196050), ``Mode(l)s of Verification and Monitorability'' (MoVeMnt) (grant no~217987) of the Icelandic Research Fund, and the Basic Research Program PEVE 2020 of the National Technical University of Athens.}%optional, to capture a funding statement, which applies to all authors. Please enter author specific funding statements as fifth argument of the \author macro.

\acknowledgements{The authors would like to thank Stathis Zachos and Aris Pagourtzis for fruitful discussions and Luca Aceto for sound advice.}%optional

\nolinenumbers %uncomment to disable line numbering

\EventEditors{John Q. Open and Joan R. Access}
\EventNoEds{2}
\EventLongTitle{42nd Conference on Very Important Topics (CVIT 2016)}
\EventShortTitle{CVIT 2016}
\EventAcronym{CVIT}
\EventYear{2016}
\EventDate{December 24--27, 2016}
\EventLocation{Little Whinging, United Kingdom}
\EventLogo{}
\SeriesVolume{42}
\ArticleNo{23}
\begin{document}

\maketitle

\begin{abstract}
We present quantitative logics with two-step semantics based on the framework of quantitative logics introduced by Arenas et al.\ (2020) and the two-step semantics defined in the context of weighted logics by Gastin \& Monmege (2018). 
We show that some of the fragments of our logics augmented with a least fixed point operator capture interesting classes of counting problems.
Specifically, we answer an open question in the area of descriptive complexity of counting problems by providing logical characterizations of two subclasses of \sP, namely \spanl{} and \totp, that play a significant role in the study of approximable counting problems. Moreover, we define logics that capture \fpspace{} and \spanpspace, 
 which are 
% contain 
counting versions of \PSPACE. 
\end{abstract}

\section{Introduction}
In this paper, we examine counting problems from the viewpoint of descriptive complexity. We present a quantitative logic with a least fixed point operator and 
%intermediate
two-step semantics, 
similar to the ones introduced by Gastin and Monmege in \cite{GastinM18} for weighted versions of automata and Monadic Second-Order logic.
% , where I
In the first step, 
the interpretation of a formula on a structure generates a set.
%a formula is interpreted  on a structure as the set it generates. 
In the second step, a quantitative interpretation results from the cardinality of that set.
%We present a quantitative logic with a least fixed point operator, where a formula is interpreted on a structure as the set it generates.
These semantics allow us to use a fairly uniform approach to identify fragments of this logic that capture several counting complexity classes.

In 1979, Valiant introduced the complexity class \sP in his seminal paper~\cite{Valiant79}  and used it to characterize the complexity of computing the permanent function. \sP is the class of functions that count accepting paths of non-deterministic poly-time Turing machines, 
%(NPTMs), 
or, equivalently, the number of solutions to problems in \NP. For example, \textsc{\#Sat} is the function that, with input a  formula $\varphi$ in CNF, returns the number of satisfying assignments of $\varphi.$ Since then, counting complexity has played an important role in computational complexity theory. 

% In this paper, we examine counting problems from the viewpoint of descriptive complexity. 
Descriptive complexity provides characterizations of complexity classes in terms of the logic needed to express their problems. We refer the reader to \cite{Gradel07} for a survey. The B\"uchi--Elgot--Trakhtenbrot theorem~\cite{Bchi1960,Elgot1962,Trakhtenbrot1961} characterizing regular languages in terms of
Monadic Second-Order logic and Fagin's theorem \cite{Fa74}, which states that Existential Second-Order logic captures \NP, are two
fundamental results in this area. Another prominent result was the introduction of the class \MaxSNP~\cite{PY91}, which has played a central role in the study of the hardness of approximation for optimization problems~\cite{AroraLMSS92}. Moreover, descriptive complexity is an interesting and active research field with more recent results in the logical characterization of the class \cP~\cite{GradelP19}, dynamic complexity~\cite{VortmeierZ21}, symmetric linear programs~\cite{AtseriasDO21}, and counting complexity~\cite{Arenas,DurandHKV21}, among others.%\\

% \noindent\textbf{Related work.}
% Arenas et al. and Saluja et al. give logical characterisations of \sP{}
% % Logical characterizations of \sP{} have been provided 
% in~\cite{Saluja, Arenas}. The authors of~\cite{Saluja} 
% substitute 
% existential quantification over second-order variables of $\ensuremath{\pmb{\exists}\mathbf{SO}}$ 
% % was substituted 
% with counting second-order variables. The 
% % second approach
% paper~\cite{Arenas} incorporated counting into the syntax of the logic by introducing Quantitative Second-Order logic, denoted by \qso, a logic for quantitative functions, which is  based on the framework of  Weighted logics~\cite{DG07,GastinM18,AchilleosP21}.
% % There have been many attempts to find a suitable
% %extension of MSO to describe quantitative properties
% %which captures the expressive power of weighted automata.

As in the case  of optimization problems, an 
% interesting 
important
and long-standing question has been the logical characterization of approximable counting problems. This is also a meaningful line of research since very few counting problems can be exactly computed in polynomial time. In the case of counting problems, the appropriate notion of approximability is the existence of a fully polynomial-time randomized approximation scheme (fpras). We denote 
% the class of approximable counting problems 
by \FPRAS
the class of counting problems that admit an fpras~\cite{DGGJ04,BCP20}.

A counting class is considered to be \emph{robust} if  it has either natural complete problems or nice closure properties. 
Two robust subclasses of \sP{} defined in terms of Turing machines (TMs), are of great significance in the quest for 
% either a logical or a structural 
a characterization of approximable counting problems. The first one is \totp, which contains all self-reducible counting problems whose decision version is in \cP. It is noteworthy that \totp is not contained in \FPRAS, unless $\RP=\NP$~\cite{BCP20}, but almost all known approximable counting problems belong to \totp (see e.g.~\cite{KLM89,JSV04,PZ06}). The second class, namely \spanl~\cite{AJ93}, is contained in \totp, and it consists of the functions that count different outputs of non-deterministic log-space \emph{transducers}, i.e.\  TMs with output. To the best of our knowledge, \spanl is the only counting class so far defined in terms of TMs that is a subclass of \FPRAS~\cite{Arenas20}, despite containing \sP-complete problems~\cite{AJ93}.
% Logical characterizations of \totp{} and \spanl{} have not been determined yet. This was posed as an open question in~\cite{Arenas}.

% Logics that capture superclasses of \sP{} have also been provided:  \spanp{}, which has been defined in~\cite{KST89}, was first characterized in~\cite{CG96}, and recently in~\cite{Arenas}. Moreover, a logic that captures \fpspace{}~\cite{Ladner89}, i.e.\ the class of poly-space computable functions, was given in~\cite{Arenas}. Finally, in~\cite{DurandHKV21}, a framework for the descriptive complexity of arithmetic circuit classes  was introduced.\\

\begin{figure}
   \centering
\includegraphics[scale=1]{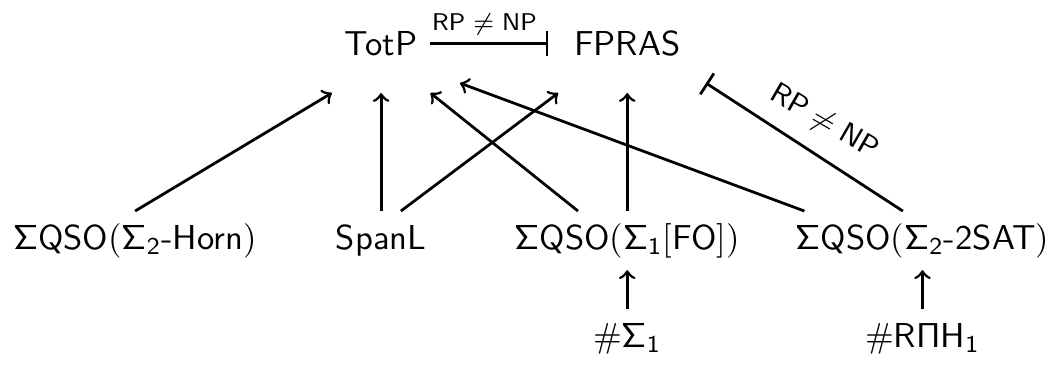}
    \caption{Subclasses of \totp{} and their relationship to \FPRAS. The following notation is used: 
$A \rightarrow B$ denotes $A \subseteq B$, and $A\mydashv B$ denotes $A \not\subseteq B$ under the assumption a.}
   \label{fig:hasse}
\end{figure}

\noindent\textbf{Our contribution.}
 % Our motivation has been to 
 Our main objective is to 
 provide logical characterizations of the classes \spanl and \totp, %Logical characterizations of \totp{} and \spanl{} have not been determined yet. This 
 which was posed as an open question in~\cite{Arenas}. To this end, we introduce a variant of the quantitative logics that are defined in~\cite{Arenas}. 
 Our two-step semantic definition is the key difference between our approach and that in~\cite{Arenas}.
 % The main difference is that we define semantics in two steps.
The first step is an \emph{intermediate semantics}, where the meaning of a
formula is given as a set of strings. These strings consist of either elements of the universe or relations over the universe,
and,
% . They, 
intuitively, represent computation paths. In the second step, a concrete semantics associates with each formula the size of the set resulting from the intermediate semantics.
%The second step is \emph{concrete semantics}, where one can translate the intermediate semantics into $\mathbb{N}$ by taking the size of the set. 
Gastin et al.\ follow an analogous approach for weighted logics 
% was followed 
in~\cite{GastinM18}, where
% . In~\cite{GastinM18},  
% they correspond 
% runs of
% automata 
% are related 
% to 
the evaluation of formulae
corresponds to automata runs, 
whereas in this paper 
the semantic interpretation of formulae
% we relate 
corresponds to 
computation runs or outputs of Turing machines.

In Section~\ref{spanl section}, we introduce logics equipped with least fixed point formulae that capture `span-classes' of restricted space, namely \spanl and \spanpspace, in a natural way (Theorems~\ref{spanl theorem} and~\ref{spanpspace theorem}). When we consider such classes, we are interested in counting the number of different outputs produced by a transducer. So far, semantics that is defined as a mapping from the set of quantitative formulae to $\mathbb{N}$ can interpret every accepting path as contributing one unit.
% the existence of an accepting path as the natural number $1$. 
Then, by evaluating the sum of formulae as the sum of natural numbers, one can sum up the accepting paths of a TM. 
On the other hand, when accepting paths with the same output collectively contribute only one to the sum, then it is more appropriate to evaluate a formula as a set of output strings and the sum of formulae as the union of sets. 
% is a more suitable choice.  

We also consider two classes, namely \spspace and \totp, which contain functions that count the number of accepting or the total number of paths of TMs with restricted resources, respectively. Using the alternative semantics introduced here, a computation path can be encoded as a sequence of configurations visited by the TM along that path---in other words, a path can be encoded  by 
%either its whole or a representative part of 
its computation history---so that different paths are mapped to different sequences.
%a computation path can be mapped to a sequence of configurations visited by the TM along the computation of the path, so that different paths are mapped to different sequences. 
Therefore, when taking
the union of sets that contain encodings of computation paths, the result 
% results in 
is
a set that contains all the
distinct encodings,
% se distinct sequences, 
and the size of the resulting set is 
% equal to 
the number of computation paths that we intend to count. In Section~\ref{fpspace section}, we provide a logical characterization of  the class of functions that count the number of accepting paths of poly-space TMs, namely \spspace~\cite{Ladner89} (Theorem~\ref{fpspace corollary}),  which  coincides with \fpspace, i.e.\ the class of functions that are computable in polynomial space. \fpspace has already been characterized by a quantitative logic with a partial fixed point~\cite{Arenas}
%over relations. 
Interestingly, the logic we define here includes  a least fixed point. 
%over functions.
%which was first introduced in~\cite{Arenas}. 
In Section~\ref{totp section}, we prove that two different quantitative logics 
%that combines two different kinds of recursion---a least fixed point over relations as defined in~\cite{Immerman82,Vardi82}, and 
%one 
%over functions---
capture \totp (Theorems~\ref{totp main theorem} and~\ref{totp main theorem-two}). 
%  
% On top of that, they can be either extended or restricted to capture other interesting classes. 
In Section~\ref{conclusions-section}, we discuss the fact that by specializing the semantics, we obtain two least fixed point logics that capture \NL and \PSPACE, respectively. We believe that the semantics we propose in this paper can contribute insight to the study of counting complexity classes.

\noindent\textbf{Related work.}
Arenas et al.\ and Saluja et al.\ give logical characterisations of \sP
% Logical characterizations of \sP{} have been provided 
in~\cite{Saluja, Arenas}. The authors of~\cite{Saluja} 
substitute 
existential quantification over second-order variables of \existsSO{} 
% was substituted 
with counting second-order variables. The 
% second approach
% paper
work in~\cite{Arenas} incorporated counting into the syntax of the logic by introducing Quantitative Second-Order logic, denoted by \qso, a logic for quantitative functions, which is  based on the framework of  weighted logics~\cite{DG07,GastinM18,AchilleosP21}.
% There have been many attempts to find a suitable
%extension of MSO to describe quantitative properties
%which captures the expressive power of weighted automata.

There has been progress in characterizing 
counting classes with respect  to their approximability in the context of descriptive complexity. Saluja et al.\ defined the classes \sigmaone and \rsigmatwo
% and proven 
in~\cite{Saluja}, and proved that they contain only problems that admit an fpras. A more recent variant of \sigmaone~\cite{DurandHKV21} 
% defined more recently in~\cite{DurandHKV21}, 
is also a subclass of \FPRAS. The class \rpihone~\cite{DGGJ04} is conjectured to contain problems which are neither as hard to approximate as \#\textsc{Sat} nor admit an fpras, and it has been used to classify Boolean \#CSP with respect to their approximability~\cite{DGJ10}. 
Since \NP-complete problems cannot have approximable counting versions unless $\RP=\NP$~\cite{DGGJ04}, Arenas et al.\ suggested in~\cite{Arenas} 
% it was suggested 
that robust classes of counting problems with an easy decision version should be examined. The papers~\cite{Arenas,BCP20} defined such counting classes 
% were defined 
and examined them with respect to the approximability of their problems. 
% A counting class is defined to be robust if either it has natural complete problems or nice closure properties. 
Some of the aforementioned classes and their relationship to \FPRAS are depicted in Figure~\ref{fig:hasse}. 
% One of them, namely $\mathsf{\Sigma QSO(\Sigma_1[FO])}$, is a subclass of \FPRAS, whereas
% $\mathsf{\Sigma QSO(\Sigma_2\text{-}Horn)}$ is of unknown approximability status. Another two subclasses of \totp{} with well-known natural complete problems, namely
% $\mathsf{\Sigma QSO(\Sigma_2\text{-}2SAT)}$ and $\#\mathsf{\Pi_2}\text{-}\mathsf{1VAR}$~\cite{BCP20}, contain problems that are as hard to approximate as \#\textsc{Sat}. Notably, all the aforementioned classes are subclasses of \totp. 

There is also work on logics that capture superclasses of \sP, namely \spanp~\cite{KST89} and \fpspace~\cite{Ladner89}.
% have also been provided:  
Compton and Gr\"{a}del were the first to characterize 
\spanp in~\cite{CG96}, 
%which was defined in~\cite{KST89}, 
followed by Arenas et al.\ 
% was first characterized in~\cite{CG96}, and recently 
in~\cite{Arenas}, where they also introduced 
% . Moreover, 
a logic that captures \fpspace.
%~\cite{Ladner89}.
%, i.e.\ the class of poly-space computable functions.
% , was given in~\cite{Arenas}. 
Finally, in~\cite{DurandHKV21}, Durand et al.\ introduced a framework for the descriptive complexity of arithmetic circuit classes.  

% \totp\ is the class of functions that count the total number of paths of NPTMs. 
% Except for this simple structural characterization, \totp\ has an alternative definition. Interestingly, it is the class  of all self-reducible problems with a decision version in \cP, which is also closed under parsimonious reductions~\cite{PZ06}. \totp\ is  a \textit{robust} class; a class is considered to be robust if either it is closed under addition, multiplication and subtraction by one or it has natural complete problems~\cite{Arenas}. In particular, \totp\ satisfies both these requirements (see Section~\ref{preliminaries}). Notably, almost all known counting problems that admit an fpras belong to \totp.

% Saluja et al.~\cite{Saluja} gave logical characterizations of \sP\ and some classes below \sP. However, in this setting, counting problems are not themselves expressed by formulas in some logic, but are rather problems that count the number of variables that satisfy some formula in a structure (see for example the second row of Table~\ref{descriptive examples}). Recently Arenas et al.~\cite{Arenas} incorporated counting into the syntax of the logic by introducing Quantitative Second-Order Logic, denoted by \qso, a logic for quantitative functions, which is  based on the framework of  Weighted Logics~\cite{DG07,AchilleosP21}. Natural syntactic fragments of \qso\ capture counting complexity classes such as \sP, \textsf{SpanP}, and other classes of functions like \FP\ and \textsf{FPSPACE}.

\section{Preliminaries}\label{preliminaries}

\subsection{Turing machines}

A (\emph{two-tape non-deterministic}) \emph{Turing machine} (TM) $N$ is a quintuple $N=(\mathcal{Q},\Sigma,\delta,q_0,q_F)$, where $\mathcal{Q}$ is a set of states, $\Sigma=\{0,1\}$ is the alphabet, $\delta \subseteq (\mathcal{Q}\times(\Sigma\cup\{\vartextvisiblespace\})^2) ~\times~ (\mathcal{Q}\times(\Sigma\cup\{\vartextvisiblespace\})\times\{L,R\}^2)$ is the transition relation, $q_0$ is the initial state, and $q_F$ is the final accepting state. 
The elements of $\delta$ are called \emph{transitions}.
The TM $N$ is assumed to have a read-only input tape, and a work tape that can be read and written on.
 $L$ and $R$ in a transition designate that the tape head moves to the left or right, respectively, in each tape.
%We also assume that once the machine accepts, it clears its tape, it moves its cursor all the way to the left and enters the unique final state $q_F$.
% A \emph{configuration} $c$ of $N$ encodes a snapshot of the computation of $N$ and is defined in the usual way.
A \emph{configuration} $c$ of $N$ is a pair from $(\mathcal{Q} \cup \Sigma)^2$ that describes a snapshot of the computation of $N$, so exactly one state appears in $c = (t_I,t_W)$, exactly once in each $t_I, t_W$. 
The configuration $c = (uqv,u'qv')$, where $q \in \mathcal{Q}$, denotes that the current contents of the input tape are $uv$,  the  contents of the work tape are $u'v'$,  the current state of $N$ is $q$ and the tape head is on the first symbol of $v$ and $v'$ in the respective tapes.
We can apply a compatible transition to a configuration to result in a new configuration in the expected way.

A \emph{run} of $N$ is a (possibly infinite) sequence $c_0c_1\cdots$ of configurations of $N$, such that for every $c_i, c_{i+1}$ in the run, $c_{i+1}$ results from $c_{i}$ and a transition in $\delta$.
Given an initial configuration $c_0$, we observe that the runs of $N$ from $c_0$ form a computation tree with each node labelled by a configuration, each edge corresponding to a transition in $\delta$, and each branch corresponding to a run.
W.l.o.g.\ we  assume that every Turing machine has a binary computation tree. This means that at any configuration on the tree, the computation is either deterministic (only one transition is applied) or the machine makes a choice between exactly two transitions. We call the two transitions that result from a non-deterministic choice, the \emph{left} and \emph{right} non-deterministic transition. Therefore, the transition relation $\delta$ maps a triple in $\mathcal{Q}\times(\Sigma\cup\{\vartextvisiblespace\})^2$ to at most two tuples in $\mathcal{Q}\times(\Sigma\cup\{\vartextvisiblespace\})\times\{L,R\}^2$.
% , where $L$ and $R$ designate that the cursor moves to the left or right, respectively. 
A \emph{transducer} $T$ is a Turing machine that has also a write-only output tape, on which a string over $\Sigma$ is written from left to right. In this case, the transition relation $\delta \subseteq (\mathcal{Q}\times(\Sigma\cup\{\vartextvisiblespace\})^2) ~\times~ (\mathcal{Q}\times(\Sigma\cup\{\vartextvisiblespace\})\times\{L,R\}^2)\times(\Sigma \cup \{\varepsilon \})$. 
The output of a run of $T$ is called \emph{valid} if it stops in the accepting state.

Let $M$ be a TM or a transducer. For every $x\in\Sigma^*$, let $t_M(x)$ be the maximum number of 
transitions 
% time steps,
%the height of the computation tree of $M$ on input $x$, 
and $s_M(x)$ be the maximum number of work tape cells used by $M$ with respect to all runs on input $x$, respectively. The worst-case time (resp.\ space) complexity of $M$ is the function $T_M:\mathbb{N}\rightarrow\mathbb{N}\cup\{+\infty\}$ (resp.\ $S_M:\mathbb{N}\rightarrow\mathbb{N}\cup\{+\infty\}$) defined by $T_M(n)=\max\{t_M(x)~\mid~ x\in\Sigma^*, |x|=n\}$ (resp.\ $S_M(n)=\max\{s_M(x)~\mid~ x\in\Sigma^*, |x|=n\}$).

\begin{definition}
    A Turing machine or a transducer $M$ is 
    \begin{itemize}
        \item \emph{polynomial-time} if for some $c\in\mathbb{N}$, $T_M (n)\in\mathcal{O}(n^c)$.
        \item \emph{log-space} if $S_M(n)\in\mathcal{O}(\log n)$.
        \item \emph{poly-space} if for some $c\in\mathbb{N}$, $S_M(n)\in\mathcal{O}(n^c)$.
    \end{itemize}
\end{definition}

A TM or tranducer is called
\emph{deterministic} if at every configuration at most one transition can be applied. 
%Otherwise, it is called \emph{non-deterministic}.
We use standard abbreviations for TMs or transducers, such as NPTM (non-deterministic poly-time Turing machine), NL-transducer (non-deterministic log-space transducer) etc.
We say that $f$ is \emph{computable in polynomial time} (resp. logarithmic/polynomial space), if there is a deterministic polynomial-time (resp. log-space/poly-space) transducer $M$, such that for every $x \in \Sigma^*$, $f(x)$ is the valid output of $M$ on input $x$.

We define the functions that count paths (resp.\ outputs) of a Turing machine (resp.\ transducer) as follows.

\begin{definition}
Let $M$ be a Turing machine and $T$ a tranducer. We define
\begin{enumerate}[(a)]
\item $acc_M:\Sigma^*\rightarrow \mathbb{N}\cup\{+\infty\}$ such that $acc_M(x)= \#(\text{accepting computation paths of } M \text{ on input } x)$, for every  $x\in\Sigma^*$.
\item $tot_M:\Sigma^*\rightarrow \mathbb{N}\cup\{+\infty\}$ such that $tot_M(x)= \#($%all
computation paths of $M$ on input $x)-1$, for every  $x\in\Sigma^*$.
% \end{enumerate}
% Let $M$ be a transducer. We define
% \begin{enumerate}[(a)]
% \setcounter{enumi}{2}
\item $span_T:\Sigma^*\rightarrow \mathbb{N}\cup\{+\infty\}$ such that $span_T(x)= \#($different valid outputs of $T$ on input $x)$, for every  $x\in\Sigma^*$.
\end{enumerate}
\end{definition}

\subsection{Classes of counting problems}

% Formal definitions of the counting classes we examine, are provided in Definition~\ref{classes}. 
Given a function $f:\Sigma^*\rightarrow \mathbb{N}$, $L_f:=\{x\in\Sigma^* \mid f(x)>0\}$ is the problem of deciding whether $f$ is non-zero on input $x$. Hence, $L_f$ is called the \emph{decision version} of $f$.  
We now present formal definitions for the counting classes we examine, in Definition~\ref{classes}. 

\begin{definition}[\cite{Valiant79,PZ06,KST89,AJ93}]\label{classes}
\begin{enumerate}[(a)]
    \item 
 $\sP=\{acc_M:\Sigma^*\rightarrow \mathbb{N} ~\mid~  M\text{ is an NPTM}\}$, 
 %where 
%$acc_M(x)= \#(\text{accepting computation paths of } M \text{ on input } x)$.
\item $\FP= \{f:\Sigma^*\rightarrow \mathbb{N} ~\mid~  f\text{ is computable in polynomial time}\}$.
\item 
$\sPE=\{f: \Sigma^*\rightarrow \mathbb{N}  ~\mid~ f\in\sP \text{ and } L_f \in \cP \}$, 
%where $L_f=\{x\in \Sigma^*  ~\mid~ f(x)>0\}$ is the decision version of the function $f$.
\item 
$\totp =\{tot_M:\Sigma^*\rightarrow \mathbb{N}  ~\mid~  M\text{ is an NPTM}\}$, 
%where 
%$tot_M(x)= \#($all computation paths of $M$ on input $x)-1$.
\item 
$\spanp =\{span_M:\Sigma^*\rightarrow \mathbb{N}  ~\mid~  M\text{ is an NP-transducer}\}$, 
%where 
%$span_M(x)= \#($different valid outputs of $M$ on input $x)$.
\item 
$\spanl =\{span_M:\Sigma^*\rightarrow \mathbb{N}  ~\mid~  M\text{ is an NL-transducer}\}$.
%where 
%$span_M(x)= \#($different valid outputs of $M$ on input $x)$.
\end{enumerate}  
\end{definition}

\begin{remark}\label{branchings}
Note that in the definition of \totp, one is subtracted from the total number of paths so that a function can take the zero value. Since a \totp function $f$ can be associated with an NPTM $M$ that has a binary computation tree, $f(x)=tot_M(x)=\#\text{(branchings of } M \text{ on input } x )$, where a branching is 
% a  time step 
an occurrence of a configuration on the computation tree, where
% at which 
% the TM 
$M$
makes a non-deterministic choice.

For the class \spanl, note that, by the pigeonhole principle, an NL-transducer has infinitely many accepting paths if and only if the length of its accepting runs is not bounded by a polynomial. It then makes sense to attach a clock that imposes a polynomial-time bound to each NLTM, as suggested in~\cite{AJ93}. In this way, every NLTM is also an NPTM with a finite number of computation paths.
%For the class \spanl, note that although an NL-transducer can contain cycles, and so infinitely many different valid outputs, a clock that imposes a polynomial-time bound can be attached to an NLTM, as suggested in~\cite{AJ93}. In this way, every NLTM is also an NPTM with a finite number of computation paths.
\end{remark}

\begin{definition}[\cite{Ladner89}]
\begin{enumerate}[(a)]
\item $\fpspace=\{f:\Sigma^*\rightarrow \mathbb{N} ~\mid~  f\text{ is computable in polynomial space}\}$,
\item $\spspace=\{acc_M:\Sigma^*\rightarrow \mathbb{N} \ | \  M\text{ is a non-deterministic poly-space TM}\}$.
\end{enumerate}  
\end{definition}

\begin{remark}\label{clock-poly-space}
As in the case of NLTMs (see Remark~\ref{branchings}), we assume that a clock that imposes an exponential-time bound can be attached to a non-deterministic polynomial-space TM so that the TM has a finite number of computation paths.
\end{remark}

Propositions \ref{prop:FPtoSpanP} and \ref{Ladner} provide
basic relationships among the aforementioned classes of functions.

\begin{proposition}[\cite{AJ93,PZ06,HV95}]\label{prop:FPtoSpanP}
$\FP \subseteq \spanl \subseteq \totp \subseteq \sPE \subseteq \sP\subseteq\spanp$. The first inclusion is proper unless $\cP=\NP=\PH=\cP^{\sP}$. The last inclusion is proper unless $\up=\NP$. All other inclusions are proper unless $\cP=\NP$.
\end{proposition}

\begin{proposition}[\cite{Ladner89}]\label{Ladner}
$\sP\subsetneq\fpspace=\spspace$.
\end{proposition}

\spanl-complete problems were known since the seminal work of \`Alvarez and Jenner~\cite{AJ93}. The most significant one is the \snfa{} problem, where an NFA $M$ and a natural number $m$ in unary are given and the output is  the number of words of length $m$ accepted by $M$. A quasi-polynomial randomized approximation scheme for \#\textsc{NFA} was known for about 25 years~\cite{KSM95}, before an fpras was designed for the problem~\cite{Arenas20}. The latter result yields an fpras for every problem in the class \spanl. It also makes \spanl the first and only class so far, to the best of our knowledge, with a TM-based definition that is a subclass of \FPRAS and is not contained in \FP (under standard assumptions). 

The definition of \totp  provides a structural characterization of its problems. However, this class has an alternative useful characterization, which is given in Proposition~\ref{PZmt}. The definitions of (poly-time) self-reducibility, and parsimonious reductions
precede the proposition.

% \begin{definition}
% The \emph{decision version} of a function $f:\Sigma^*\rightarrow \mathbb{N}$ is the language $L_f=\{x\in\Sigma^* ~\mid~ f(x)>0\}$.
% \end{definition}

\begin{definition}[\cite{ABCP22}]
A function $f : \Sigma^*\rightarrow \mathbb{N}$ is \emph{(poly-time) self-reducible} if 
%there exist polynomials 
%$r$ and $q$, and poly-time computable functions $h : \Sigma^*\times\mathbb{N} \rightarrow \Sigma^*$, 
%$g :\Sigma^*\times\mathbb{N} \rightarrow \mathbb{N}$, and $t :\Sigma^*\rightarrow\mathbb{N}$ such that 
for all $x\in\Sigma ^*$:
\begin{enumerate}[(a)]
\item $f$ can be processed recursively by reducing $x$ to a polynomial number of instances $h(x,i)$, where $h\in\FP$ and $0 \le i\le r(|x|)$ for some polynomial $r$. Formally, for every $x\in\Sigma^*$,  $$f(x) = t(x) +\sum_{i=0}^{r(|x|)} g(x,i)f(h(x,i)), \text{ where } t,g\in\FP.$$
\item  The recursion terminates after at most polynomial depth. Formally, the depth of the recursion is $q(|x|)$, for some polynomial $q$, and for every $x\in\Sigma^*$ and $\Vec{j}\in \{0,\dots ,r(|x|)\}^{q(|x|)}$,
%$j_1,j_2,\dots,j_{q(|x|)}\in \{0,\dots ,r(|x|)\}$,
%$f\big(h(\dots h(h(x,j_1),j_2)\dots ,j_{q(|x|)})\big)$
$$f(\tilde{h}(x,\Vec{j})) \text{ can be computed in polynomial time w.r.t.\ } |x|,$$
 where $\tilde{h}$ is the extension of $h$ such that $\tilde{h}(x,\varepsilon)=x$ and $\tilde{h}(x,j_1...j_k)=h(\tilde{h}(x,j_1...j_{k-1}),j_k)$.
\item Every instance invoked in the recursion is of polynomial size in $|x|$. Formally, there is a polynomial $p$, such that $|\tilde{h}(x,\Vec{j})|\in\mathcal{O}\big(p(|x|)\big)$, for every $x\in\Sigma^*$, $k\leq q(|x|)$ and $\Vec{j} \in \{0,\dots ,r(|x|)\}^k$.
\end{enumerate}
\end{definition}

Informally, a function is (poly-time) self-reducible if its value on an instance can be recursively computed by evaluating the same function on a polynomial number of smaller instances. 
%The formal and more general definition of a poly-time self-reducible function follows.

\begin{example}\label{self-reducibility}
 The problem of counting satisfying assignments of a formula $\varphi$ in disjunctive normal form, denoted by $\#\textsc{DNF}$, is self-reducible, since $\#\textsc{DNF}(\varphi)=\#\textsc{DNF}(\varphi_0)+\#\textsc{DNF}(\varphi_1)$, where $\varphi_0$ and $\varphi_1$ are the formulae obtained from $\varphi$, by setting the value of a variable, let's say $x_1$, to false and true, respectively.
\end{example}

%For example, $\sSAT$ is self-reducible, since the number of satisfying assignments for a propositional formula $\varphi$ equals the sum of the number of satisfying assignments for $\varphi_0$ and $\varphi_1$, which are obtained from $\varphi$ by setting the value of a variable to false and true, respectively.

A reduction from a counting function $f$ to $g$ is called parsimonious when no post-computation is required. In other words, parsimonious reductions preserve the number of solutions of the two involved functions.

\begin{definition}
Let $f$, $g:\Sigma ^*\rightarrow \mathbb{N}$ be two counting functions.
We say that there is a \emph{parsimonious reduction} from $f$ to $g$, denoted by $f \le_{pars}^p g$,  if there is $h\in\FP$, such that  $f(x) = g(h(x))$, for every $x\in\Sigma^*$.
% (b) We say that there is a product reduction from $f$ to $g$, denoted by  $f\le_{pr} g$, if there are poly-time computable functions $h_1,h_2$ such that  $f(x) = g(h_1(x))\cdot h_2(x)$, for every $x\in\Sigma^*$.~\footnote{Note that we slightly changed the definition of a product reduction given in~\cite{Saluja}. We replaced $h_2(|x|)$ by $h_2(x)$, i.e.\ the value of $h_2$ depends on the input and not only on the size of the input. This small change is not crucial for any of the results obtained in~\cite{Saluja}.}
%For a class $\mathsf{F}$ of functions, we denote by $\mathrm{Closure}_{\le_m^p}(\mathsf{F})$ (resp. $\mathrm{Closure}_{\le_{pr}}(\mathsf{F})$) the closure of the class $\mathsf{F}$ under parsimonious (resp. product) reductions.
\end{definition}

\begin{proposition}[\cite{PZ06}]\label{PZmt}
\totp is the closure under parsimonious reductions of the class of self-reducible \sPE functions.
\end{proposition}

%The class \totp\ is the Karp closure of all self-reducible problems in \sP\ with easy decision version. For example, consider the problem \textsc{\#IS}, that corresponds to counting all independent sets (of all sizes) of an input graph.  Clearly this problem has an easy decision version, since every non empty graph has at least one independent set. It is also self-reducible since the number of independent sets of a (non empty) graph $G$ equals the number of independent sets containing some vertex $v$ plus the number of independent sets not containing vertex $v$. Computing the two latter numbers is equivalent to counting independent sets of two subgraphs $G_1$ and $G_0$, respectively. $G_1$ results from $G$ by removing vertex $v$, all its neighbours, and all edges adjacent to them. $G_0$ results from $G$ by removing vertex $v$ and its adjacent edges.

For counting problems, self-reducibility appears to be the rule and not the exception.  \totp contains several well-studied problems such as counting satisfying assignments of a formula in DNF, computing the permanent of a matrix, counting perfect matchings of a graph, counting independent sets (of all sizes) of a graph, counting $q$-colorings  with $q$ greater than or equal to the maximum degree of the graph, computing the partition function of several models from statistical physics (e.g.\ the Ising and the hard-core model), counting bases of a matroid, computing the volume of a convex body, and many more. The following example shows how self-reducibility and the easy-decision property of a counting problem imply its membership in \totp.

\begin{example}\label{totp example 1}
\begin{figure}
   \centering
   \includegraphics[scale=1]{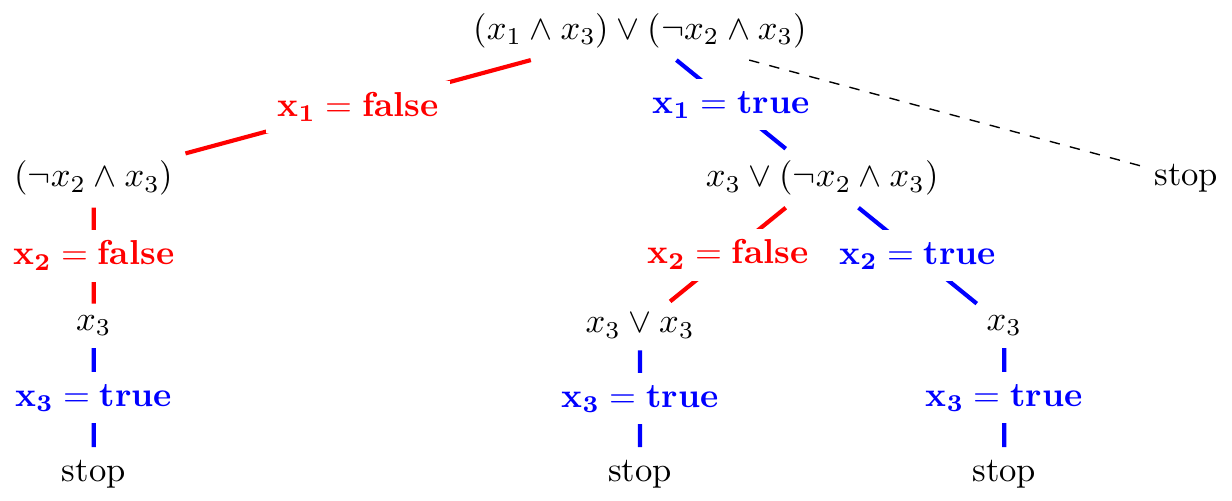}
    \caption{The computation tree of
    NPTM $M$ 
    on $(x_1\wedge x_3)\vee (\neg x_2 \wedge x_3)$, where 
    % for which 
    % it holds that 
    $tot_M(\enc(\varphi)) = \#\textsc{DNF}(\varphi)$, 
    % where 
    and
    $\enc(\varphi)$ is a binary encoding of 
    $\varphi$.
    % $\varphi = (x_1\wedge x_3)\vee (\neg x_2 \wedge x_3)$.
    }
   \label{fig:totp}
\end{figure}
\begin{enumerate}[(a)]
\item Consider the problem \textsc{\#DNF}, which is a self-reducible counting problem (see Example~\ref{self-reducibility}) with a decision version in \cP.
Let $\varphi$ be a formula in disjunctive normal form and $x_1, ...,x_n$ be an enumeration list of its variables. Consider an NPTM $M$ that, at its first step, determines in polynomial time whether $\varphi$ is satisfiable. If the answer is no, it halts. Otherwise, it generates a dummy path and starts a recursive computation as follows. If the enumeration list is empty, then $M$ halts. If it is not empty, $M$ picks the first variable appearing in the list, let's say $x_i$, removes $x_i$ from the list, and checks whether formulae  $\varphi_0$ and $\varphi_1$, i.e.\ $\varphi$ with $x_i$ assigned to false and true, respectively, are satisfiable. 
\begin{itemize}
    \item If the answer is yes for both cases, $M$ chooses non-deterministically to set $x_i$ to either  false or true, and proceeds recursively with $\varphi_0$ or $\varphi_1$, respectively.
    \item If the answer is yes for only one case, $M$ deterministically proceeds recursively with the corresponding formula, i.e.\ either $\varphi_0$ or $\varphi_1$. 
\end{itemize} 
Since $M$ removes at least one variable from the list at each step, the recursion depth is polynomial in the size of $\varphi$. Finally, note that every sequence of non-deterministic choices of $M$ corresponds to a satisfying assignment of $\varphi$ and so the definition of \totp is satisfied;  the number of satisfying assignments of $\varphi$ equals the number of all paths of $M(\enc(\phi))$ minus one. The computation of $M$ on input $(x_1\wedge x_3)\vee (\neg x_2 \wedge x_3)$ is depicted in Figure~\ref{fig:totp}.
\item Consider the problem  of counting independent sets of all sizes in a graph $G$, denoted by \textsc{\#IS}.  Clearly, this problem has an easy decision version, since every non-empty graph has at least one independent set. It is also self-reducible: the number of independent sets of a (non-empty) graph $G$ equals the number of independent sets containing some vertex $v$ plus the number of independent sets not containing vertex $v$. Computing the two latter numbers is equivalent to counting independent sets of two subgraphs $G_1$ and $G_0$, respectively. $G_1$ results from $G$ by removing vertex $v$, all its neighbours, and all edges adjacent to them. $G_0$ results from $G$ by removing vertex $v$ and its adjacent edges.
Now consider an NPTM $N$ that finds an independent set of $G$ by non-deterministically choosing at step $i$, to add vertex $v_i$ to the independent set or not, and proceeds recursively with the remaining graph. Since at each step, $N$ removes at least one vertex, the recursion depth is polynomial in the size of $G$. Moreover, every sequence of non-deterministic choices of $N$ corresponds to an independent set. Finally, by adding a dummy path to $N$, we have that $\#\textsc{IS}(G)=\#$(paths of $N$ on input $G)-1$.
\end{enumerate}
\end{example}

\totp is a \emph{robust} class. We adopt here the notion of robustness suggested in~\cite{Arenas}: a counting class is robust if either it has natural complete problems, or it is closed under addition, multiplication and subtraction by one. \totp satisfies both properties~\cite{ABCP22}. Note that the classes \sP and \sPE are not closed under subtraction by one (under widely-believed assumptions)~\cite{OH93, PZ06}, and \sPE is not known to have complete problems. In specific, closure under subtraction guarantees that \totp does not contain \sPE problems like $\#\textsc{Sat}_{+1}$, which outputs the number of satisfying assignments of a formula plus one, unless $\sSAT\in\totp$, and that we can manipulate witnesses (paths of an NPTM) and in some cases remove them.

\begin{lemma}\label{closure}
\totp is closed under addition, multiplication, and subtraction by one.
\end{lemma}
\begin{proof} We prove here closure under subtraction by one. We show that if $f\in\totp$, then $h = f\dot- 1$ also belongs to \totp, where $h:\Sigma^*\rightarrow \mathbb{N}$ is defined as follows $$h(x)=\begin{cases}
f(x)-1, &\text{ if }f(x)\neq 0 \\
f(x), &\text{ if } f(x)=0
\end{cases}.$$

Let $M_f$ be an NPTM such that for every $x\in\Sigma^*$, $f(x)=tot_{M_f}(x)=\#($paths of  $M_f$ on  $x)-1$. We construct $M_h$ such that $h(x)=tot_{M_h}(x)=\#(\text{paths of } M_h \text{ on } x)-1$.
 $M_h(x)$ simulates $M_f(x)$ until the first non-deterministic choice is made or $M_f(x)$ terminates. If $M_f(x)$ has only one path, then $M_h(x)$ halts. If $M_f(x)$ makes at least one non-deterministic choice, $M_h(x)$ copies the behavior of $M_f(x)$, but while simulating the leftmost path, before making a non-deterministic choice, it checks whether one of the choices leads to a deterministic computation. The first time $M_h(x)$ detects such a choice, it eliminates the path corresponding to the deterministic computation, and continues the simulation of $M_f(x)$. Notice that $M_h(x)$ can recognize the leftmost path since computation paths can be lexicographically ordered. In this case, $M_h(x)$ has one path less than $M_f(x)$. In both cases,  $h(x) = tot_{M_h}(x) = tot_{M_f}(x)\dot-1 = f(x)\dot- 1$.
\end{proof}

The following counting versions of natural decision problems belong to \spspace and they were introduced along with the definition of the class in~\cite{Ladner89}: the problem of counting the number of words not accepted by a given NFA and the problem of counting the number of \emph{verifying trees} for a quantified Boolean formula.

% which we denote here  by \snfa{} (if the set $NA$ of non-accepted words by NFA $M$ is finite, then $|NA|<s^{2^q-1}+1$, so in the case of $NA$ being infinite \snfa($M$) is defined to be equal to $s^{2^q-1}+1$, where $s$ is the number of input symbols of $M$ and $q$ is the number of states in $M$). 
% %
% which given a \textbf{QBF} formula $\varphi=\exists x_1\forall x_2\dots Q x_m B(x_1,x_2,\dots,x_m)$, $Q\in\{\exists,\forall\}$, . A verifying tree for $\varphi$ is a tree, the root of which is labeled with a truth assignment for $x_1$. The root has two children, one for each possible truth assignment for $x_2$. Each of them has a child labeled with one assignment for $x_3$, and so on and so forth. Each leaf corresponds to an assignment of all the variables and $\varphi$ must be true for all the assignments associated to the leaves of the verifying tree.

Below we introduce three classes that are variants of \spanl and \totp, and they are defined by transducers and Turing machines, repsectively, of polynomial space.

\begin{definition}
\begin{enumerate}[(a)]
\item $\spanpspace =\{span_M:\Sigma^*\rightarrow \mathbb{N}  ~\mid~  M$ is a non-deterministic poly-space transducer$\}$.
\item $\spanpspacep =\{span_M:\Sigma^*\rightarrow \mathbb{N}  ~\mid~  M$ is a non-deterministic poly-space transducer every output of which is of polynomial size$\}$.
\item $\totpspace =\{tot_M:\Sigma^*\rightarrow \mathbb{N} \ | \  M$ is a non-deterministic poly-space TM$\}$.
\end{enumerate}
\end{definition}

We show that the class \totpspace coincides with \fpspace, and so with \spspace, whereas $\spanpspacep\subseteq\spspace\subseteq\spanpspace$.

\begin{theorem}\label{totpspace theorem}
$\fpspace =  \totpspace=\spspace$.
\end{theorem}
\begin{proof}
$\fpspace \subseteq  \totpspace$: Consider $f\in \fpspace$ with a corresponding TM $M$.  Ladner~\cite{Ladner89} describes in detail how a non-deterministic TM $M'$ can simulate $M$ so that it computes $f(x)$ bit by bit and generates $f(x)$ accepting computation paths. It is not hard to see that a slight modification of this procedure allows $M'$ to generate only the accepting paths, and so the number of accepting paths of $M'(x)$ is equal to the total number of paths of $M'(x)$. By generating an additional dummy path, $M'$ has as many paths as needed and so $f\in \totpspace$.

$\totpspace \subseteq \fpspace$: Let $f\in \totpspace$ and $M$ be such that $f(x)=$\#(paths of $M$ on $x)-1=$\#(accepting paths of $M$  on  $x)+$\#(rejecting paths of  $M$  on  $x)-1$. There are deterministic poly-space TMs $M_{acc}$ and $M_{rej}$ such that they compute the number of accepting and rejecting paths of $M$, respectively, as described by Ladner~\cite{Ladner89}. Since, \fpspace is closed under sum and subtraction by one, $f\in\fpspace$.

$\fpspace=\spspace$ is true by Proposition~\ref{Ladner}.
\end{proof}

\begin{theorem}
$\spanpspacep\subseteq\spspace\subseteq\spanpspace$.
\end{theorem}
\begin{proof}
$\spanpspacep\subseteq\spspace$: Let $M$ be a poly-space  transducer that on any input of size $n$, it generates outputs of length at most $n^k$. Assume that $M$, before entering the accepting state, it erases its work tape, and then accepts. As a result, since there are $2^{n^k}$ different possible outputs, $M$ has at most $2^{n^k}$ different accepting configurations. By Savitch's theorem~\cite[Section 7.3]{Pap94}, there is a deterministic poly-space TM $M_{con}$ that, ggiven the initial configuration $c_0$ and some other configuration $c$ of $M$, determines whether there is a computation of $M$ starting  from $c_0$ and ending at $c$. Since \fpspace{} is closed under exponential sum, there is a deterministic poly-space TM that computes the number of different accepting configurations of $M$ by simulating $M_{con}$ on $(c_0,c)$, for every accepting configuration $c$, reusing space. So $\spanpspacep\subseteq\fpspace$, which implies that $\spanpspacep\subseteq\spspace$.

$\spspace\subseteq\spanpspace$: Let $f\in\spspace$ and $M$ be a non-deterministic poly-space TM such that $f(x)=\#(\text{accepting paths of } M \text{ on } x)$. Define the non-deterministic poly-space transducer $M'$ that on input $x$ simulates $M(x)$ and on any  path $p$ outputs the encoding of the non-deterministic choices made by $M$ on $p$. Then, $f(x)=\#(\text{different valid outputs of }$\\
$M \text{ on } x)$.
\end{proof}

%\subsection{Quantitative Second-Order Logic}
\subsection{Logics}

A relational vocabulary $\sigma=\{\mathcal{R}_1^{k_1},...,\mathcal{R}_m^{k_m}\}$ is a finite set of relation symbols. Each relation symbol $\mathcal{R}_i$ has a positive integer $k_i$ as its designated arity. 

\begin{definition}
A \emph{finite structure} $\mathcal{A}=\langle A, R_1,...,R_m\rangle$ over $\sigma$ consists of a finite set $A$, which is called the \emph{universe} of $\mathcal{A}$ and relations $R_1$,...,$R_m$ of arities $k_1,..,k_m$ on $A$, which are interpretations of  the corresponding relation symbols. We may write that $\arity(R_i)=k_i$ or that $R_i$ is a $k_i$-ary relation. We define the \emph{size of the structure}, denoted by $|\mathcal{A}|$ or $|A|$, to be the size of its universe.

A  \emph{finite ordered structure} is a  finite structure with an extra relation $\leq$, which is interpreted as a total order on the elements of the universe.
\end{definition}

In the sequel, $\mathcal{A}$ denotes a finite ordered structure unless otherwise specified. For convenience, we use letters $B, C, P, R, S$, and so on, to denote both relation symbols and their interpretations. 

\begin{example}
 % For example, t
 The vocabulary of binary strings is $\sigma_{bs} = \{\leq^2, B^1\}$. Binary string $x=00101$ corresponds to the  structure $\mathcal{A} = \langle \{0,1,...,4\},\leq, B=\{2,4\}\rangle$, where relation $B$ represents the positions where $x$ is one, and $\leq$ is the restriction of the usual linear order of the naturals on $\{0,1,...,4\}$. Moreover, $|\mathcal{A}|=5$.    
\end{example}
 
\subsubsection{\fo{} and \SO} 
The following grammar defines first-order formulae over $\sigma$:
\begin{align*}
    \varphi::= &~  R(x_1,\dots, x_k) &\mid~& (x_1 = x_2) &\mid~&\top &\mid~& \perp&\mid~& (\neg\varphi) &\mid~& (\varphi\wedge \varphi) &\mid~& (\varphi\vee \varphi)  &\mid~& (\varphi\rightarrow \varphi) \\
 &\mid~ (\forall x \varphi) &\mid~& (\exists x \varphi)
 \end{align*}
 % $$\begin{aligned}
 % \varphi::= & ~  R(x_1,\dots, x_k) ~\mid~ (x_1 = x_2) ~\mid~\top ~\mid~ \perp~\mid~ (\neg\varphi) ~\mid~ (\varphi\wedge \varphi) ~\mid~ (\varphi\vee \varphi)  ~\mid~ (\varphi\rightarrow \varphi) \\
 % &~\mid~ (\forall x \varphi) ~\mid~ (\exists x \varphi)
 % \end{aligned}$$
 where $x_1,\dots, x_k$ are first-order variables, and $R\in\sigma$ is a relation symbol of arity $k$. For convenience and clarity, we omit function and constant symbols from the syntax of \fo. We include $\top$ and $\perp$, which are the logical constants for truth and falsehood.
An occurrence of a variable $x$ is said to be bound if that occurrence of $x$ lies within the scope of at least one of either $\exists x$ or $\forall x$.  Finally, $x$ is bound in $\varphi$ if all occurrences of $x$ in $\varphi$ are bound. A first-order formula with no free variable occurrences is called a first-order \emph{sentence}.

In addition to the syntax of \fo, \SO{} includes a new sort of variables, namely second-order variables, that range over relations, are denoted by uppercase letters, and each of them  has an arity. If $X$ is a second-order variable of arity $k$, and $x_1,\dots, x_k$ are first-order variables, then $X(x_1,\dots, x_k)$ is  a second-order formula. In \SO{}, existential and universal quantification over second-order variables is also allowed.  The fragment of \SO{} consisting only of existential second-order formulae is called existential second-order logic and is abbreviated as \existsSO.

We use the usual $\mathcal{A}, v, V\models\varphi$ interpretation of an \SO-formula $\varphi$, given a structure $\mathcal{A}$ and first- and second-order assignments $v$ and $V$, respectively. When $\varphi$ has no free first- or second-order variables, 
%is an \fo-formula, then we can omit $V$, and when $\varphi$ is a sentence, 
we can omit $v$ or $V$, respectively.
We refer the reader to~\cite{enderton1972} for 
% the semantics 
a more extensive presentation 
of \fo and \SO{}.

\subsubsection{Quantitative Second-Order logic}

The logical symbols of Quantitative Second-Order logic, denoted by \qso, include all the logical symbols of \fo, an infinite set of second-order variables, and the quantitative quantifiers \quantsigma and \quantpi for sum and product quantification, respectively. 
%In most cases, first- and second-order variables are denoted by lowercase letters $x,y,z,\dots$, and uppercase letters $X,Y,Z,\dots$, respectively. However, we also use other letters to denote them. 
The arity of a second-order variable $X$ is denoted by  $\arity(X)$.
The set of \qso formulae over $\sigma$ are defined by the following grammar:
\begin{equation}
\alpha::= ~ \varphi ~\mid~ s ~\mid~ (\alpha+\alpha) ~\mid~ (\alpha\cdot\alpha) ~\mid~ \quantsigma x. \alpha ~\mid~ \quantpi x.\alpha ~\mid~ \quantsigma X.\alpha  ~\mid~ \quantpi X.\alpha
\label{alpha1}
\end{equation}
where $\varphi$ is an \SO{} formula over $\sigma$, $s\in\mathbb{N}$, $x$ is a first-order variable, and $X$ is a second-order variable.

As above, the definitions of structures and formulae are parameterized with respect to a vocabulary $\sigma$. 
%From now on, unless we specify otherwise, we consider a fixed such $\sigma$. 
When we write logic \loglam over $\sigma$, we mean the set of  \loglam formulae over $\sigma$. 
A formula $\alpha$ in \qso is a sentence if every variable occurrence in $\alpha$ is bound
 by a first-order, second-order, or quantitative quantifier.
%A formula $\alpha$ in \qso is a sentence if every variable occurrence in $\alpha$ is bound, where an occurrence of variable $x$ is bound when it is in the scope of at least one of $\forall x$, $\exists x$, $\quantsigma x$, $\quantpi x$.

The syntax of \qso formulae is divided in two levels: the first level is composed by \SO formulae over $\sigma$ and the second level is made by counting operators of addition and multiplication. By parameterizing one or both of these levels, different set of formulae and different counting classes are defined.  \sigqso denotes the fragment of \qso formulae where $\quantpi$ is not allowed; \sigqsofo is the set of \sigqso formulae obtained by restricting $\varphi$ in (\ref{alpha1}) to be an \fo formula.
    
Let $\mathcal{A}$ be a finite ordered structure over $\sigma$, $v$ and $V$ be a first- and a second-order assignment, respectively, for $\mathcal{A}$. Then the evaluation of a \qso formula $\alpha$ over $\mathcal{A}, v,$ and $V$ is defined as a function $\llbracket \alpha\rrbracket $ that on input $\mathcal{A}, v,$ and $V$ returns a number in $\mathbb{N}$. We refer the reader to~\cite[p.\ 5]{Arenas} for the definition of the semantics of \qso formulae. In the case that $\alpha$ is a sentence, the term $\llbracket \alpha\rrbracket (\mathcal{A})$ is used to denote $\llbracket \alpha\rrbracket (\mathcal{A}, v, V)$ for some arbitrary $v$ and $V$. An example can be seen in the second row of Table~\ref{descriptive examples}.
%The function $\llbracket \alpha\rrbracket $ is recursively defined in Table \ref{semantics1}. 

\begin{table}[H] 
\begin{tabular}{ | m{3.5cm} | m{9.6cm}| } 
\hline
\makecell{Fagin~\cite{Fa74}\\ $\existsSOclass = \NP$} & \makecell{$G$ contains a clique of any size iff \\ \vspace{2mm}
$\mathcal{G}\models\exists X \forall x\forall y \big(X(x)\wedge X(y)\wedge x\neq y)\rightarrow E(x,y)$}\\
% \hline
% \makecell{Saluja et al.~\cite{Saluja}\\ $\#\textsf{FO}=\sP$} & \makecell{$\#(\text{cliques of } G)=$ \\ \vspace{2mm}
% $|\{ X  ~\mid ~ \mathcal{G}\models\forall x\forall y \big(X(x)\wedge X(y)\wedge x\neq y)\rightarrow E(x,y)\}|$}\\
\hline
\makecell{Arenas et al.~\cite{Arenas}\\$\sigqsofoclass=\sP$} & \makecell{$\#(\text{cliques of } G)=$ \\ \vspace{2mm}
$\llbracket \quantsigma X.\forall x\forall y \big(X(x)\wedge X(y)\wedge x\neq y)\rightarrow E(x,y)\rrbracket (\mathcal{G})$}\\
\hline
\end{tabular}
\vspace{1mm}
\caption{The decision and counting versions of the \textsc{Clique} problem expressed in the logics \existsSO
and \sigqsofo, respectively. The input structure $\mathcal{G}$ is over vocabulary $\langle E^2\rangle$ with a binary relational symbol $E$ representing the edge relation. $X$ is a unary second-order variable that represents the possible subsets of vertices. The quantifier $\quantsigma$ adds 1 for every interpretation of $X$ that encodes a clique. }
%\textcolor{orange}{X is a second order variable that represents the possible subsets of vertices. <X> means... and $\Sigma X$ means...}}
\label{descriptive examples}
\end{table}

% A \qso formula $\alpha$ is said to be a sentence if it does not have any free variable, that is every variable in $\alpha$ is under the scope of a usual or a quantitative quantifier. Notice that if $\alpha$ is a \qso sentence over $\sigma$, then for every finite ordered structure $\mathcal{A}$,
% first-order assignments $v_1$, $v_2$ and second-order assignments $V_1$, $V_2$ for $\mathcal{A}$, it holds
% that $\llbracket \alpha\rrbracket (\mathcal{A}, v_1, V_1) = \llbracket \alpha\rrbracket (\mathcal{A}, v_2, V_2)$. Thus, in such a case we use the term $\llbracket \alpha\rrbracket (\mathcal{A})$ to denote $\llbracket \alpha\rrbracket (\mathcal{A}, v, V)$ for some arbitrary first- and second-order assignments $v$ and $V$, respectively, for $\mathcal{A}$. 

% For example, the problem of counting cliques in a graph $\mathcal{G}$ can be expressed as $\llbracket \alpha_{clique}\rrbracket (\mathcal{G})$, where $\alpha_{clique} = \quantsigma X. \forall x \forall y (X(x)\wedge X(y)\wedge x\neq y)\rightarrow E(x,y)$.

There is a standard mapping from finite ordered structures to strings over $\{0,1\}$, which is used to encode any $\mathcal{A}$ (see for example~\cite[Chapter 6]{Libkin04}). This mapping can be extended to encode triples $(\mathcal{A},v,V)$ using space polynomial in $|A|$. We denote by $\enc(\mathcal{A})$ (resp.\  $\enc(\mathcal{A},v,V)$) the encoding of $\mathcal{A}$ (resp.\ $(\mathcal{A},v,V)$). We always assume that a TM $M$ takes as input the encoding of $\mathcal{A}$ (or $(\mathcal{A},v,V)$), even if we write $M(\mathcal{A})$ (or $M(\mathcal{A},v,V)$) for the shake of brevity.

\begin{definition}\label{class def}
 We say that  $f\in\sigqsofoclass$ (resp.\ $f\in\qsoclass$)  if there exists a \sigqsofo (resp.\ \qso) formula $\alpha$ such that $f(\enc(\mathcal{A}))=\llbracket \alpha\rrbracket (\mathcal{A})$, for every  $\mathcal{A}$.
 %over $\sigma$.
\end{definition}

\begin{remark}
 \sigqsofo (resp.\ \qso etc) is a set of logical formulae, whereas \sigqsofoclass (resp.\ \qsoclass etc.) is a class of functions. For every logic \loglam, we can define a corresponding class of functions as above, and denote it by \classlam. 
 %We say that a logic $\ensuremath{\mathbf{L}}$ captures a computational class $\textsf{C}$ over finite ordered structures, if $\textsf{L}=\textsf{C}$ provided that the inputs to the functions are encodings of  finite ordered structures. 
\end{remark}

% There are logics that can express all \totp{} problems, such as \sigqsofo, which captures \sP\ over finite ordered structures. Also, other logics may define subclasses of \totp, e.g.\ since $\textsf{TQFO(FO)}=\sL$~\cite{Arenas} and $\sL\subseteq\totp$.
% In this section, we define a logic for which we prove that captures \totp.

\begin{definition}\label{def:L-captures-C}
A logic \loglam \emph{captures a  complexity class} \classC, and equivalently $\classlam=\classC$, over finite ordered structures over $\sigma$, if the following two conditions hold:
\begin{enumerate}
    \item For every $f\in \classC$, there is a sentence $\alpha\in\loglam$, such that $f(\mathrm{enc}(\mathcal{A}))=\llbracket \alpha\rrbracket (\mathcal{A})$  for every finite ordered structure $\mathcal{A}$ over $\sigma$.
    \item For every sentence $\alpha\in\loglam$, there is a function $f\in\classC$, such that $\llbracket \alpha\rrbracket (\mathcal{A})=f(\mathrm{enc}(\mathcal{A}))$ for every finite ordered structure $\mathcal{A}$ over $\sigma$.   
\end{enumerate}
Moreover, \loglam captures \classC over finite ordered structures if \loglam captures \classC over finite ordered structures over $\sigma$, for every $\sigma$.
\end{definition}

\begin{proposition}[\cite{Arenas}]
$\sigqsofoclass=\sP$ over finite ordered structures.
\end{proposition}

In all cases that we consider in this paper, the initial configuration of a TM is \fo{} definable~\cite{Immerman} and therefore, to prove that \loglam captures \textsf{C}, it suffices to verify conditions 1 and 2 in the definition above for $f(\mathrm{enc}(\mathcal{A},v, V))=\llbracket \alpha\rrbracket (\mathcal{A},v, V)$, where $v, V$ encode the initial configuration of a TM that corresponds to $f$.

Finally, in the sequel, we use the fact that $\mathcal{A},v,V\models \varphi$ can be decided in deterministic logarithmic space, if $\varphi$ is an $\fo$ formula, and in deterministic polynomial space, if $\varphi\in\SO$, for every finite structure $\mathcal{A}$~\cite{Immerman}.

% \begin{theorem}[\cite{Arenas}]
% $\sP=\ensuremath{\mathsf{\Sigma QSO(FO)}}$ over finite ordered structures.
% \end{theorem}

% \begin{table}[h!]

% \centering

% $\llbracket \varphi\rrbracket (\mathcal{A},v,V)=\begin{cases}
% 1, \text{ if } \mathcal{A}\models\varphi\\
% 0, \text{ otherwise}
% \end{cases}$

% \vspace{3mm}
% $\llbracket s\rrbracket (\mathcal{A},v,V)=s$

% \vspace{3mm}

% $\llbracket \alpha_1+\alpha_2\rrbracket (\mathcal{A},v,V)=\llbracket \alpha_1\rrbracket (\mathcal{A},v,V)+\llbracket \alpha_2\rrbracket (\mathcal{A},v,V)$

% \vspace{3mm}

% $\llbracket \alpha_1\cdot\alpha_2\rrbracket (\mathcal{A},v,V)=\llbracket \alpha_1\rrbracket (\mathcal{A},v,V)\cdot\llbracket \alpha_2\rrbracket (\mathcal{A},v,V)$

% \vspace{3mm}

% $\displaystyle\llbracket \quantsigma x. \alpha\rrbracket (\mathcal{A},v,V)=\sum_{a\in A}\,\llbracket \alpha\rrbracket (\mathcal{A},v[a/x],V)$

% \vspace{3mm}

% $\displaystyle\llbracket \quantpi x. \alpha\rrbracket (\mathcal{A},v,V)=\prod_{a\in A}\,\llbracket \alpha\rrbracket (\mathcal{A},v[a/x],V)$

% \vspace{3mm}

% $\displaystyle \llbracket \quantsigma X. \alpha\rrbracket (\mathcal{A},v,V)=\sum_{B\subseteq A^{\arity(X)}}\llbracket \alpha\rrbracket (\mathcal{A},v,V[B/X])$

% \vspace{3mm}

% $\displaystyle \llbracket \quantpi X. \alpha\rrbracket (\mathcal{A},v,V)=\prod_{B\subseteq A^{\arity(X)}}\llbracket \alpha\rrbracket (\mathcal{A},v,V[B/X])$
% \vspace{3mm}
% \caption{The semantics of \qso formulae defined in~\cite{Arenas}.}
% \label{semantics1}
% \end{table}

\section{The quantitative logic \sigmasol}\label{relational semantics}

\subsection{The syntax of the logic \sigmasol}\label{syntax}

%Given a vocabulary $\sigma$, t
The set of \sigmasol formulae over $\sigma$ are defined by the following grammar.
% \begin{equation}
% \textcolor{red}{\alpha:= ~ \varphi 
% ~\mid~ \varphi(\Vec{\underline{X}}) 
% ~\mid~ (\alpha+\alpha) ~\mid~ (\alpha\cdot\alpha) 
% %~\mid~ \quantsigma x. \alpha 
% ~\mid~ \quantsigma \Vec{X}.\alpha}
% \end{equation}
\begin{equation}
\alpha::= ~ x ~ \mid ~ X ~\mid~ \varphi 
%~\mid~ \varphi(\Vec{\underline{B}}) 
~\mid~ (\alpha+\alpha) ~\mid~ (\alpha\cdot\alpha) 
~\mid~ \quantsigma y. \alpha 
~\mid~ \quantsigma Y.\alpha
\label{alpha2}
\end{equation}
% \begin{equation}
% \textcolor{red}{\alpha:=  ~ X ~\mid~ \varphi 
% ~\mid~  (\alpha+\alpha) ~\mid~ (\alpha\cdot\alpha) 
% ~\mid~ \quantsigma x. \alpha 
% ~\mid~ \quantpi y. \alpha 
% ~\mid~ \quantsigma Y.\alpha}
% \end{equation}
where $\varphi$ is in \loglam, $x$, $y$ are  first-order variables, and $X$, $Y$ are second-order variables.

The syntax of logic \sigmasol is the same as that of \sigmaqsol, where a formula can also be a first- and second-order variable, but not a natural number $s\in\mathbb{N}$. The logic \sigmafol is the fragment of \sigmasol in which $\quantsigma$ is not allowed over second-order variables. Furthermore, we say that a \sigmasol formula is \emph{$x$-free} (resp.\  \emph{$X$-free}) if it is given by  grammar~(\ref{alpha2}) without $x$ (resp.\ $X$).

\begin{notation}\label{not:underline}
We denote $X\cdot\varphi(X)$ (or $\varphi(X)\cdot X$)  by $\varphi(\underline{X})$. 
\end{notation}

\subsection{The semantics of the logic \sigmasol}

We define the semantics of the logic \sigmasol in two phases: a formula $\alpha$ is mapped to a set of strings. Then, the semantic interpretation of formula $\alpha$ is defined to be the size of this set. Formally, $\llbracket \alpha\rrbracket (\mathcal{A},v,V)=|\rel[\alpha](\mathcal{A},v,V)|$, where $\rel[\alpha](\mathcal{A},v,V)$ is recursively defined in Table~\ref{semantics2}. $\rel$ stands for Explicit and  we call $\rel[\alpha](\mathcal{A},v,V)$  the \emph{intermediate semantic interpretation} of formula $\alpha$. Note that $\cup$ and $\circl$ between sets have replaced sum and multiplication of natural numbers, respectively, in the semantics of \qso. $S_1\cup S_2$ is the union of $S_1$ and $S_2$, whereas $S_1\circl S_2$ is \emph{concatenation} of sets of strings lifted from the concatenation operation on strings, that is $S_1\circl S_2=\{x\circl y~\mid~ x\in S_1, y\in S_2\}$. For example, $\{\varepsilon, a_1, a_2a_3\}\circl \{\varepsilon,a_2a_3\} =\{\varepsilon, a_2a_3, a_1, a_1a_2a_3,a_2a_3a_2a_3\}$, where $\varepsilon$ denotes the empty string. In specific, if one of $S_1$, $S_2$ is $\emptyset$, then $S_1\circl S_2=\emptyset$. 
% We call $+$ and $\cdot$ that appear in the syntax of the logic, addition and multiplication, respectively, and their interpretations $\cup$ and $\circl$, \emph{union} and \emph{concatenation}, respectively.
% 
% 
\begin{table}[H]    %% not necessarily for arXiv, but it is usually good to use t, to place the table to the top of the page 
%\footnotesize
\begin{center}
% \centering
%\vspace{2mm}
% $\rel[x](\mathcal{A},v,V)=\{v(x)\}$

% \vspace{3mm}
% $\rel[X](\mathcal{A},v,V)=\{V(X)\}$

% \vspace{3mm}

% $\rel[\varphi](\mathcal{A},v,V)=\begin{cases} \{\varepsilon\}, &\text{if } \mathcal{A}\models\varphi\\
% \emptyset, &\text{otherwise}
% \end{cases}$

% %\vspace{3mm}
% %$\llbracket s\rrbracket (\mathcal{A},v,V)=s$

% \vspace{3mm}

% $\rel[\alpha_1+\alpha_2](\mathcal{A},v,V)=\rel[\alpha_1](\mathcal{A},v,V)\cup\rel[\alpha_2](\mathcal{A},v,V)$

% \vspace{3mm}

% $\rel[\alpha_1\cdot\alpha_2](\mathcal{A},v,V)=\rel[\alpha_1](\mathcal{A},v,V)\circl\rel[\alpha_2](\mathcal{A},v,V)$

% \vspace{3mm}

% $\displaystyle\rel[\quantsigma y. \alpha](\mathcal{A},v,V)=\bigcup_{a\in A}\,\rel[\alpha](\mathcal{A},v[a/y],V)$

% %\vspace{3mm}
% %
% %$\displaystyle\llbracket \quantpi x. \alpha\rrbracket (\mathcal{A},v,V)=\prod_{a\in A}\,\llbracket \alpha\rrbracket (\mathcal{A},v[a/x],V)$

% \vspace{3mm}

% $\displaystyle \rel[\quantsigma Y. \alpha](\mathcal{A},v,V)=\bigcup_{B\subseteq A^{k}}\rel[\alpha](\mathcal{A},v,V[B/Y])$

\begin{align*}
\rel[x](\mathcal{A},v,V)&=\{v(x)\}\\
\rel[X](\mathcal{A},v,V)&=\{V(X)\}\\
\rel[\varphi](\mathcal{A},v,V)&=\begin{cases} \{\varepsilon\}, &\text{if } \mathcal{A}, v, V\models\varphi\\
 \emptyset, &\text{otherwise}
 \end{cases}\\
\qquad\qquad\quad    %% to move it towards the center
\rel[\alpha_1+\alpha_2](\mathcal{A},v,V)&=\rel[\alpha_1](\mathcal{A},v,V)\cup\rel[\alpha_2](\mathcal{A},v,V)\\
\rel[\alpha_1\cdot\alpha_2](\mathcal{A},v,V)&=\rel[\alpha_1](\mathcal{A},v,V)\circl\rel[\alpha_2](\mathcal{A},v,V)\\
\rel[\quantsigma y. \alpha](\mathcal{A},v,V)&=\bigcup_{a\in A}\,\rel[\alpha](\mathcal{A},v[a/y],V)\\
\rel[\quantsigma Y. \alpha](\mathcal{A},v,V)&=\bigcup_{B\subseteq A^{k}}\rel[\alpha](\mathcal{A},v,V[B/Y])
\end{align*}
% 
%\vspace{3mm}
%
%$\displaystyle \llbracket \quantpi X. \alpha\rrbracket (\mathcal{A},v,V)=\prod_{B\subseteq A^{\arity(X)}}\llbracket \alpha\rrbracket (\mathcal{A},v,V[B/X])$
\end{center}
% \vspace{3mm}
\caption{Intermediate semantics of \sigmasol formulae.}
\label{semantics2}
\end{table}

\begin{notation}\begin{enumerate}[(a)]
    \item Let $K$ be a finite set. $\displaystyle K^*:=\bigcup_{n\in\mathbb{N}} K^n$ denotes the set of strings over $K$, $\mathcal{P}(K^*)$ the powerset of $K^*$, and $\varepsilon$ the empty string. 
    %Of course, $\varepsilon=a^0$, for every $a\in K$. 
    \item Let  $\mathcal{A}$ be a finite ordered structure over $\sigma$. Then, $\mathcal{R}_k:=\mathcal{P}(A^k)$  denotes the set of relations over $k$-tuples of the universe (relations of arity $k$).
    \item Given $s\in K^*$, $s[i]$, $1\leq i\leq |s|$, denotes the element in the $i$-th position of $s$, $s[i:]$,  $1\leq i\leq |s|$, denotes the substring of $s$ that starts with the element in the $i$-th position of $s$ and continues to the end of $s$, whereas $s[i:]$,  $i > |s|$, denotes the empty string.
    %s(1) and $s(\mathrm{last})$ denote the element of $K$ that is in the $i$-th and last position of $s$, respectively.
    % \item We write $s_1\preceq_p s_2$ if $s_1$ is a prefix of $s_2$, e.g.\ $B_1B_2\preceq_p B_1B_2B_3$, and $s_1\preceq_s s_2$  if $s_1$ is a suffix of $s_2$, e.g.\ $B_1\preceq_s B_2B_1$.
\end{enumerate}  
\end{notation}

% \begin{definition}\label{concatenation def}
% Let $K$ be a finite set and $S_1,S_2\in \mathcal{P}(K^*)$. We define the \emph{product} of $S_1$ and $S_2$, denoted by $S_1\circl S_2$, to be a set in $\mathcal{P}(K^*)$ such that it contains string $s=a_1\dots a_n$, $n\in\mathbb{N}$, iff there is $i\in\{1,\dots, n\}$ such that $a_1\dots a_i\in S_1$ and $a_{i+1}\dots a_n\in S_2$.
% \end{definition}

\subsubsection{Discussion on the choice of the logics} 
Intuitively, the sets of strings that are values of the intermediate semantics can express sets of computation paths: each string encodes a computation path. As the reader has probably already noticed, union (resp.\ concatenation) of two sets $S_1$ and $S_2$ may result in a set the size of which is not the sum (resp.\ the product) of $|S_1|$ and $|S_2|$. In specific, given two sets that contain encodings of computation paths, their union may be a set with an incorrect number of elements if the same string encodes two different paths. This will not be a problem, since formulae that express problems in the classes \spspace and \totp, yield encodings so that only distinct strings encode distinct paths. Moreover, as it will become clear in Section~\ref{spanl section}, union and concatenation are more suitable than addition and multiplication, respectively, for counting different valid outputs of transducers, which is needed for the classes \spanl and \spanpspace. In that case, sets of strings express sets of outputs. When the union (resp.\ concatenation) of such sets is computed, identical outputs will contribute  one string to the resulting set.

\subsection{The logic \sigmasol with recursion}\label{recursion semantics}

To start with, we add a function symbol $f$ to the syntax of \sigmasol. In this way, we obtain formulae defined by the following grammar:
\begin{equation}
\beta::= ~ x ~ \mid ~ X ~\mid~ \varphi 
~\mid~ f(x_1,\dots,x_k)
~\mid~ (\beta +\beta) ~\mid~ (\beta\cdot\beta) 
~\mid~ \quantsigma y. \beta 
~\mid~ \quantsigma Y.\beta,
\label{alpha function}
\end{equation}
where $f$ is a function symbol and $x_1,\dots,x_k$ are first-order variables. We say that $f$ is a \emph{first-order function symbol} and it has arity $k\in\mathbb{N}$, denoted by $\arity(f)$. In the sequel, a sequence $x_1,\dots,x_k$ of first-order variables is often denoted by $\Vec{x}$.

In like manner, we can define \sigmasol equipped with a \emph{second-order function symbol}, i.e.\ of the form $f(X_1,\dots,X_k)$, where $X_1,\dots, X_k$ are second-order variables.
%, where $\arity(X_i)=l\in\mathbb{N}$, for every $i\in\{1,\dots,k\}$. 
In specific, we consider only second-order function symbols of arity 1, which means of the form $f(X)$.
%, where $\arity(X)=l$.

A \sigmasol formula $\beta(X,f)$ equipped with a second-order function symbol $f(Y)$ is called \emph{arity-consistent} when it has at most one free second-order variable $X$, where $X$ has the same arity as $Y$. We fix an arity $k$ for the first-order function symbol, or the argument of the second-order function symbol.

Let $\mathcal{A}$ be a finite ordered structure over $\sigma$. To extend the semantics of \sigmasol to the case of a formula of
the form $f(x_1,\dots,x_k)$, we say that $F$ is a \emph{first-order function assignment} for $\mathcal{A}$, if  
$F(f):A^k\rightarrow\mathcal{P}(A^*)$. In the case of formula $f(X)$, where $\arity(X)=l$, we say that $F$ is a \emph{second-order function assignment} for $\mathcal{A}$, if $F(f):\mathcal{R}_l\rightarrow\mathcal{P}(K^*)$, where $K$ can be either $A$ or $\bigcup_{i\in\mathbb{N}}\mathcal{R}_i$.
%$\mathcal{R}_l$.

We define $\mathcal{FOF}$ to be the set of functions $h:A^k\rightarrow\mathcal{P}(A^*)$, $\mathcal{SOF}$ the set of  functions $h:\mathcal{R}_k\rightarrow\mathcal{P}(A^*)$, and $\mathcal{RSOF}$ the set of functions 
$h:\mathcal{R}_k\rightarrow\mathcal{P}((\bigcup_{i\in\mathbb{N}}\mathcal{R}_i)^*)$.
%$h:\mathcal{R}_k\rightarrow\mathcal{P}(\mathcal{R}_k^*)$.

 Given first- and second-order assignments $v$ and $V$, respectively, we define:
 $$\rel[f(\Vec{x})](\mathcal{A},v,V,F)=F(f)(v(\Vec{x})) \text{ and } \llbracket f(\Vec{x})\rrbracket (\mathcal{A},v,V,F)=|F(f)(v(\Vec{x}))|.$$ 
The semantics of $f(X)$ are defined in an analogous way.

Now we can add to the syntax of \sigmasol formulae of the form $[\lfp_f\beta](\Vec{x})$ (resp.\  $[\lfp_f\beta](X)$), where $\beta$ is a (resp.\ arity-consistent) \sigmasol formula equipped with a first-order (resp.\ second-order) function symbol $f$.

To define the semantics of $[\lfp_f  \beta](\Vec{x})$, we first define the following two lattices:
\begin{enumerate}
    \item The first lattice is $(\mathcal{P}(A^*),\subseteq)$, i.e.\ it contains all sets of strings over $A$. The bottom element is $\emptyset$ and the top element is the set $A^*$, 
    \item The second lattice is $(\mathcal{F},\leq_F)$: $\mathcal{F}$ is the set of functions $g:A^k\rightarrow\mathcal{P}\big(A^*)$ and for  $g,h\in \mathcal{F}$, $g\leq_F h$ iff $g(\Vec{x})\subseteq h(\Vec{x})$, for every $\Vec{x}$. The bottom element is $g_0$ which takes the  value $\emptyset$ for every $\Vec{x}$, and the top element is $g_{max}$, which is equal to $A^*$ for every $\Vec{x}$.
\end{enumerate}

For an infinite increasing sequence of functions $h_1 \leq_F h_2 \leq_F h_3 \leq_F  \cdots $ from $\mathcal{F}$, we define 
$\lim_{n\to+\infty} h_n := h$, where for every $x \in A^*$, $h(x) = \bigcup_{i\in \mathbb{N} } h_i(x)$. 

We interpret $\beta(\Vec{x}, f)$ as an operator $T_\beta$ on $\mathcal{FOF}$. For every $h\in\mathcal{FOF}$ and $\Vec{x}\in A^k$:
$$T_\beta (h)(\Vec{a})=\rel[\beta(\Vec{x},f)](\mathcal{A},v,V,F)$$
\noindent where $v$ is a first-order assignment for $\mathcal{A}$ such that $v(\Vec{x})=\Vec{a}$ and $F$ is a first-order function assignment for $\mathcal{A}$ such that $F(f)=h$.

In this paper, we introduce formulae of the form $[\lfp_f  \beta](\Vec{x})$ such that the operator $T_\beta$ is monotone on the complete lattice $(\mathcal{F},\leq_F)$. Thus, by the Knaster--Tarski theorem, $T_\beta$ has a least fixed point.

\begin{proposition}\label{monotonicity}
Let $f$ be a first-order function symbol with  $\arity(f)= k$ and $\beta$ be a formula over $\sigma$ defined by grammar~(\ref{alpha function}), such that if $\beta$ contains a function symbol, then this function symbol is $f$. Let also $\mathcal{A}$ be a finite ordered structure over $\sigma$, $\displaystyle h,g:A^k\rightarrow  \mathcal{P}(A^*)$ and $H,G$ be function assignments such that $H(f)=h$ and $G(f)=g$. If $h\leq_F g$, then for every first- and second-order assignments $v$ and $V$, respectively:
$$\rel[\beta](\mathcal{A},v,V,H)\subseteq \rel[\beta](\mathcal{A},v,V,G).$$
\end{proposition}
\begin{proof}
We prove the proposition by induction on the structure of $\beta$. 
    \begin{itemize}
        \item If $\beta$ does not contain a function symbol, then $\rel[\beta](\mathcal{A},v,V,H) = \rel[\beta](\mathcal{A},v,V,G)$.
        \item If $\beta$ is $f(\Vec{x})$, then $\rel[\beta](\mathcal{A},v,V,H)=h(\Vec{a})$ and $\rel[\beta](\mathcal{A},v,V,G)=g(\Vec{a})$,  where $\Vec{a}=v(\Vec{x})$. Since $h\leq_F g$, it holds that $h(\Vec{a})\subseteq g(\Vec{a})$.
        \item If $\beta= \beta_1+\beta_2$ or $\beta= \beta_1\cdot\beta_2$ then by inductive hypothesis, $\rel[\beta_1](\mathcal{A},v,V,H)\subseteq \rel[\beta_1](\mathcal{A},v,V,G)$ and $\rel[\beta_2](\mathcal{A},v,V,H)\subseteq \rel[\beta_2](\mathcal{A},v,V,G)$. Then, we have that $\rel[\beta_1](\mathcal{A},v,V,H)\cup \rel[\beta_2](\mathcal{A},v,V,H)\subseteq\rel[\beta_1](\mathcal{A},v,V,G)\cup \rel[\beta_2](\mathcal{A},v,V,G)$ and $\rel[\beta_1](\mathcal{A},v,V,H)\circl \rel[\beta_2](\mathcal{A},v,V,H)\subseteq\rel[\beta_1](\mathcal{A},v,V,G)\circl \rel[\beta_2](\mathcal{A},v,V,G)$ which means that $\rel[\beta](\mathcal{A},v,V,H)\subseteq\rel[\beta](\mathcal{A},v,V,G)$ in both cases.
        \item  If $\beta=\quantsigma y.\beta'$ or $\beta=\quantsigma Y.\beta'$, then it can be shown as in the previous case that $\rel[\beta](\mathcal{A},v,$\\ $V,H)\subseteq\rel[\beta](\mathcal{A},v,V,G)$. \qedhere
    \end{itemize}
\end{proof}

\begin{corollary}\label{monotone operator}
    For every formula $[\lfp_f  \beta](\Vec{x})$, where $\beta$ is in \sigmasol equipped with a first-order function symbol, operator $T_{\beta}$ is monotone on the complete lattice $(\mathcal{FOF},\leq_F)$. In other words, for every $h,g\in \mathcal{FOF}$, if $h\leq_F g$, then $T_\beta(h)\leq_F T_\beta(g)$.
\end{corollary}

To compute the least fixed point of $T_\beta$, let us consider the sequence of functions $\{h_i\}_{i\in\mathbb{N}}$, $h_i:A^k\rightarrow\mathcal{P}\big(A^*)$, where $h_0(\Vec{a})=\emptyset$ for every $\Vec{a}\in A^k$, and $h_{i+1}:=T_\beta(h_i)$, for every $i\in\mathbb{N}$. 
%Since $T_\beta$ has a least fixed point, there exists some $n\in\mathbb{N}$ such that $h_{n+1}(\Vec{x})=h_n(\Vec{x})$ for every $\Vec{x}\in A^k$, and thus $h_j=_F h_n$ for every $j \geq n$. 
We define $\mathrm{lfp}(T_\beta):=\lim_{n\to+\infty} h_n$.
%such that $h_n=_F h_{n+1}$.
Finally,
$$\rel[ \,[\lfp_f  \beta](\Vec{x})\,](\mathcal{A},v,V):= \mathrm{lfp} (T_\beta)(v(\Vec{x}))=\lim_{n\to+\infty} h_n(v(\Vec{x})) \text{ and}$$
$$\llbracket \,[\lfp_f  \beta](\Vec{x})\,\rrbracket (\mathcal{A},v,V)=|\lim_{n\to+\infty} h_n(v(\Vec{x}))|.$$
%\text{ such that }  h_n=_{F}h_{n+1}.$$  

The semantics of $[\lfp_f\beta](X)$ are defined in a completely analogous way, where the first lattice is $(\mathcal{P}(A^*),\subseteq)$ (resp.\ $(\mathcal{P}((\bigcup_{i\in\mathbb{N}}\mathcal{R}_i)^*),\subseteq)$), and $T_\beta$ is defined on $\mathcal{SOF}$ (resp.\ $\mathcal{RSOF}$). $T_\beta$ can be proven to be monotone on $(\mathcal{SOF},\leq_F)$ (resp.\ $(\mathcal{RSOF},\leq_F)$) and the semantics of $[\lfp_f  \beta](X)$ is defined to be
$\rel[ \,[\lfp_f  \beta](X)\,](\mathcal{A},v,V):= \mathrm{lfp} (T_\beta)(V(X))=\lim_{n\to+\infty} h_n(V(X)) \text{ and}$
$\llbracket \,[\lfp_f  \beta](X)\,\rrbracket (\mathcal{A},v,V)=|\lim_{n\to+\infty} h_n(V(X))|$.
%such that  $h_n=_F h_{n+1}$.  

The logics we define below are fragments of \sigmaqsou with recursion. Given a formula $[\lfp_f\beta](\Vec{x})$ or $[\lfp_f\beta](X)$ in any of them, operator $T_\beta$ is monotone on the complete lattice $(\mathcal{F},\leq_F)$, where $\mathcal{F}$ can be $\mathcal{FOF}$, $\mathcal{SOF}$, or $\mathcal{RSOF}$.

\begin{remark}
The name of a logic with recursion will be of the form \genlog, where $\lone\in\{\folower,\solower\}$ indicates that function symbol $f$ is over first- or second-order variables, respectively, $\ltwo\in\{\folower,\solower\}$ means that quantifier $\quantsigma$ is over first- or second-order variables, respectively, and $(\lthree)$ means that $\varphi$ in~(\ref{alpha2}) is in \lthree.
\end{remark}

\subsubsection{Discussion on the choice of the logics (cont'd)} In~\cite{Arenas} only first-order function symbols were considered and they were interpreted as functions $h:A^k\rightarrow\mathbb{N}$. Then, lattice $(\mathcal{F},\leq_F)$ is not complete and the least fixed point of $T_\beta$ was defined by considering the supports of functions in $\mathcal{F}$. For more details we refer the reader to~\cite[Section 6]{Arenas}. By defining functions in $\mathcal{F}$ to take values in $\mathcal{P}(A^*)$ (or $\mathcal{P}((\bigcup_{i\in\mathbb{N}}\mathcal{R}_i)^*)$), lattice $(\mathcal{F},\leq_F)$ becomes complete, and the definition of the least fixed point of $T_\beta$ is straightforward.

Note that in~\cite{Arenas}, the class of counting versions of \NL problems, namely \sL, was characterized by the logic \qfo with recursion defined by an operator called \emph{path}. Operator path can be seen as the counting version of the transitive closure operator that was used to capture \NL in~\cite{Immerman86,Immerman88}. It was then conjectured that by using the path operator over second-order function symbols would provide an alternative logical characterization of \spspace (equivalently, \fpspace), which accords also with the fact that  $\PSPACE=\sotcclass$ over finite ordered structures~\cite{HarelP84}. Likewise, adding second-order function symbols and a least fixed point on them to the logics defined presently, leads to a logical characterization of \spspace in a natural way.

\subsection{The length of strings mapped to a \sigmasol formula}

Let $\mathcal{A}$ be a finite ordered structure over $\sigma$ and $\alpha\in\sigmasol$. In this subsection we show that any string $s \in \rel[\alpha](\mathcal{A},v,V)$ is of bounded length. 

The length of $\alpha$, denoted by $|\alpha|$, is defined recursively as follows: $|x|=|X|=|\varphi|:=1$, $|\alpha_1+\alpha_2|=|\alpha_1\cdot\alpha_2|:=|\alpha_1|+|\alpha_2|+1$, and $|\Sigma y.\alpha'|= |\Sigma Y.\alpha'|:=|\alpha'|+1$. 

The length of $s\in A^*\cup (\bigcup_{i\in\mathbb{N}}\mathcal{R}_i)^*$, denoted by $|s|$, is also defined recursively: $|\varepsilon| := 0$, $|v(x)|=|V(X)| := 1$, and $|s_1\circl s_2|:=|s_1|+|s_2|$. Moreover, $s$ can be encoded as follows: for any first-order variable $x$, $v(x)$ is an element of the universe $A$, and so it can be encoded using $\log |A|$ bits. For any second-order variable $X$ with $\arity(X)=k$, $V(X)$ is a set of $k$-tuples over $A$. Consider the lexicographic order on $k$-tuples over $A$ induced by the total order on the elements of $A$. Then $V(X)$ can be encoded by the binary string of length $|A|^k$ that its  $i$-th position is equal to $1$ iff the $i$-th smallest $k$-tuple belongs to $V(X)$. Finally, $s_1\circl s_2$ can be encoded by concatenating the encoding of $s_1$ by that of $s_2$. We denote the \emph{encoding} of $s$ by $\enc(s)$. It is also an immediate consequence that $|\enc(s)|\leq |s|\cdot \log |A|$, if $s\in A^*$, and $|\enc(s)|\leq |s|\cdot |A|^k$, if $s\in (\bigcup_{1\leq i\leq k}\mathcal{R}_i)^*$.

\begin{lemma}\label{lem:bounded-length-a}
Let $\alpha$ be a \sigmasol formula over $\sigma$. 
For every finite ordered structure $\mathcal{A}$ over $\sigma$, $v$, and $V$, and every $s \in \rel[\alpha](\mathcal{A},v,V)$, 
%it holds that 
$|s| \leq |\alpha|$. Moreover,
\begin{enumerate}[(a)]
    \item if $\alpha$ is an $X$-free formula, then $|\enc(s)|\leq |\alpha|\cdot \log |A|$, and
    \item if $\alpha$ is an $x$-free formula,
    %that contains only second-order variables of arity $k$,
    then $|\enc(s)|\leq |\alpha|\cdot \mathrm{poly}(|A|)$.
\end{enumerate} 
\end{lemma}
\begin{proof}
It can be proven that $|s| \leq |\alpha|$ by straightforward structural induction on $\alpha$. Claim~(a) 
%(resp.\ (b)) 
is a direct consequence of the fact that $s \in A^*$. For claim~(b), note that any second-order variable in $\alpha$ has arity at most $m\in\mathbb{N}$, for some $m\in\mathbb{N}$, which implies that $s\in(\bigcup_{1\leq i\leq m} \mathcal{R}_i)^*$, and hence $|\enc(s)|\leq |\alpha|\cdot |A|^m$.
\end{proof}

% \begin{lemma}\label{lem:space-bound-tocheck-a}
% Let $\alpha$ be a $\ensuremath{\mathbf{\Sigma SO(\underline{L})}}$ formula over $\sigma$. 
% There is a deterministic TM $M_\alpha$, such that takes as input 
% $(\mathcal{A},v,V)$, and a string $s$, and decides if $s \in \rel[\alpha](\mathcal{A},v,V)$, using at most $\mathcal{O}(|\enc(s)| + T)$ space, where $T$ is the time needed to check whether $\mathcal{A}\models\varphi$ for any $\varphi\in\ensuremath{\mathbf{L}}$.
% \end{lemma}
% \begin{proof}
% $M_\alpha$ can be defined recursively on $\alpha$ and $|s|$.
% If $|s|>|\alpha|$ then $M_\alpha$ can reject immediately by Lemma \ref{lem:bounded-length-a};  as such, we can assume that $|s|$ is constant, so the recursion has constant depth.
% Taking this into account, the construction of $M$ is straightforward. 
% We only describe the case where $\alpha =\alpha_1 \cdot \alpha_2$. 
% In that case, $M_\alpha$ uses a for-loop:\\
% %\begin{verbatim}
% \verb|for all| $pq$ \verb|==| $s$ \verb|do:|\\
% \verb|    if | $M_{\alpha_1}$\verb|(|$p$,$\mathcal{A},v,V$\verb|) and| $M_{\alpha_2}$\verb|(|$q$,$\mathcal{A},v,V$\verb|) then accept|\\
% \verb|reject|
% %\end{verbatim}

% Since $|s|$ is constant, the loop only takes constant time. Space is needed to store the elements of $s$, and in the base case of $\alpha=\varphi$, to check whether $\mathcal{A}\models \varphi$. 
% \end{proof}

\section{Logics that capture \spanl{} and \spanpspace}\label{spanl section}

\subsection{The logic \spanlogic}

The definition of logic \sigmaqfou{} that is used below, can be found in Subsection~\ref{syntax}.
% \begin{equation}
% \alpha::=  ~ x 
% %~\mid~  X 
% ~\mid~ \varphi 
% ~\mid~ (\alpha+\alpha) ~\mid~ (\alpha\cdot\alpha) 
% ~\mid~ \quantsigma y. \alpha 
% \label{alpha3}
% \end{equation}
% where $\varphi$ is an \fo{} formula, and $x$, $y$ are  first-order variables.

\begin{definition}\label{def:spanl}
We define the logic \spanlogic{} over $\sigma$ to be the set of formulae  $[\lfp_f\beta](\Vec{x})$, where $\beta$ is defined by the following grammar:
 \begin{equation}
\beta::=  ~ \alpha 
~\mid~  f(x_1,\dots, x_k)
~\mid~ (\beta+\beta)
~\mid~ (\alpha\cdot\beta) 
~\mid~ \quantsigma y. \beta
\label{beta1}
\end{equation}
where $\alpha$ is an $X$-free \sigmaqfou{} formula over $\sigma$, $x_1,\dots, x_k,y$ are first-order variables, and $f$ is a first-order function symbol.  
\end{definition}

\begin{remark}\label{rem:spanl-clock}
    Notice that for a formula $[\lfp_f\beta](\Vec{x})\in\spanlogic$, it may be the case that $\llbracket \, [\lfp_f\beta](\Vec{x})\,\rrbracket (\mathcal{A},v,V)=+\infty$ analogously to the fact that the computation of an NLTM  may contain cycles. For the sake of simplicity, to proceed with the proofs of this section, we assume that an NL-transducer $M$ can have infinitely many accepting paths, $acc_M$ can take the value $+\infty$, and \spanl contains functions from $\Sigma^*$ to $\mathbb{N}\cup\{+\infty\}$.
    
    %To address this issue
    To be in accordance with the literature, we can adjust the syntax of \spanlogic{} formulae to express the operation of the clock attached to NLTMs as discussed in Remark~\ref{branchings}. Since the clock imposes a polynomial-time bound on an NLTM, its contents can be encoded by tuples of the universe $A$.
    %a clock which is attached to NLTMs as suggested in~\cite{AJ93}. The clock imposes a polynomial-time bound on an NLTM, and so its contents can be encoded by tuples of the universe $A$.
    Let $\beta(\Vec{x},f)$ be given by grammar~(\ref{beta1}) and contain a function symbol $f(\Vec{y})$. We define formula $\beta_{\textsf{cl}}
    (\Vec{x},\Vec{cl},f)$ to be $\beta(\Vec{x},f)$ where $f(\Vec{y})$ is replaced by $\textsf{clock}(\Vec{x},\Vec{cl},f):=\quantsigma \Vec{cl}'.(\Vec{cl}<\mathrm{max})\cdot (\Vec{cl}'=\Vec{cl}+1)\cdot f(\Vec{y},\Vec{cl}')$, where $\Vec{cl},\Vec{cl'}$ are $k$-tuples of  first-order variables for some $k\in\mathbb{N}$. The lexicographic order on $k$-tuples over $A$ induced by the total order on elements of $A$ can be defined in \fo. So, formula $\Vec{cl}'=\Vec{cl}+1$ which describes that $\Vec{cl}'$ is the successor of $\Vec{cl}$, and max that expresses the maximum $k$-tuple with respect to the lexicographic order on $k$-tuples, are \fo{} definable. We can define a restricted version of 
    \spanlogic{} as the set of formulae such that recursion is on $\textsf{clock}(\Vec{x},\Vec{cl},f)$ instead of $f$. The proof of Theorem~\ref{spanl theorem}  can  then be  easily extended to show that this subset of
    %restricted version of
    \spanlogic{} captures \spanl over finite ordered structures.  
    % For simplicity, to proceed with the proofs of this section, we consider the logic \spanlogic{} to be the one defined in Definition~\ref{def:spanl}. In alignment with that, we assume that a transducer $M$ can have infinitely many accepting paths, $acc_M$ can take the value $+\infty$ as well, and \spanl contains functions from $\Sigma^*$ to $\mathbb{N}\cup\{+\infty\}$.   
\end{remark}

\subsection{\spanlogic{} captures \spanl{} over finite ordered structures} 

Let $N$ be an NL-transducer and $\mathcal{A}$ be a finite ordered structure over $\sigma$. 
%A configuration of $N$ consists of the cursor position on the work tape, the contents of the work tape, and the current state. 
The number of all different configurations of $N$ is at most $n^k-1$ for some $k\in\mathbb{N}$, where $n:=|A|$. To encode them, we use $k$-tuples of elements of the universe $A$. Moreover, we need to encode the output symbol, if any, that is produced at each configuration. Since we assume that the output alphabet is $\Sigma=\{0,1\}$, it suffices to use two distinct elements of the universe; we use the minimum element and the successor of the minimum element, which are both definable in \fo.
Below, we 
informally 
write $\varphi(c)$ 
to denote $\varphi(x)$ 
interpreted in $\mathcal{A}$ 
where first-order variable $x$ is assigned $c\in A$.

Then,  formula $[\lfp_f \mathsf{span_L}](\Vec{x})$ expresses the problem of counting the different valid outputs of $N$, where formula $\mathsf{span_L}(\Vec{x},f)$ is given below:

$$\begin{aligned}
\mathsf{span_L}(\Vec{x},f):= \mathsf{acc}(\Vec{x})\,+
\quantsigma \Vec{y}.\quantsigma z. \big(&\mathsf{output}_0(\Vec{x},\Vec{y},\underline{z}) + \mathsf{output}_1(\Vec{x},\Vec{y},\underline{z}) + \\
&\mathsf{next}_0(\Vec{x},\Vec{y}) + \mathsf{next}_1(\Vec{x},\Vec{y}
)\big)\cdot f(\Vec{y})
\end{aligned}$$
 where $z$ is a first-order variable, and $\Vec{x},\Vec{y}$ are $k$-tuples of first-order variables. 

  \begin{wrapfigure}[17]{l} {0.43\textwidth}
    \centering
    \includegraphics[scale=1]{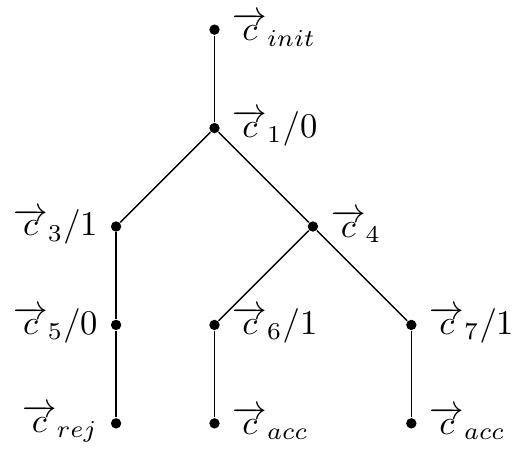}
    \caption{The computation tree of transducer $N$ on input $\enc(\mathcal{A})$. $c/b$ represents that $N$ has entered the configuration encoded by $c$ and has written bit $b$ on the output tape.}
     \label{machine}
\end{wrapfigure}

Interpretations of $z$ and $\Vec{x},\Vec{y}$ will encode  a bit of the output, and configurations of $N$, respectively. Formulae $\mathsf{next}_i(\Vec{c},\Vec{c'})$, $i=0,1$, say that if $N$ is in configuration $\Vec{c}$ and makes non-deterministic choice $i$, then it is in $\Vec{c'}$, and no output symbol is produced. Formulae $\mathsf{output}_i(\Vec{c},\Vec{c'},b)$, $i=0,1$, state that $N$ makes choice $i$ and so it transitions from configuration $\Vec{c}$ to $\Vec{c'}$ and writes the bit encoded by $b$ on the next output cell. When $N$ is in some $\Vec{c}$ that only a deterministic transition can be made, then exactly one of $\mathsf{next}_i(\Vec{c},\Vec{c'})$, $\mathsf{output}_i(\Vec{c},\Vec{c'},b)$, $i=0,1$, is satisfied in $\mathcal{A}$ for a $\Vec{c'}\in A^k$ (and a $b\in A$). Formula $\mathsf{acc}(\Vec{c})$ states that $\Vec{c}$ is the accepting configuration. All aforementioned formulae can be expressed in \fo.

We note that for any $\mathcal{A}$, $v$, and $V$,  $\rel[\,[\lfp_f \mathsf{span_L}](\Vec{x})\,](\mathcal{A},v,V)$ is a set of strings in $A^*$, that encode outputs of $N$. Moreover, identical outputs correspond to the same string in $\rel[\,[\lfp_f \mathsf{span_L}](\Vec{x})\,](\mathcal{A},v,V)$.

\begin{example} 
Consider the computation tree shown in Figure~\ref{machine} which corresponds to a transducer $N$ that on input $\enc(\mathcal{A})$ has three outputs, and $span_N(\enc(\mathcal{A}))=1$. Let $\mathbf{0},\mathbf{1}$ denote the minimum and the successor of the minimum element of $A$ which encode the two bits $0,1$, respectively, that can be written on the output tape of $N$. Then, 
\begin{itemize}\setlength\itemsep{1em}
    \item $\rel[ \,[\lfp_f  \mathsf{span_L}](\Vec{x})\,](\mathcal{A},v[\Vec{c}_{acc}/\Vec{x}])=\{\varepsilon\}$ and  
    $\rel[ \,[\lfp_f  \mathsf{span_L}](\Vec{x})\,](\mathcal{A},v[\Vec{c}_{rej}/\Vec{x}])=\emptyset$,
    \item $\rel[ \,[\lfp_f  \mathsf{span_L}](\Vec{x})\,](\mathcal{A},v[\Vec{c}_1/\Vec{x}])=\emptyset\cup\{\mathbf{1}\}\circl f(\Vec{c}_3)\cup f(\Vec{c}_4) = \{\mathbf{10}\}\circl\emptyset \cup \{\mathbf{1}\}\circl\{\varepsilon\}\cup \{\mathbf{1}\}\circl\{\varepsilon\}= \{\mathbf{1}\}$, and
    \item $\rel[ \,[\lfp_f  \mathsf{span_L}](\Vec{x})\,](\mathcal{A},v[\Vec{c}_{init}/\Vec{x}])=\emptyset\cup\{\mathbf{0}\}\circl f(\Vec{c}_1)=\{\mathbf{01}\}$.
\end{itemize}
Intuitively, the intermediate interpretation of $[\lfp_f  \mathsf{span_L}](\Vec{c})$ is the set of the different valid outputs that are produced during the computation of  $N$ that starts from the configuration encoded by $\Vec{c}$.
\end{example}

\begin{proposition}\label{spanl first inclusion}
Given an NL-transducer $N$, 
%it holds that 
$span_N(\enc(\mathcal{A}))=\llbracket \,[\lfp_f \mathsf{span_L}](\Vec{x})\,\rrbracket(\mathcal{A},v,V)$, for every $\mathcal{A}$, $v$, and $V$, such that $v(\Vec{x})$ encodes  the starting configuration of $N$.
\end{proposition}
\begin{proof}
Let $\mathcal{A}$ be a finite ordered structure over $\sigma$; let also $h_i:A^k\rightarrow\mathcal{P}\big(A^*)$, $i\in\mathbb{N}$, be such that $h_0(\Vec{a})=\emptyset$ for every $\Vec{a}\in A^k$, and $h_{i+1}(\Vec{a})=\rel[\mathsf{span_L}(\Vec{x},f)](\mathcal{A},v[\Vec{a}/\Vec{x}],V,F[h_i/f])$, or in other words $h_{i+1}:=T_\mathsf{span_L}(h_i)$.

% First note that for every encoding $\Vec{c}$ of a configuration $c$ that does not reach the accepting configuration in a finite number of steps,  $h_i(\Vec{c})=\emptyset$ for every $i\in\mathbb{N}$. We omit the proof of this claim, since it is similar to and simpler than the proof that follows.

%%% alternative suggestion:
 % We prove by induction on $i$ that for every encoding $\Vec{c}_{i}$ of a configuration $c_i$ of $N$,  
 % $|h_{i+1}(\Vec{c}_i)|$ is equal to the number of different valid outputs written by $N$ on its output tape, that then reach the accepting state, within up to $i$ transitions from when the computation starts from $c_i$.

We prove by induction on $i$ that for every encoding $\Vec{c}$ of a configuration $c$, $h_{i+1}(\Vec{c})$ is the set of (the encodings of) the different outputs that $N$ writes on its output tape on the runs that start from $c$ and reach an accepting configuration within at most $i$ steps.  

% We prove by induction on $i$ that for every encoding $\Vec{c}_{i}$ of a configuration $c_i$ of $N$ that needs at most $i$ time steps to reach the accepting state along all of its accepting branches, both the following conditions hold: 
% \begin{enumerate}
%     \item $\mathrm{lfp} (T_\mathsf{span_L})(\Vec{c}_i) = h_{i+1}(\Vec{c}_i)$ and
%     \item $|h_{i+1}(\Vec{c}_i)|$ is equal to the number of different valid outputs written by $N$ on its output tape when the computation starts from $c_i$.
% \end{enumerate}
% The proof is as follows.
\begin{description}
    \item[Let $i=0$.] By the definitions of formula $\mathsf{span_L}$ and $h_1$, $h_{1}(\Vec{c})=\{\varepsilon\}$, if $\Vec{c}$ encodes an accepting configuration, and $h_{1}(\Vec{c})=\emptyset$, otherwise.
    % $$h_{1}(\Vec{c})=\begin{cases}
    % \{\varepsilon\}, &\text{if }  \Vec{c}=\Vec{c}_{acc}\\
    % \emptyset, &\text{otherwise}
    % \end{cases}$$   
    \item[Let $i=k$.] By the definitions of formula $\mathsf{span_L}$ and $h_{k+1}$, if $\Vec{c}$ encodes an accepting configuration, then $h_{k+1}(\Vec{c})=h_k(\Vec{c})=\{\varepsilon\}$. If $\Vec{c}$ does not encode an accepting configuration and $N$ cannot make any transition from $c$, then $h_{k+1}(\Vec{c})=\emptyset$. Otherwise, $N$ can make one or two transitions from $c$, and so one of the following holds: 
    \begin{enumerate}[(a)]
        \item $\displaystyle h_{k+1}(\Vec{c})=\{b_0\}\circl h_k(\Vec{c}_0) \cup \{b_1\}\circl h_k(\Vec{c}_1)$,
        \item $\displaystyle h_{k+1}(\Vec{c})= h_k(\Vec{c}_0) \cup \{b_1\}\circl h_k(\Vec{c}_1)$,
        \item $\displaystyle h_{k+1}(\Vec{c})=\{b_0\}\circl h_k(\Vec{c}_0) \cup h_k(\Vec{c}_1)$,
        \item $\displaystyle h_{k+1}(\Vec{c})= h_k(\Vec{c}_0) \cup  h_k(\Vec{c}_1)$, 
        \item $\displaystyle h_{k+1}(\Vec{c})=\{b_0\}\circl h_k(\Vec{c}_0)$,
        or
        \item $\displaystyle h_{k+1}(\Vec{c})=h_k(\Vec{c}_0)$,
    \end{enumerate}
     where $\Vec{c}_j$ denotes the configuration $c_j$ that $N$ transitions to when it makes non-deterministic choice $j$ and $b_j$ denotes the bit that $N$ writes on its output tape when it makes this transition. If no output is produced during the transition from $c$ to $c_j$, for some $i=0,1$, then we are in cases (b)--(d) where concatenation with $\{b_j\}$ is discarded. If $N$ makes a deterministic transition from $c$, then we are in one of the last two cases. We assume case (a), which is the most general. By the inductive hypothesis, both $h_k(\Vec{c}_0)$ and $h_k(\Vec{c}_1)$ contain the different outputs that $N$ writes on its output tape on the runs that start from $c_0$, $c_1$, respectively, and reach an accepting configuration within at most $k-1$ steps. It is straightforward that $h_{k+1}(\Vec{c})$ contains the different outputs produced by $N$ on the runs that start from $c$ and reach an accepting configuration  within at most $k$ steps.
    \end{description}
Consequently,
\begin{itemize}
    \item if the starting configuration $c_{init}$ needs at most $n$ time steps to reach an accepting configuration along all of its accepting branches, then $\mathrm{lfp} (T_\mathsf{span_L})(\Vec{c}_{init}) = h_{n+1}(\Vec{c}_{init})$, and $|h_{n+1}(\Vec{c}_{init})|=span_N(\enc(\mathcal{A}))$,  
    \item  if there is no finite path from  $c_{init}$ to an accepting configuration, then $\mathrm{lfp} (T_\mathsf{span_L})(\Vec{c}_{init})=\emptyset$, which accords with the fact that $span_N(\enc(\mathcal{A}))=0$ in this case, and
    \item if an infinite number of paths that start from $c_{init}$, reach an accepting configuration, then since $N$ has a computation tree with a finite maximum degree, the length of these paths is growing to infinity. Then, $\mathrm{lfp} (T_\mathsf{span_L})(\Vec{c}_{init})=\lim_{n\to +\infty} h_n (\Vec{c}_{init})=\bigcup_i h_i(\vec{c}_{init})$, and from the claim we prove above, $|\lim_{n\to +\infty} h_n (\Vec{c}_{init})|=span_N(\enc(\mathcal{A}))$. Note that although the number of accepting paths is infinite, $span_N(\enc(\mathcal{A}))$ can be either in $\mathbb{N}$ or equal to $+\infty$.\qedhere
    \end{itemize}
% Consequently, if the starting configuration $c_{init}$ needs at most $n$ time steps to reach the accepting configuration, then $\mathrm{lfp} (T_\mathsf{span_L})(\Vec{c}_{init}) = h_{n+1}(\Vec{c}_{init})$, where $\Vec{c}_{init}$ is the encoding of $c_{init}$, and  $|h_{n+1}(\Vec{c}_{init})|=span_N(\enc(\mathcal{A}))$.
 
% Finally, if an infinite number of paths that start from $c_{init}$, reach $c_{acc}$, then since $N$ has a computation tree with a finite maximum degree, 
% %the branching of $N$ is finite,
% the length of these paths is growing to infinity. In an analogous way, we can prove that in this case, $\mathrm{lfp} (T_\mathsf{span_L})(\Vec{c}_{init})=\lim_{n\to +\infty} h_n (\Vec{c}_{init})$, and $|\lim_{n\to +\infty} h_n (\Vec{c}_{init})|=span_N(\enc(\mathcal{A}))$. Note that although the number of accepting paths is infinite, $span_N(\enc(\mathcal{A}))$ can be either in $\mathbb{N}$ or equal to $+\infty$.
% \qedhere 
\end{proof}

Proposition~\ref{spanl first inclusion} is equivalent to $\spanl\subseteq\spanclass$ over finite ordered structures.  The following example demonstrates how two specific \spanl problems are expressed in \spanlogic.

\begin{example}
\begin{enumerate}[(a)]
    \item Let $\mathcal{G}=\langle V, E, \leq\rangle$ be a finite ordered structure that represents a directed graph with a source. Then, $\llbracket\,[\lfp_f \beta](x)\,\rrbracket (\mathcal{G},v,V)$ is equal to the number of sinks in the graph, where $\beta(x,f):=\forall y \neg E(x,y)\cdot x+\quantsigma y.E(x,y)\cdot f(y)$, and $v(x)$  encodes the source of the graph.
    \item Let $\mathcal{N}= \langle Q=\{q_0,\dots, q_{n-1},\ell_1,\dots,\ell_m\}, L, E_0, E_1, \leq \rangle$ be a finite ordered structure that represents an NFA $N$ over the input alphabet $\{0,1\}$, together with $1^m$; $Q$ is the universe, $L=\{\ell_1,\dots,\ell_m\}$ is a relation that distinguishes states of $N$ from the encoding of $1^m$, and $E_i$, $i=0,1$, contains pairs of states of $N$ that are connected through an edge labelled by $i$. Define $\beta(x,y,f)$ to be the following formula:
    $$\mathsf{acc}(x)+ (y\leq \mathrm{max})\cdot\quantsigma x'.\quantsigma y'.(y'=y+1)\cdot \big(E_0(x,x')\cdot \mathrm{min_0}+ E_1(x,x')\cdot \mathrm{min_1}\big)\cdot f(x',y')$$
    where $\mathrm{min_0}$, $\mathrm{min_1}$, and $\mathrm{max}$ express the minimum, the successor of the minimum, and the maximum element of $Q$, respectively. Then, $\llbracket\,[\lfp_f \beta](x,y)\,\rrbracket (\mathcal{N},v,V)$ is equal to the number of strings of length at most $m$ accepted by $N$, where $v(x)$ encodes the starting state of $N$, and $v(y)$ encodes the minimum element that belongs to relation $L$. This problem was defined in~\cite{AJ93} under the name of the \emph{census function} of an NFA, and was shown to be \spanl-complete.
\end{enumerate}
\end{example}

% \begin{lemma}\label{lem:bounded-length-a}
% Let $\alpha$ be an $X$-free \sigmaqfou{} formula over $\sigma$. 
% For every $\mathcal{A}$, $v$, and $V$, and every $s \in \rel[\alpha](\mathcal{A},v,V)$, $|s| \leq |\alpha|$  and  $|\enc(s)|\leq |\alpha|\cdot \log |A|$.
% \end{lemma}
% \begin{proof}
% The first claim can be proven by straightforward structural induction on $\alpha$, and the second one is a direct consequence, observing that $s \in A^*$.
% \end{proof}

Lemma \ref{lem:space-bound-tocheck-a}, Proposition \ref{prop:alpha-in-sharpL-mathching-outputs}, and Corollary \ref{cor:alpha-in-sharpL} demonstrate that log-space Turing machines can verify and evaluate $X$-free \sigmaqfou{} formulae.

\begin{lemma}\label{lem:space-bound-tocheck-a}
Let $\alpha$ be an $X$-free \sigmaqfou{} formula over $\sigma$. 
There is a deterministic TM $M_\alpha$ that takes as input 
 $\enc(\mathcal{A},v,V)$, and a string $s \in A^*$, and decides if $s \in \rel[\alpha](\mathcal{A},v,V)$, using at most $\mathcal{O}(\log |A|)$ space.
\end{lemma}
\begin{proof}
$M_\alpha$ can be defined recursively on $\alpha$ and $|s|$.
If $|s|>|\alpha|$ then $M_\alpha$ can reject immediately by Lemma \ref{lem:bounded-length-a}, using at most $\mathcal{O}(\log\log |A|)$ space;  as such, we can assume that $|s|$ is constant, so the recursion has constant depth.
Taking this into account, the construction of $M_\alpha$ is straightforward. 
We only describe two cases. Recall that $\alpha'(a)$ denotes formula $\alpha'(y)$ interpreted in $\mathcal{A}$, such that $y$ is assigned $a\in A$.
\begin{itemize}
    \item $\alpha =\alpha_1 \cdot \alpha_2$: in that case, $M_\alpha$ uses the following for-loop.\\
%\begin{verbatim}
\verb|for all| $pq$ \verb|==| $s$ \verb|do:|\\
\verb|    if | $M_{\alpha_1}$\verb|(|$p$,$\mathcal{A},v,V$\verb|) and| $M_{\alpha_2}$\verb|(|$q$,$\mathcal{A},v,V$\verb|) then accept|\\
\verb|reject|
%\end{verbatim}
\item $\alpha =\quantsigma y.\alpha'$: $M_\alpha$ proceeds as follows.\\
\verb|for all| $a\in A$ \verb|do:|\\
\verb|    if | $M_{\alpha'(a)}$\verb|(|$s$,$\mathcal{A},v,V)$\verb| then accept|\\
\verb|reject|
\end{itemize}

Since $|s|$ is constant, the first loop only takes constant time. In the case of $\alpha =\quantsigma y.\alpha'$, $M_\alpha$ reuses space; it stores an element $a\in A$, runs $M_{\alpha'(a)}$, and it clears its work tape before moving to the next element of $A$. It is not hard to see that in all other cases logarithmic space suffices.   
\end{proof}

\begin{proposition}\label{prop:alpha-in-sharpL-mathching-outputs}
For every $X$-free \sigmaqfou{} formula $\alpha$ over $\sigma$, there is an NL-transducer $M$, that on input $\enc(\mathcal{A},v,V)$ has exactly one accepting run for each $s \in \rel[\alpha](\mathcal{A},v,V)$, on which it outputs $enc(s)$, and no other accepting runs.
\end{proposition}
\begin{proof}
Using Lemmata \ref{lem:bounded-length-a} and \ref{lem:space-bound-tocheck-a}, $M$ can non-deterministically guess every string $s \in A^*$ of length at most $|\alpha|$ and deterministically check whether $s \in \rel[\alpha](\mathcal{A},v,V)$. If $s \in \rel[\alpha](\mathcal{A},v,V)$, it accepts and outputs $\enc(s)$.
\end{proof}

\begin{corollary}\label{cor:alpha-in-sharpL}
Let $\alpha$ be an $X$-free \sigmaqfou{} formula over $\sigma$. There is an NLTM $M$, such that $acc_{M}(\enc(\mathcal{A},v,V))=\llbracket \alpha\rrbracket (\mathcal{A},v,V)$ for every $\mathcal{A},v$ and $V$.
\end{corollary}

We now prove that $\spanclass \subseteq \spanl$.

\begin{proposition}\label{spanl second inclusion}
Let $[\lfp_f\beta](\Vec{x})$ be an \spanlogic{} formula over $\sigma$. There is an NL-transducer $M_\beta$, such that  $span_{M_\beta}(\enc(\mathcal{A},v,V))=\llbracket \,[\lfp_f\beta](\Vec{x})\,\rrbracket (\mathcal{A},v,V)$, for every 
$\mathcal{A},v$ and $V$.
%holds.
\end{proposition}
\begin{proof}
%Let $\alpha=[\lfp_f \beta](\Vec{x})$ in $\spanlogic$; 
Let $[\lfp_f\beta](\Vec{x}) \in\spanlogic$. The corresponding NL-transducer $M_\beta(\mathcal{A},v,V)$ calls $MSp_\beta^{sub}(\beta,\mathcal{A},v,V)$,
which is defined in Algorithm \ref{alg:spanL}. 
If $\beta$ does not contain a function symbol, then $\llbracket \,[\lfp_f\beta](\Vec{x})\,\rrbracket (\mathcal{A},v,V)=\llbracket \beta\rrbracket (\mathcal{A},v,V)$. By Proposition~\ref{prop:alpha-in-sharpL-mathching-outputs}, there is an NL-transducer $M$, such that $span_M(\enc(\mathcal{A},v,V))=\llbracket \beta\rrbracket (\mathcal{A},v,V)$. In this case, define $M_\beta$ to be identical to $M$.
Similarly, for any subformula $\alpha$ of $\beta$ without function symbols, we can define 
$M_\alpha$ to be the NL-transducer associated with $\alpha$ from the proof of Proposition~\ref{prop:alpha-in-sharpL-mathching-outputs}.

\begin{algorithm}
\caption{NLTM $MSp_\beta^{sub}$}\label{alg:spanL}
\DontPrintSemicolon
\KwIn{$\gamma, \mathcal{A},v,V$}
\If{$\gamma==\alpha$ has no function symbol}{
simulate 
 transducer  $M_\alpha$ from Proposition~\ref{prop:alpha-in-sharpL-mathching-outputs}\;
}
\If{$\gamma==f(\vec{y})$}{
simulate $MSp_\beta^{sub}(\beta, \mathcal{A},v[v(\vec{y})/\vec{x}],V)$\;
}
\If{$\gamma==\gamma_1 + \gamma_2$}{
non-deterministically choose $\gamma'\in{\gamma_1,\gamma_2}$\;
simulate $MSp_\beta^{sub}(\gamma', \mathcal{A},v,V)$\;
}
\If{$\gamma==\alpha \cdot \gamma'$}{
\For{$s \in A^*$ where $|s| \leq |\alpha|$}{
\If{$s \in \rel[\alpha](\mathcal{A},v,V)$}{
simulate $MSp_\beta^{sub}(\gamma', \mathcal{A},v,V)$\;
}
}
}
\If{$\gamma==\sum y. \gamma'$}{
non-deterministically choose $a\in A$\;
simulate $MSp_\beta^{sub}(\gamma', \mathcal{A},v[a/y],V)$\;
}
\end{algorithm}

Let $\gamma$ be a subformula of $\beta$. We observe that 
%the input of 
$MSp_\beta^{sub}$($\gamma, \mathcal{A},v,V$) requires space logarithmic with respect to $|A|$, and each call does not need to retain any information from previous calls. Therefore, $M_\beta$($\mathcal{A},v,V$) runs using logarithmic space with respect to its input (i.e.\ the size of the encoding of $(\mathcal{A},v,V)$).

Let
$h_i:A^k\rightarrow\mathcal{P}\big(A^*)$, $i\in\mathbb{N}$, be such that  $h_0(\Vec{a})=\emptyset$ for every $\Vec{a}\in A^k$, and $h_{i+1}(\Vec{a})=\rel[\mathsf{\beta}(\Vec{x},f)](\mathcal{A},v[\Vec{a}/\Vec{x}],V,F[h_i/f])$, or in other words $h_{i+1}:=T_\mathsf{\beta}(h_i)$.
%Let $k \geq 0$ be minimal, such that $h_k = h_{k+1}$.

We observe that $M_\beta$ only outputs encodings of strings in $A^*$. Furthermore, for the purposes of this proof and for our convenience, we define the recursion depth of a call of $MSp_\beta^{sub}$ by only taking into account the recursive calls of 
$MSp_\beta^{sub}$\verb|(|$\beta, \mathcal{A},v[v(\vec{y})/\vec{x}],V$\verb|)| (the case of $\gamma==f(\vec{y})$ in the description of $MSp_\beta^{sub}$).
To complete the proof of the proposition, we prove that for every $o \in A^*$, $o \in \rel[\beta](\mathcal{A},v,V,F[h_i/f])$ if and only if $\enc(o)$ is an output of an accepting run of $MSp_\beta^{sub}(\beta,\mathcal{A}, v, V)$
at recursion depth at most $i$.
%To do so, we prove the following claim: 
%
%    \noindent 
%    \emph{Claim:} For every $o \in A^*$, subformula $\gamma$ of $\beta$, and $i \geq 0$, $o \in \rel[\gamma](\mathcal{A},v,V,F[h_{i}/f])$ if and only if $\enc(o)$ is an output of an accepting run of $M_\beta(\gamma,\mathcal{A}, v, V)$ at recursion depth at most $i$.
%
We prove this claim by induction on $i$ and $\gamma$. 
\begin{description}
    \item[The case of $i=0$:] If $o \in \rel[\gamma](\mathcal{A},v,V,F[h_{0}/f])$, then $\gamma$ is not of the form $f(\Vec{y})$, and we prove that 
    $MSp_\beta^{sub}(\gamma,\mathcal{A}, v, V)$ outputs $\enc(o)$ in an accepting run that does not go through the case of $\gamma==f(\vec{y})$ in the description of $MSp_\beta^{sub}$. We proceed by induction on $\gamma$. As we see above, it cannot be the case that  $\gamma = f(\Vec{y})$; if $\gamma=\alpha$, a formula without the function symbol $f$, then the argument is complete by Proposition \ref{prop:alpha-in-sharpL-mathching-outputs}; the remaining cases for $\gamma$ are straightforward.
    The converse direction is similar.
%   
%    Conversely, we prove that if $M_\beta(\gamma,\mathcal{A}, v, V)$ outputs $\enc(o)$ in an accepting run that does not go through the case of $\gamma$\verb|==|$f(\vec{y})$ in the description of $M_\beta$, then $o \in \rel[\gamma](\mathcal{A},v,V,F[h_{0}/f])$.
%    We proceed by induction on $\gamma$. 
%    From our assumptions, it cannot be the case that $\gamma = f(\Vec{y})$; 
%    if $\gamma=\alpha$, 
    \item[Assuming that the statement holds for $i$, we prove it for $i+1$:] This is similar to the base case, except for when $\gamma = f(\vec{y})$, in which case we use the inductive hypothesis for $i$.
    \qedhere
\end{description}
\end{proof}

\begin{theorem}\label{spanl theorem}
$\spanclass=\spanl$ over finite ordered structures.
\end{theorem}
\begin{proof}
$\spanl\subseteq \spanclass$ follows from Proposition~\ref{spanl first inclusion} and the fact that $[\lfp_f \mathsf{span_L}](\Vec{x})\in\spanlogic$. $\spanclass\subseteq \spanl$ is an immediate corollary of Proposition~\ref{spanl second inclusion}.
\end{proof}

% \begin{remark}\label{number of paths}
% In the proof of  $\spanclass\subseteq\spanl$, if a path of $N_\alpha$ does not stop after a finite number of steps, then this is because formula $\alpha$ does not define a reasonable recursion. For example, for every finite structure $\mathcal{A}$,  $\llbracket \,[\lfp_f \quantsigma x. f(x)]\,\rrbracket (\mathcal{A})=0$, and the corresponding transducer generates computation paths that never terminate.

% Regarding the number of paths of $M_\alpha$, there are two cases. 
% \begin{enumerate}
%     \item $\rel[\alpha](\mathcal{A},v,V)$ is an infinite set. Then, $M_\alpha$ has infinitely many different valid outputs, and so infinitely many accepting paths. For example, let $\mathcal{A}=\langle \{0,1\},\leq\rangle$. It holds that $\llbracket \,[\lfp_f \quantsigma x. \big( x\cdot f(x)\big)+\top]\,\rrbracket (\mathcal{A})$ is the set of all strings over $\{0,1\}$. The respective transducer $M_\alpha$ generates infinitely many different valid outputs.
%     \item In the case that $\rel[\alpha](\mathcal{A},v,V)$ is finite, the number of different outputs of $M_\alpha$ is equal to $|\rel[\alpha](\mathcal{A},v,V)|$, but the number of accepting paths can be either infinite or finite. For example, let $\mathcal{A}=\langle \{0\},\leq\rangle$ and $\alpha(y)=[\lfp_f \quantsigma x. f(x) + \underline{y}=\mathrm{min}]$, where $\mathrm{min}$ denotes the minimum element. Then, $\rel[\alpha](\mathcal{A},v[0/y])=\{0\}$, but $M_\alpha$ generates infinitely many accepting paths that all output $0$.
% \end{enumerate} 
% \end{remark}

\subsection{The logic \spanpspacelogic}

To capture the class \spanpspace, we proceed similarly to the case of \spanl, except we need to use a second-order version of our logic, to account for the exponential increase in the space constraints.

%  The $X$-free \sigmaqsou{}  formulae over $\sigma$
%  are defined by the following grammar:
% \begin{equation}
% \alpha::=  ~ x   ~\mid~ \varphi 
% ~\mid~ (\alpha+\alpha) ~\mid~ (\alpha\cdot\alpha) 
% ~\mid~ \quantsigma y. \alpha ~\mid~ \quantsigma Y. \alpha
% \label{alpha4}
% \end{equation}
% where $\varphi$ is an \SO{} formula, $x$, $y$ are  first-order variables, and $Y$ is a second-order variable.

\begin{definition} 
We define the logic \spanpspacelogic{} over $\sigma$ to be the set of formulae  $[\lfp_f\beta](X)$,  where $\beta$ is defined by the following grammar:
 \begin{equation}
\beta::=  ~ \alpha 
~\mid~  f(X)
~\mid~ (\beta+\beta)
~\mid~ (\alpha\cdot\beta) 
~\mid~ \quantsigma y. \beta
~\mid~ \quantsigma Y. \beta
\label{beta2}
\end{equation}
where $\alpha$ is an $X$-free \sigmaqsou{} formula over $\sigma$, $y$ is a first-order variable, $X,Y$ are  second-order variables, and $f$ is a second-order function symbol. 
\end{definition}

\begin{remark}\label{rem:spanpspace-clock}
To avoid formulae $[\lfp_f\beta](X)\in\spanpspacelogic$ with $\llbracket \, [\lfp_f\beta](X)\,\rrbracket (\mathcal{A},v,V)=+\infty$,
%and computations of non-deterministic poly-space transducers that contain cycles, we impose an exponential-time bound to this kind of TMs by attaching a clock, and 
we adjust the syntax of \spanpspacelogic{} similarly to Remark~\ref{rem:spanl-clock}. The only difference is that in the case of polynomial space, the clock imposes an exponential-time bound, and so the contents of the clock need to be encoded by a relation. Thus, here $f$ is replaced by $\textsf{Clock}(X,Cl,f):=\quantsigma Cl'.(Cl<\mathrm{Max})\cdot (Cl'=Cl+1)\cdot f(Y,Cl')$, where $Cl,Cl'$ are second-order variables of arity $k$. An order on relations of arity $k$ induced by the lexicographic order on $k$-tuples can be defined in \fo. The same holds for $Cl'=Cl+1$ and Max.
\end{remark}

\begin{remark}\label{encoding of relations}  
Let $\mathcal{A}$ be a finite ordered structure over $\sigma$. Relations $R_1,\dots, R_{m}$ on $A$ with $\arity(R_j)=k$, for every $1\leq j\leq m$, can be encoded by one relation $R$ on $A$ of arity $k+\lceil \log m\rceil$, by defining $R(\Vec{i},\Vec{a})$ iff $R_i(\Vec{a})$, for every $\Vec{a}\in A^k$, 
where $\Vec{i}$ is the $i$-th smallest $\lceil \log m\rceil$-tuple over $A$. 
%The lexicographic order between such sequences can be defined in \fo. 
We use this observation to show that a second-order function symbol $f$ with $\arity(f)=1$, suffices to capture \spanpspace.
\end{remark}

% \subsubsection{The semantics}
% For any formula $\alpha\in\sigmaqsou$ and any $\mathcal{A},v,V$, we define $\rel[\alpha](\mathcal{A},v,V)$ as in Table~\ref{semantics2}, and $\llbracket \alpha\rrbracket (\mathcal{A},v,V)=|\rel[\alpha](\mathcal{A},v,V)|$. 

% The semantics of $[\lfp_f  \beta](X)$ are defined as in Subsection~\ref{recursion semantics}, where operator $T_\beta$ is defined on $\mathcal{SOF}$ and can be proven to be monotone.

% \begin{proposition}\label{monotonicity2}
% Let $f$ be a function symbol of arity 1 over a second-order variable that has arity $k$. Consider also a \sigmaqsou{} formula $\beta$ over $\sigma$, such that if $\beta$ contains a function symbol, then this function symbol is $f$, a finite ordered structure $\mathcal{A}$ over $\sigma$, $h,g:\mathcal{R}_k\rightarrow  \mathcal{P}(A^*)$, and  the function assignments $H,G$ such that $H(f)=h$ and $G(f)=g$. If $h\leq_F g$, then for every first- and second-order assignments $v$ and $V$, respectively, it holds that:
% $$\rel[\beta](\mathcal{A},v,V,H)\subseteq \rel[\beta](\mathcal{A},v,V,G).$$
% \end{proposition}
% \begin{proof} The proof is completely analogous to that of Proposition~\ref{monotonicity1}.
% \end{proof}

% \begin{corollary}\label{monotone operator 2}
%     For every formula $[\lfp_f  \beta](X)$, where $\beta$ is in \sigmaqsou{}, operator $T_{\beta}$ is monotone on the complete lattice $(\mathcal{SOF},\leq_F)$. In other words, for every $h,g\in \mathcal{SOF}$, if $h\leq_F g$, then $T_\beta(h)\leq_F T_\beta(g)$.
% \end{corollary}

\subsection{\spanpspacelogic captures \spanpspace over finite ordered structures}\label{span formula}

Let $\mathcal{A}$ be a finite ordered structure over $\sigma$  with $|A|=n$ and $M=(\mathcal{Q},\Sigma,\delta,q_0,q_F)$ be a non-deterministic poly-space transducer that uses $n^c-1$ space. Let also $k=\max\{c,\lceil \log |\mathcal{Q}|\rceil\}$.  We can use $k$-tuples of the $n$ elements of $A$, to encode  $n^c-1$ tape cells and $|\mathcal{Q}|$ states. The lexicographic order on them can be defined in \fo, and it will be denoted by $\leq$, which is also used to represent the total order on the elements of $A$. W.l.o.g.\ assume that $M$ has a single tape. 
A configuration of $M$ can be encoded by the tuple of 
$k$-ary relations $\Vec{C}=(T,E,P,Q)$:  $T(\Vec{c})$ iff cell $c$ encoded by $\Vec{c}$ contains symbol $1$ (tape contents), $E(\Vec{c})$ denotes that all cells greater than $c$ contain the symbol $\vartextvisiblespace$ (end of zeros and ones on the tape), $P(\Vec{c})$ indicates that the head is on cell $c$ (head's position), and $Q(\Vec{c})$ means that $N$ is in state $q$ that is encoded by $\Vec{c}$.
%When we write $E(\Vec{x})$, it is a shorthand for $E(\Vec{x})\wedge\forall \Vec{u}(\Vec{u}\neq \Vec{x}\rightarrow \neg E(\Vec{u}))$. The same holds for $P(\Vec{x})$ and $Q(\Vec{x})$.
As in the case of \spanl, we encode a bit that $M$ outputs at some time step using two elements of $A$.

We  informally write $\varphi(C)$  
to denote $\varphi(X)$ 
interpreted in structure $\mathcal{A}$ 
where  $X$ is assigned  relation $C\in \mathcal{R}_{\arity(X)}$. Let $\mathsf{Next}_i(\Vec{X},\Vec{Y})$, $i=0,1$, be two formulae with free second-order variables $\Vec{X},\Vec{Y}$, such that $\mathsf{Next}_i(\Vec{C},\Vec{C'})$ expresses that $\Vec{C'}$ is a configuration following $\Vec{C}$ when $M$ makes non-deterministic choice $i$. These two formulae can be expressed in \fo{} in a similar way to the formulae that describe the computation of an NPTM in the proof of Fagin's theorem~\cite{Immerman}. Analogously, $\mathsf{Output}_i(\Vec{C},\Vec{C'},b)$, $i=0,1$, express the same as $\mathsf{Next}_i(\Vec{C},\Vec{C'})$ and also  $b$ encodes the bit that is written on the output tape when $M$ makes this transition. Finally, $\mathsf{Acc}(\Vec{C})$ is an \fo{} formula that expresses that $\Vec{C}$ is the accepting configuration.
According to Remark~\ref{encoding of relations}, the aforementioned formulae can be replaced by \fo formulae  such that a unique relation is used to encode the configuration of $M$. Therefore, in this and the next section, we abuse notation and write $\mathsf{Next}_i(X,Y)$, $\mathsf{Output}_i(X,Y,x)$, and $\mathsf{Acc}(X)$.
%$\wedge\forall \Vec{s}(\Vec{s}\neq \Vec{s_0}\wedge \Vec{s}\neq \Vec{s_1})\rightarrow \neg Path'(\Vec{s})$
% Because of Remark~\ref{encoding of relations}, in sequel, we consider that a configuration can be encoded by one relation, instead of a sequence of relations, and we denote by $C$ this unique relation.

\begin{theorem}\label{spanpspace theorem}
$\spanpspace=\spanpspaceclass$ over finite ordered structures.
\end{theorem}
\begin{proof}
$\spanpspace\subseteq\spanpspaceclass$: For a non-deterministic poly-space transducer $M$ consider the following formula:
$$\begin{aligned}
\mathsf{span_{pspace}}(X,f):= \mathsf{Acc}(X)+\quantsigma Y.\quantsigma x.\big(&\mathsf{Output}_0(X,Y,\underline{x}) + \mathsf{Output}_1(X,Y,\underline{x}) +\\
&\mathsf{Next}_0(X,Y) + \mathsf{Next}_1(X,Y
)\big)\cdot f(Y).
\end{aligned}$$
As in the proof of Proposition~\ref{spanl first inclusion}, we can show that $\llbracket \,[\lfp_f  \mathsf{span_{pspace}} ](X)\,\rrbracket (\mathcal{A},v,V)=span_M(\enc(\mathcal{A}))$, for every $\mathcal{A}$, $v$, and $V$, such that $V(X)$ encodes $M$'s  initial configuration. 

$\spanpspaceclass\subseteq\spanpspace$: The proof is analogous to that of Proposition~\ref{spanl second inclusion}.
\end{proof} 

\section{A logic that captures \fpspace and \spspace} \label{fpspace section}

\subsection{The logic \pspacelogic}

In this subsection, we define the logic \sigmaqsou{} equipped with a second-order function symbol and a restricted form of recursion. Superscript \supr in the name of the logic stands for the fact that recursion is restricted.

\begin{definition} We define \pspacelogic{} over $\sigma$ to be the set of formulae $[\lfp_f  \beta](X)$, where $\beta$ is defined by the following grammar: 
    \begin{equation}
    %\begin{aligned}
       \beta::= ~ \alpha ~\mid~ \quantsigma Y.\varphi(X,\underline{Y})\cdot f(Y)  ~\mid~ (\alpha+\beta)
   % \end{aligned}
     \label{beta3}
    \end{equation} 
where $X,Y$ are second-order variables, $\varphi$ is an \SO{} formula over $\sigma$, $\alpha$ is an $x$-free \sigmaqsou{} formula over $\sigma$, and $f$ is a second-order function symbol. 
\end{definition}

\begin{remark}\label{rem:spspace-clock}
In the following subsections, we prove that \pspacelogic captures  \spspace and \fpspace. To this end, we can 
%attach a clock to non-deterministic poly-space TMs, and 
restrict the syntax of \pspacelogic 
%accordingly, 
as in Section~\ref{spanl section}, to encode the clock attached to poly-space TMs. An alternative approach is the following: we prove that for every $\beta\in\pspacelogic$, $\llbracket \beta\rrbracket $ is in \fpspace in the sense that there is a deterministic poly-space TM $N$ such that on input $\enc(\mathcal{A},v,V)$ outputs $\llbracket \beta\rrbracket (\mathcal{A},v,V)$, if $\llbracket \beta\rrbracket (\mathcal{A},v,V)\in\mathbb{N}$, and it outputs the symbol $\perp$, if $\llbracket \beta\rrbracket (\mathcal{A},v,V)=+\infty$. Thus, although $\pspaceclass$ contains functions $f:\Sigma^*\rightarrow\mathbb{N}\cup\{+\infty\}$, they can all be computed in deterministic polynomial space. By Theorem~\ref{Ladner}, $\pspaceclass\subseteq\spspace$, in the sense that for any $\llbracket \beta\rrbracket \in\pspacelogic$ there is a non-deterministic poly-space TM $M$ such that it outputs symbol $\perp$ and halts, if $\llbracket \beta\rrbracket (\mathcal{A},v,V)=+\infty$, and otherwise $acc_M(\enc(\mathcal{A},v,V))=\llbracket \beta\rrbracket (\mathcal{A},v,V)$. This alternative approach is described in Subsection~\ref{fpspace subsection}.
\end{remark}

\subsection{\pspacelogic captures \spspace over finite ordered structures}

% First we show that \pspacelogic{} captures \spspace{} over finite ordered structures.
We first prove that \pspacelogic captures \spspace over finite ordered structures.

\begin{proposition}\label{spspace first inclusion}
$\spspace\subseteq \pspaceclass$ over finite ordered structures.
\end{proposition}
\begin{proof}
Let $M$ be a non-deterministic poly-space TM. Consider the formula $\mathsf{acc_{pspace}}(X,f):= \mathsf{Acc}(X)+\quantsigma Y.\big(\mathsf{Next}_0\vee \mathsf{Next}_1\big)(X,\underline{Y}) \cdot f(Y)$, where $\mathsf{Acc}$, $\mathsf{Next}_0$, and $\mathsf{Next}_1$ have been defined in Subsection~\ref{span formula}. Similarly to the proof of Lemma~\ref{spanl first inclusion}, it can be proven that $acc_M(\enc(\mathcal{A})) = \llbracket \,[\lfp_f  \mathsf{acc_{pspace}} ](X)\,\rrbracket (\mathcal{A},v,V)$ for every $\mathcal{A}$, $v$ and $V$, such that $V(X)$ encodes the initial configuration of $M$.
\end{proof}

Note that in contrast to $\rel[\,[\lfp_f  \mathsf{span_{pspace}} ](X)\,](\mathcal{A},v,V)$ that contains encodings of outputs, $\rel[\,[\lfp_f  \mathsf{acc_{pspace}} ](X)\,](\mathcal{A},v,V)$ contains encodings of computation paths; a computation path of $M$ is encoded as the sequence of configurations that $ M$ visits along this path. Intuitively,
%the intermediate interpretation of 
$\rel[\,[\lfp_f  \mathsf{acc_{pspace}} ](X)\,](\mathcal{A},v,V)$ with $V(X)=C$, 
%$[\lfp_f  \mathsf{acc_{pspace}}](C/X)$ 
is the set of encodings of accepting paths that are generated by $M$ when it starts its computation from the configuration encoded by $C$. 

% \begin{lemma}\label{lem:poly-bounded-length-a}
% Let $\alpha$ be an $x$-free \sigmaqsou{} formula over $\sigma$ that contains only second-order variables of arity $k$. 
% For every $\mathcal{A}$, $v$, and $V$, and every $s \in \rel[\alpha](\mathcal{A},v,V)$, $|s| \leq |\alpha|$  and  $|\enc(s)|\leq |\alpha|\cdot |A|^k$.
% \end{lemma}

Lemmata~\ref{lem:poly-space-bound-tocheck-a} and \ref{cor:alpha-in-sp} state that $x$-free \sigmaqfou{} formulae can be verified and evaluated by polynomial-space Turing machines. 

\begin{lemma}\label{lem:poly-space-bound-tocheck-a}
Let $\alpha$ be an $x$-free \sigmaqsou{} formula over $\sigma$.  
There is a deterministic TM $M$ that takes as input 
 $\enc(\mathcal{A},v,V)$, and a string $s \in (\bigcup_{i\in\mathbb{N}}\mathcal{R}_i)^*$, and decides if $s \in \rel[\alpha](\mathcal{A},v,V)$, using at most $\mathcal{O}(\mathrm{poly}(|A|))$ space.
\end{lemma}

\begin{lemma}\label{cor:alpha-in-sp}
Let $\alpha$ be an $x$-free \sigmaqsou{} formula over $\sigma$. There is a non-deterministic poly-space TM $M$, such that $acc_M(\enc(\mathcal{A},v,V))=\llbracket \alpha\rrbracket (\mathcal{A},v,V)$ for every $\mathcal{A}$, $v$, and $V$.
\end{lemma}
 \begin{proof}
Let $m$ denote the maximum arity of any second-order variable that appears in $\alpha$. Using Lemmata \ref{lem:bounded-length-a} and \ref{lem:poly-space-bound-tocheck-a},  $M$ can non-deterministically guess every string $s \in (\bigcup_{1\leq i\leq m}\mathcal{R}_i)^*$ of length at most $|\alpha|$, and then deterministically verify that $s \in \rel[\alpha](\mathcal{A},v,V)$.
%
% Let $m$ denote the maximum arity of any second-order variable that appears in $\alpha$. Using Lemmata \ref{lem:bounded-length-a} and \ref{lem:poly-space-bound-tocheck-a}, $M$ can check whether  a string $s$ is in $\rel[\alpha](\mathcal{A},v,V)$ for all $s \in (\bigcup_{1\leq i\leq m}\mathcal{R}_i)^*$ of length at most $|\alpha|$. If the answer is positive, $M$ increases a counter by one. 
%and stores $s$ in its work tape.
\end{proof}

As shown in Lemma~\ref{lem:poly-space-bound-tocheck-a-2}, a string $s$ can be verified to be in $\rel[\,[\lfp_f\beta](X)\,](\mathcal{A},v,V)$ for a \pspacelogic{} formula $[\lfp_f\beta](X)$ in polynomial space w.r.t.\ $|A|$ and $|\enc(s)|$. 

\begin{lemma}\label{lem:poly-space-bound-tocheck-a-2}
Let $[\lfp_f\beta](X)$ be a \pspacelogic{} formula over $\sigma$.
There is a deterministic TM $M_\beta$, such that on input 
$(\mathcal{A},v,V)$, and a string $s\in(\bigcup_{i\in\mathbb{N}}\mathcal{R}_i)^*$, $M_\beta$ decides if $s \in \rel[\,[\lfp_f\beta](X)\,](\mathcal{A},v,V)$ in space $\mathcal{O}(\mathrm{poly}(|A|),|\enc(s)|)$.
\end{lemma}
\begin{proof}
If $\beta$ is an $x$-free \sigmaqsou{} formula, then there is such a TM by Lemma~\ref{lem:poly-space-bound-tocheck-a}, since $\rel[\beta](\mathcal{A},v,V)=\rel[\,[\lfp_f\beta](X)\,](\mathcal{A},v,V)$.  If $\beta=\quantsigma Y.\varphi(X,\underline{Y})\cdot f(Y)$, then the lemma is trivially true, since $\rel[\,[\,[\lfp_f\beta](X)\,].(\mathcal{A},v,V)=\emptyset$.  In the case of $\beta=\alpha+\quantsigma Y.\varphi(X,\underline{Y})\cdot f(Y)$, $M_\beta$ is described in Algorithm~\ref{alg:spspace-membership}. Let $k$ denote $\arity(X)=\arity(Y)$ and $B$ denote the relation encoded by $V(X)$, where $V$ is the input second-order assignment. 
%We also assume that the empty string is also encoded by some tuple of elements of $A$. 
\begin{algorithm}
\caption{$M_\beta$ when $\beta=\alpha+\quantsigma Y.\varphi(X,\underline{Y})\cdot f(Y)$}\label{alg:spspace-membership}
\DontPrintSemicolon
\KwIn{$s, \mathcal{A},v,V$}
{simulate $M_\alpha(s,\mathcal{A},v,V)$ from Lemma~\ref{lem:poly-space-bound-tocheck-a}}\;
\lIf{$M_\alpha(s,\mathcal{A},v,V)$ accepts}{accept}
\For{$C\in\mathcal{R}_k$}{
\If{$(s[1] == C)$ and $(\mathcal{A},V[B/X,C/Y]\models\varphi(X,Y))$}{
%\lIf{$|s|>1$}
$s:=s[2:]$\;
%\lElse{$s:=\varepsilon$}
simulate $M_\beta(s,\mathcal{A},v,V[C/X])$
}}
reject
\end{algorithm}
% \verb|def| $M_\beta(s,\mathcal{A},v,V)$\\
% \verb|    if| $M_\alpha(s,\mathcal{A},v.V)$ \verb|accepts then accept|\\
% \verb|    for all| $B\in\mathcal{R}_k$ \verb|do reusing space|:\\
% \verb|     if| ($s(1)$ \verb|==| $B$) \verb|and| ($\mathcal{A}\models\varphi(V(X)/X,B/Y)$): \\
% \verb|       if| $|s|>1$ \verb|then| $s:=s(2)\circl\dots\circl s(\mathrm{last})$ \verb|else| $s:=\varepsilon$\\
% \verb|       simulate| $M_\beta(s,\mathcal{A},v,V[B/X])$ \verb|reusing space|\\
% \verb|    reject|\\
$M_\alpha$ uses at most $\mathcal{O}(\mathrm{poly}(|A|))$ space from Lemma \ref{lem:poly-space-bound-tocheck-a}, and $\mathcal{A},V[B/X,C/Y]\models\varphi(X,Y)$ can be checked using $\mathcal{O}(\mathrm{poly}(|A|))$ space, as $\varphi$ is an \SO{} formula. 
The for-loop is executed reusing space, and the if statement in line 3 is true for at most one relation $C\in\mathcal{R}_k$. 
Moreover, at any time the machine stores one string of length at most $|s|$; therefore, $\mathcal{O}(|\enc(s)|)$ space is also required.
\end{proof}

As a result, a formula in \pspacelogic{} can be evaluated by a non-deterministic polynomial-space TM as shown in the following proposition. 

\begin{proposition}\label{spspace second inclusion}
$\pspaceclass\subseteq\spspace$ over finite ordered structures.
\end{proposition}
\begin{proof}

\begin{algorithm}
\caption{Non-deterministic poly-space $M_\beta$}\label{alg:spspace}
\DontPrintSemicolon
\KwIn{$\mathcal{A},v,V$}
\If{$\beta==\alpha$ has no function symbol}{
simulate $M_\alpha(\mathcal{A},v,V)$ from Lemma~\ref{cor:alpha-in-sp}\;
}
\lIf{$\beta==\quantsigma Y.\varphi(X,\underline{Y})\cdot f(Y)$}{
reject%\;
}
\If{$\beta==\alpha + \quantsigma Y.\varphi(X,\underline{Y})\cdot f(Y)$}{
non-deterministically 
go to line $6$ or $12$\; 
%choose to execute lines $7-12$ or $13-15$ \;
non-deterministically choose $s\in(\bigcup_{1\leq i\leq m} \mathcal{R}_i)^*$ s.t. $|s|\leq|\alpha|$\;
\lIf{$N_\alpha(s,\mathcal{A},v,V)$ rejects}{reject}
\Else{$t:=s[2:]$\;
\lIf{$N_\beta(t,\mathcal{A},v,V[s[1]/X])$ accepts}{reject}
\lElse{accept}}
non-deterministically choose $C\in \mathcal{R}_k$\;
\lIf{$\mathcal{A},V[B/X,C/Y]\models \varphi(X,Y)$}{simulate $M_\beta(\mathcal{A},v,V[C/X])$} 
\lElse{reject}}
\end{algorithm}

Let $[\lfp_f \beta](X) \in \pspacelogic$. 
% From Lemmata \ref{lem:poly-space-bound-tocheck-a} and \ref{lem:poly-space-bound-tocheck-a-2}, there are deterministic TMs $N_\alpha$ for every $\alpha$ with no function symbol, and $N_\beta$, such that 
% on input 
% $(\mathcal{A},v,V)$, and a string $s\in(\bigcup_{i\in\mathbb{N}}\mathcal{R}_i)^*$, 
% the TMs decide if $s \in \rel[\,[\lfp_f\alpha](X)\,](\mathcal{A},v,V)$ or $s \in \rel[\,[\lfp_f\beta](X)\,](\mathcal{A},v,V)$, respectively.
% $N_\alpha$ uses polynomial space in $|A|$ and $N_\beta$ that uses $\mathcal{O}(\mathrm{poly}(|A|),|\enc(t)|)=\mathcal{O}(\mathrm{poly}(|A|))$ space.
%
We describe a non-deterministic poly-space TM $M_\beta$ such that $\llbracket \,[\lfp_f \beta](X)\,\rrbracket (\mathcal{A},v,V)=acc_{M_\beta}(\enc(\mathcal{A},v,V))$, for every $\mathcal{A}$, $v$, and $V$. Let $k$ denote $\arity(Y)=\arity(X)$, and $m$ denote the maximum arity of any second-order variable that appears in $\beta$.  Let also $N_\alpha$ be the deterministic poly-space TM associated with $\alpha\in\sigmaqsou$ from Lemma~\ref{lem:poly-space-bound-tocheck-a} and $N_\beta$ be the deterministic TM associated with $[\lfp_f\beta](X)$ from Lemma~\ref{lem:poly-space-bound-tocheck-a-2}.
$M_\beta(\mathcal{A},v,V)$ 
% calls $M_\beta^{sub}(\beta,\mathcal{A},v,V)$, which 
is defined in Algorithm~\ref{alg:spspace}, where $B$ denotes $V(X)$, for the input second-order assignment $V$.

In the case of $\beta\in\sigmaqsou$ (line 1), $\llbracket \beta\rrbracket (\mathcal{A},v,V)=\llbracket \,[\lfp_f\beta](X)\,\rrbracket (\mathcal{A},v,V)$, and the proposition is true by Lemma~\ref{cor:alpha-in-sp}. If $\beta=\quantsigma Y.\varphi(X,\underline{Y})\cdot f(Y)$ (line 3), then $\llbracket \,[\lfp_f\beta](X)\,\rrbracket (\mathcal{A},v,V)=0$ and the proposition holds trivially. The only interesting case is when $\beta=\alpha+\quantsigma Y.\varphi(X,\underline{Y})\cdot f(Y)$, where $\alpha\in\sigmaqsou$ (line 4). Then, $M_\beta$ 
uses non-determinism to branch between the two summands.
For the second summand, the machine introduces branches for all relations $C$, verifies with $\varphi$, and recurses.
For the first summand, the machine branches for every string $s$ that has a compatible length with $\alpha$; it verifies with $N_\alpha$ that it is in $\rel[\,[\lfp_f\alpha](X)\,](\mathcal{A},v,V)$; and finally, before it accepts, it verifies using $N_\beta$ that the second summand does not also generate $s$.
% makes a straightforward recursive computation and before accepting, it checks whether the last sequence of relations that guessed, can also be guessed and lead to acceptance along a different path; if such a path exists, then it rejects. 

Regarding the space used by $M_\beta$, two strings of length at most $|\alpha|$ have to be stored because of lines $6$ and $9$; since their length is constant, their encodings are of polynomial length with respect to $|A|$. $N_\alpha$ uses polynomial space in $|A|$ by Lemma~\ref{lem:poly-space-bound-tocheck-a}, and $N_\beta$ uses $\mathcal{O}(\mathrm{poly}(|A|),|\enc(t)|)=\mathcal{O}(\mathrm{poly}(|A|))$ space by Lemma~\ref{lem:poly-space-bound-tocheck-a-2} and the fact that $|t|\leq|\alpha|$. Moreover, the recursive call of $M_\beta$ in line 13 is done reusing space.  Overall, $M_\beta$ uses polynomial space in $|A|$. \qedhere
\end{proof}

\begin{theorem}\label{fpspace corollary}
   $\pspaceclass=\spspace$ over finite ordered structures. 
\end{theorem}
\begin{proof} The theorem is immediate from Propositions~\ref{spspace first inclusion} and~\ref{spspace second inclusion}.
\end{proof}

\subsection{\pspacelogic captures \fpspace over finite ordered structures}\label{fpspace subsection}

In this subsection we prove what we promised in Remark~\ref{rem:spspace-clock}: for every $\beta\in\pspacelogic$, $\llbracket \beta\rrbracket \in\fpspace$ in the sense that there is a deterministic poly-space TM $N$ such that on input $\enc(\mathcal{A},v,V)$ outputs $\llbracket \beta\rrbracket (\mathcal{A},v,V)$, if $\llbracket \beta\rrbracket (\mathcal{A},v,V)\in\mathbb{N}$, and it outputs the symbol $\perp$, if $\llbracket \beta\rrbracket (\mathcal{A},v,V)=+\infty$. Thus, all functions in \pspaceclass can be computed in deterministic polynomial space.

An example of a formula $[\lfp_f\beta](X)\in\pspacelogic$ with $\llbracket \,[\lfp_f \beta](X)\,\rrbracket (\mathcal{A},v,V)=+\infty$ is provided below.

\begin{example}\label{infinite lfp}
 Consider $\beta(X)=\big(\quantsigma Y. (Y = X) \cdot Y\cdot  f(Y)\big)+X(\mathrm{min})$, where $\mathrm{min}$ expresses the minimum element. Let $\mathcal{A}=\langle \{0\},\leq\rangle$, $B_0=\emptyset$, and $B_1=\{0\}$. Then,   
 $$\rel[ \,[\lfp_f  \beta](X)\,](\mathcal{A},V[B_1/X])=\bigcup_{k\geq 0} \{{B_1}^k\}.$$ 

 The reason that infinitely many strings emerge in $\rel[ \,[\lfp_f  \beta](X)\,](\mathcal{A},V[B_1/X])$ is that  $\rel[\beta](\mathcal{A},V[B_1/X])=\{B_1\}\cdot\rel[\beta](\mathcal{A},V[B_1/X])\cup\{\varepsilon\}$, i.e.\ the intermediate semantics of $\beta$ when $X$ is evaluated to be $B_1$ must be repeatedly evaluated. Moreover, the fact that $\mathcal{A},V[B_1/X]\models X(\mathrm{min})$ implies that there are non-deterministic branches of the recursion that terminate (if we consider $+$ as non-determinism). On the contrary, for example, $\rel[\,[\lfp_f \zeta](X)\,](\mathcal{A},V[B_1/X])=\emptyset$, where $\zeta=\quantsigma Y. (Y = X) \cdot Y\cdot  f(Y)$.
\end{example}

In general,  $|\rel[ \,[\lfp_f  \beta](X)\,](\mathcal{A},v,V)|=+\infty$ when (a) $\rel[\beta(X)](\mathcal{A},v,V[B/X])$, for some relation $B$, must be evaluated infinitely often during the recursion and (b) the form of $\beta$ allows  some branches of the recursion to terminate.
We describe an algorithm that can detect in polynomial space whether both these facts are true. Essentially, we reduce this problem to solving the \textsc{Reachability} problem in a graph related to formula $\beta$ and $(\mathcal{A}, v, V)$. We first define this graph, which we call the \emph{graph of connections} of $\varphi$ with respect to $\mathcal{A}$, where $\varphi(X,\underline{Y})$ is the \sigmaqsou{} subformula  of $\beta$.
%As a result, $\pspacelogic\subseteq\fpspace$, in the sense that, for every $\alpha\in \pspacelogic$, there is a deterministic poly-space TM $N_\alpha$, such that $N_\alpha$ outputs $m$, if $\llbracket \alpha\rrbracket (\mathcal{A},v,V)=m\in\mathbb{N}$, and otherwise $N_\alpha$ outputs the symbol $\perp$, which denotes that $\llbracket \alpha\rrbracket (\mathcal{A},v,V)=+\infty$. 

\begin{definition}
For any \sigmaqsou{} formula $\varphi(X,\underline{Y})= Y\cdot \varphi(X,Y)$ (or $\varphi(X,\underline{Y})= \varphi(X,Y)\cdot Y$), where $\varphi\in \SO$ and $\arity(X)=\arity(Y)=k$, and any $\mathcal{A}$ over $\sigma$, we define the \emph{graph of connections} of $\varphi$ with respect to $\mathcal{A}$, denoted by $G_\varphi^\mathcal{A}$, as follows: 
\begin{itemize}
    \item The set of vertices of $G_\varphi^\mathcal{A}$ is $V_\varphi^\mathcal{A}:=\{B~\mid~ B\in\mathcal{R}_k \}$, and 
    \item the set of edges of $G_\varphi^\mathcal{A}$ is 
    $E_\varphi^\mathcal{A}=\{(B,C)~\mid~ \mathcal{A},V[B/X,C/Y]\models \varphi(X,Y)\}$.
\end{itemize}
Moreover, if a cycle in $G_\varphi^\mathcal{A}$ starts and ends at a vertex $B$, we say that $B$ is the \emph{starting and ending point} of the cycle.
\end{definition}

We prove below that given a \sigmaqsou{} formula $\varphi(X,\underline{Y})$, \textsc{Reachability} in $G_\varphi^\mathcal{A}$ can be solved in deterministic polynomial space with respect to $|\mathcal{A}|$.

\begin{lemma}
Let $\varphi(X,\underline{Y})= Y\cdot \varphi(X,Y)$ (or $\varphi(X,\underline{Y})= \varphi(X,Y)\cdot Y$) be a \sigmaqsou{} formula where $\varphi\in \SO$ and $\arity(X)=\arity(Y)=k$, and let $\mathcal{A}$ be a finite ordered structure over $\sigma$. Given two $k$-ary relations $B$ and $C$, the problem of deciding whether there is a path from $B$ to $C$ in $G_\varphi^\mathcal{A}$, which we denote by  $\ensuremath{\textsc{Reach}}_\varphi^\mathcal{A}$, can be solved in deterministic $\mathcal{O}(|A|^{2k})$ space.
Moreover, if there is a path from $B$ to $C$, we say that $C$ is \emph{reachable} from $B$.
\end{lemma}
\begin{proof}
By Savitch's theorem~\cite[Section 7.3]{Pap94}, given a graph with $n$ vertices, there is an algorithm for solving $\textsc{Reachability}$  in deterministic $\mathcal{O}(\log^2 n)$ space. $\textsc{Reach}_\varphi^\mathcal{A}$ can be solved by implementing this deterministic algorithm on $G_\varphi^\mathcal{A}$. Graph $G_\varphi^\mathcal{A}$ contains $2^{|A|^k}$ vertices. However, not the whole graph is stored in memory. Only a constant number of vertices are stored by the algorithm at any time, and  $|A|^k$ many bits suffice to store each of them.  Moreover, given two vertices, it can be decided whether they are connected by an edge in space polynomial in $|A|$, since $\varphi\in\SO$. So, $\textsc{Reach}_\varphi^\mathcal{A}$ can be solved in  $\mathcal{O}(\log^2(2^{|A|^k}))=\mathcal{O}(|A|^{2k})$ space.
\end{proof} 

For any formula $[\lfp_f  \beta](X)$ in \spanpspacelogic, there is a deterministic TM $M_\beta^\infty$ that can detect whether $\llbracket \,[\lfp_f  \beta](X)\,\rrbracket (\mathcal{A},v,V)=+\infty$ in polynomial space with respect to $|\mathcal{A}|$ using the polynomial space algorithm for $\textsc{Reach}_\varphi^\mathcal{A}$, where $\varphi(X,\underline{Y})$ is the \sigmaqsou{} formula that appears in $[\lfp_f  \beta](X)$. The TM $M_\beta^\infty$, which is defined in the proof of Lemma~\ref{infinity detection}, determines whether the conditions described in Example~\ref{infinite lfp} are both true.

\begin{notation}
Let $\varphi(X,\underline{Y})$ be a \sigmaqsou{} formula, where $\varphi\in \SO$ and $\arity(X)=\arity(Y)=k$; let also $\mathcal{A}$ be a finite ordered structure over $\sigma$. For any $B\in\mathcal{R}_k$, let $\mathrm{reach}(B)$ denote the set that contains $B$ and all relations in $\mathcal{R}_k$ that are reachable from $B$ in $G_\varphi^\mathcal{A}$.
\end{notation}

\begin{lemma}\label{infinity detection}
For every $[\lfp_f  \beta](X)\in \spanpspacelogic$, there is a deterministic poly-space TM $M_\beta^\infty$ that accepts an input $\enc(\mathcal{A},v,V)$ if and only if $\llbracket \,[\lfp_f  \beta](X)\,\rrbracket (\mathcal{A},v,V)=
+\infty$.
\end{lemma}
\begin{proof}  %Let $\varphi(X,\underline{Y})$ be a \sigmaqsou{} formula, where $\varphi\in \SO$ and $\arity(X)=\arity(Y)=k$; let also $\mathcal{A}$ be a finite ordered structure over $\sigma$. For any $B\in\mathcal{R}_k$, define $\mathrm{reach}(B)$ to be the set that contains $B$ and all relations in $\mathcal{R}_k$ that are reachable from $B$ in $G_\varphi^\mathcal{A}$.
The definition of $M_\beta^\infty$ is based on the following observations. 
\begin{itemize}
\item Let $\alpha$ be an $x$-free \sigmaqsou{} formula. Then $\llbracket \,[\lfp_f  \alpha](X)\,\rrbracket (\mathcal{A},v,V)\in\mathbb{N}$ for every $\mathcal{A},v$,$V$.
So, in that case $M_\beta^\infty$ is the TM that rejects any input.
\item Let $\beta$ be of the form  $\alpha(X)+\quantsigma Y.\varphi(X,\underline{Y})\cdot f(Y)$, where $\alpha(X)$ is an $x$-free \sigmaqsou{} formula and $\arity(X)=k$.
%Define $\mathcal{B}$ to be the set that contains $B$ and all relations in $\mathcal{R}_k$ that are reachable from $B$ in $G_\varphi^\mathcal{A}$. 
If the following two conditions are true, then $\llbracket \,[\lfp_f  \beta](X)\,\rrbracket (\mathcal{A},v,V)=+\infty$, and otherwise, $\llbracket \,[\lfp_f  \beta](X)\,\rrbracket (\mathcal{A},v,V)\in\mathbb{N}$.
    \begin{enumerate}[a.]
        \item  There is some $C\in\mathrm{reach}(V(X))$ such that $C$ is the starting and the ending point of a cycle in $G_\varphi^\mathcal{A}$.
        \item There is $D\in\mathcal{R}_k$ that  belongs to the cycle of (a), such that $\rel[\alpha](\mathcal{A},v, V[D/X])\neq \emptyset$.
    \end{enumerate}
In this case, $M_\beta^\infty$ is defined in Algorithm~\ref{alg:infty-detection}.
\begin{algorithm}
\caption{$M_\beta^\infty$ when $\beta=\alpha(X)+\quantsigma Y.\varphi(X,\underline{Y})\cdot f(Y)$}\label{alg:infty-detection}
\DontPrintSemicolon
\KwIn{$\mathcal{A},v,V$}
\For{$C\in\mathcal{R}_k$}{
\If{$\textsc{Reach}_\varphi^\mathcal{A}(V(X),C)$ accepts}{
\For{$D\in\mathcal{R}_k$}{
\If{$\rel[\alpha](\mathcal{A},v,V[D/X])\neq\emptyset$}{
\lIf{$(\textsc{Reach}_\varphi^\mathcal{A}(C,D)$ accepts$)$ and $(\textsc{Reach}_\varphi^\mathcal{A}(D,C)$ accepts$)$}{accept}
}}}}
reject
\end{algorithm}
% \verb|def| $M_\beta^\infty(\mathcal{A},v,V)$\\
% \verb|  for all| $C\in\mathcal{R}_k$ \verb|do reusing space|:\\
% \verb|   if| $\textsc{Reach}_\varphi^\mathcal{A}(V(X),C)$ \verb|accepts|: \\
% \verb|     for all| $D\in\mathcal{R}_k$ \verb|do reusing space|:\\
% \verb|       if| $\rel[\gamma](\mathcal{A},v,V[D/X])\neq\emptyset$:\\
% \verb|         if| ($\textsc{Reach}_\varphi^\mathcal{A}(C,D)$ \verb|accepts) and| ($\textsc{Reach}_\varphi^\mathcal{A}(D,C)$ \verb|accepts) then accept|
% \verb|   reject|\\
% % To determine whether conditions a and b are both true, $M_\alpha^\infty$ first finds all relations $C$ that are reachable from $B$ by solving $\textsc{Reachability}_\varphi^\mathcal{A}$ $(B,C)$ for every $C\in\mathcal{R}_k$ reusing space. Then, it finds all relations $D$ such that $\rel[\gamma](\mathcal{A},v, V[D/X])\neq \emptyset$ reusing space. For every such $C$ and $D$ that are currently found and stored, $M_\alpha^\infty$ checks whether there is a path from $C$ to $D$ and from $D$ to $C$ by solving $\textsc{Reachability}_\varphi^\mathcal{A}$ $(C,D)$ and $\textsc{Reachability}_\varphi^\mathcal{A}$ $(D,C)$, respectively, reusing space. If for some pair $C$, $D$ the answer is positive, then  $M_\alpha^\infty$ accepts. Otherwise, it rejects. 
The computation of $M_\beta^\infty$ requires polynomial space: lines 1 and 3 are executed reusing space; so the number of relations that need to be stored at any time is constant and $M_\beta^\infty$ can use $|A|^k$ many bits to store each of them. Moreover, $\rel[\zeta](\mathcal{A},v,V[F/X])\neq \emptyset$ can be decided in polynomial space, for any $\zeta\in\sigmaqsou$ and $F\in \mathcal{R}_k$ by Lemma~\ref{cor:alpha-in-sp} and Proposition~\ref{Ladner}.\qedhere
\end{itemize}
\end{proof}

Finally we prove the main result of this section below.

\begin{proposition}\label{fpspace proposition}
    $\pspaceclass\subseteq\fpspace$ over finite ordered structures.
\end{proposition}
\begin{proof}
For any $[\lfp_f \beta](X)$ 
in $\pspacelogic$, we describe $N_\beta(\enc(\mathcal{A},v,V))$ in Algorithm~\ref{alg:fpspace}, such that  $N_\beta(\enc(\mathcal{A},v,V))$ outputs $m$, if $\llbracket \,[\lfp_f \beta](X)\,\rrbracket (\mathcal{A},v,V)=m\in\mathbb{N}$, and otherwise it outputs the symbol $\perp$, which denotes that $\llbracket \,[\lfp_f \beta](X)\,\rrbracket (\mathcal{A},v,V)=+\infty$. 

If $\llbracket \,[\lfp_f \beta](X)\,\rrbracket (\mathcal{A},v,V)\in\mathbb{N}$, there is a  non-deterministic poly-space TM $M$, defined in the proof of Proposition~\ref{spspace second inclusion}, such that $acc_{M}(\enc(\mathcal{A},v,V))=\llbracket \,[\lfp_f \beta](X)\,\rrbracket (\mathcal{A},v,V)$. Then by Proposition~\ref{Ladner}, there is a deterministic poly-space Turing machine that outputs the value of $acc_{M}(\enc(\mathcal{A},v,V))$, which is denoted by  $M_\beta$ in Algorithm~\ref{alg:fpspace}.
\begin{algorithm}
\caption{Deterministic poly-space $N_\beta$}\label{alg:fpspace}
\DontPrintSemicolon
\KwIn{$\mathcal{A},v,V$}
\If{$\beta==\alpha$ has no function symbol}{
simulate $M_\alpha(\mathcal{A},v,V)$\;
}
\If{$\beta==\alpha + \quantsigma Y. \varphi(X,\underline{Y})\cdot f(Y)$}{
\lIf{$M_\beta^\infty(\mathcal{A},v,V)$ accepts}{output $\perp$}
\Else{
$cycle \gets \mathrm{false}$\;
\For{$C\in\mathcal{R}_k$}{
\If{$(\textsc{Reach}_\varphi^\mathcal{A}(V(X),C)$ accepts$)$ and $(\textsc{Reach}_\varphi^\mathcal{A}(C,C)$ accepts$)$}{$cycle \gets \mathrm{true}$} }
\lIf{$cycle==\mathrm{false}$}{simulate $M_\beta(\mathcal{A},v,V)$}
\lElse{simulate $M'_\beta(\mathcal{A},v,V)$}}}
\end{algorithm}

In the case of $\beta=\alpha + \quantsigma Y. \varphi(X,\underline{Y})\cdot f(Y)$, $M_\beta^\infty$ denotes the TM defined in the proof of Lemma~\ref{infinity detection}. Consider the case when $M_\beta^\infty$ rejects and $cycle=\mathrm{true}$, i.e.\  there is  $C\in\mathcal{R}_k$ which is reachable from $V(X)$ and belongs to a cycle in $G_\varphi^\mathcal{A}$. Since $M_\beta^\infty$ rejects, from the proof of Lemma~\ref{infinity detection}, we have that $\llbracket \,[\lfp_f \beta](X)\,\rrbracket (\mathcal{A},v,V)\in\mathbb{N}$ and that there is a cycle in $G_\varphi^\mathcal{A}$ that consists only of relations $D$ such that $\rel[\alpha](\mathcal{A},v,V[D/X])=\emptyset$. Intuitively, this means that there is a cycle in the recursion of $M(\mathcal{A},v,V)$, where $M$ was defined in Proposition~\ref{spspace second inclusion}, that generates no finite path, or in other words, $M(\mathcal{A},v,V)$ has an infinite path that produces no string in $\rel[\beta](\mathcal{A},v,V)$. Define $M'$ to be the TM that on any input simulates $M$ and halts at any time step after which no string is produced by doing the following:  when $M'$ is in a configuration $c$ right before a recursive call of $M$, it checks  whether $M$ reaches the accepting configuration from $c$.
%in polynomial space. 
If not, it halts. Otherwise, it continues simulating $M$. Then $M'_\beta$ is the deterministic poly-space TM obtained by Proposition~\ref{Ladner}, which outputs the value of $acc_{M'}(\enc(\mathcal{A},v,V))$.

The computation of $M'$ is of polynomial space by Savitch's theorem~\cite[Section 7.3]{Pap94}. If line 8 is executed reusing space, it is not hard to see that $N_\beta$ requires polynomial space.
\end{proof}

\begin{theorem}\label{fpspace-theorem}
$\pspaceclass=\fpspace$ over finite ordered structures.
%=\totpspace
\end{theorem}
\begin{proof}
$\fpspace\subseteq \pspaceclass$ is immediate from Propositions~\ref{spspace first inclusion} and~\ref{Ladner}. The inverse inclusion follows from Proposition~\ref{fpspace proposition}.
\end{proof}
 
\begin{corollary} $\pspaceclass=\fpspace=\totpspace=\spspace$ over finite ordered structures.
\end{corollary}
\begin{proof}
This is immediate from Theorems~\ref{fpspace-theorem} and~\ref{totpspace theorem}. Note that for the purposes of this subsection, we consider a slightly different kind of a non-deterministic poly-space TM which on input $x$, if $f(x)=+\infty$, it outputs $\perp$ and halts, and if $f(x)=m\in\mathbb{N}$, it generates $m$ accepting paths (resp.\ $m+1$ paths).
\end{proof}

%\section{A logic that captures \totp}\label{totp section}
\section{Two logics that capture \totp}\label{totp section}

We introduce two different logics, namely \totlogiclfpth and \totlogiclfpf, that both capture \totp. As a corollary, we get that these logics have the same expressive power  over finite ordered structures.

\subsection{\totlogiclfpth captures \totp over finite ordered structures}\label{subsec:totp-first-logic}

Logic \totlogiclfpth includes two kinds of recursion: a least fixed point over relations on the first-order level of the logic together with a least fixed point over functions on the quantitative level. In other words, \totlogiclfpth{} is a fragment of \sigmalfp equipped with a least fixed point on the quantitative level. \lfpu in the brackets indicates that $\varphi$ in~(\ref{alpha2}) is a formula in \folfp, which is the logic that captures \cP over finite ordered structures~\cite{Immerman82,Vardi82}. 
\folfp extends \fo{} with least-fixed-point operations of the form $[\lfpd_{P,x_1,\dots, x_k}\psi](x_1,\dots, x_k)$, where
% The following grammar defines formulae of \textbf{FO(LFP)}. Since we do no need them, we omit function symbols (and constants) to avoid confusion with the function symbols added to quantitative logics.
% $$\begin{aligned}
% \varphi::= & ~  P(x_1,\dots, x_k) ~\mid~ (x_1 = x_2) ~\mid~ [\lfpd_{P,x_1,\dots, x_k}\psi](x_1,\dots, x_k)~\mid~\top ~\mid~ \perp \\
% &~\mid~ (\neg\varphi) ~\mid~ (\varphi\wedge \varphi) ~\mid~ (\varphi\vee \varphi) ~\mid~ (\varphi\rightarrow \varphi) ~\mid~ (\forall x \varphi) ~\mid~ (\exists x \varphi)
% \end{aligned}$$
% where 
$x_1,\dots,x_k$ are first-order variables, $P$ is a relation symbol of arity $k$, and $\psi$ is an \fo{} formula where every occurrence of $P$ is positive. For the definition of the semantics of $[\lfpd_{P,x_1,\dots, x_k}\psi](x_1,\dots, x_k)$, we refer the reader to~\cite[Chapter 10]{Libkin04}. 
%For convenience and clarity, we omit function and constant symbols from the syntax of \folfp.

% In this section we present our results in the reverse order: w
% In Subsection~\ref{express-totp} we describe how to express every \totp problem using a 
% \sigmalfp formula with a least fixed point on the quantitative level.
% We observe that we do not require the full syntax of \sigmalfp
% with recursion, 
% and thus, in Subsection~\ref{lfp logic totp}, we define 
% a fragment of that logic, namely
% \totlogiclfpth{}, which captures \totp over finite ordered structures.

%\subsection{\totlogiclfpth captures \totp over finite ordered structures}\label{lfp logic totp}

%The addition of recursion to the logic \sigmalfp allows us to express \totp functions. In fact, \sigmalfp equipped with recursion on a second-order function symbol is more expressive than needed.

We now formally define \totlogiclfpth. Sums over second-order variables and least fixed point quantitative formulae will be of a specific form. Definitions~\ref{def:synt-defines} and~\ref{def:extends} are syntactic definitions that are used to restrict the operator $\quantsigma$ over second-order variables.
%Intuitively, any function that corresponds to a formula in the logic defined below, namely \totlogiclfpth, can be seen as a self-reducible problem with an easy decision version.

\begin{definition}\label{def:synt-defines}
We say that a formula $\varphi(Y)$  \emph{syntactically defines} $Y$  if $\varphi(Y)$ is of the form $\forall \Vec{y} Y(\Vec{y})\leftrightarrow \psi(\Vec{y})$, for some formula $\psi$.
\end{definition}

% \begin{lemma}
% (a) Relation $S_{old}$ is uniquely defined by an $\ensuremath{\mathbf{FO(LFP)}}$ formula.\\
% (b) Relation $S_{new}$ is uniquely defined by an \fo{} formula.
% \end{lemma}
% \begin{proof} (a) $S_{old}$ is defined uniquely by $\varphi(\Vec{ctr})=[\lfpd_{S,\Vec{xyz}}\mathsf{detstep}](\Vec{ctr})$.\\
% (b) $S_{new}$ is defined uniquely by $\varphi(\Vec{ctr}):= S_{old}(\Vec{ctr}) \vee \bigvee_{i\in\Delta_i}\mathsf{update}_i(S_{old}, \Vec{ctr})$.
% \end{proof}

\begin{definition}\label{def:extends}
We say that a formula  $\varphi(X,Y)$: 
\begin{enumerate}[(a)]
    \item \emph{extends} $X$ to $Y$ if it is of the form $$\forall\Vec{y}Y(\Vec{y})\leftrightarrow X(\Vec{y})\vee \psi(X,\Vec{y}) \text{ and }$$
    \item \emph{strictly extends} $X$ to $Y$ if it is of the form
$$\forall\Vec{y}\big(Y(\Vec{y})\leftrightarrow X(\Vec{y})\vee \psi(X,\Vec{y})\big)\wedge \exists \Vec{y}\big(\neg X(\Vec{y})\wedge Y(\Vec{y})\big)$$
for some formula  $\psi$ and  $\arity(X)=\arity(Y)$.
\end{enumerate} 
\end{definition}

\begin{notation}
(a) $Y:=\varphi\cdot \alpha$ denotes $\quantsigma Y. \varphi(Y)\cdot \alpha$, where $\varphi$ syntactically defines $Y$, and (b) $Y:=\varphi(X)\cdot f(Y)$ denotes $\quantsigma Y. \varphi(X,Y)\cdot f(Y)$, where $\varphi$ (strictly) extends $X$ to $Y$.
\end{notation}

In the following definition, we first define a fragment of \sigmalfp, which we call \sigmalfpu, and then we obtain \totlogiclfpth by adding recursion to \sigmalfpu.

\begin{definition} \label{totp definition}
\begin{enumerate}[(a)]
    \item \sigmalfpu\ formulae over $\sigma$ are defined by the following grammar:
\begin{equation}
   \alpha:=  X  ~\mid~ \varphi ~\mid~ (\alpha+\alpha) ~\mid~ (\alpha\cdot\alpha) ~\mid~ \quantsigma y.\alpha 
   ~\mid~ Y:=\psi\cdot \alpha
\label{alpha5} 
\end{equation} 
where $y$ is a first-order variable, $X,Y$ are  second-order variables, and $\varphi$, $\psi$ are \folfp formulae over $\sigma$.
\item We define the logic \totlogiclfpth{} over $\sigma$ to be the set of  formulae $[\lfp_f  \beta](X)$, where $\beta$ is defined by the following grammar:
\begin{equation}
\begin{aligned}
\beta::=~ \alpha ~\mid~
%\sum_{i=1}^r\quantsigma Y.\, \varphi(\underline{X})\cdot\big(\top+\psi_i(X,Y)\cdot f(Y)\big)
\varphi(\underline{X}) \cdot\big(\top+\sum_{i=1}^r Y:=\psi_i(X)\cdot f(Y)\big)
~\mid~ (\alpha+\beta)
\end{aligned}
\label{beta4}
\end{equation}
    \noindent where $X,Y$ are  second-order variables, $\alpha$ is a \sigmalfpu  formula over $\sigma$,  $\varphi,\psi_i$, $1\leq i\leq r$, are \folfp formulae over $\sigma$, $\psi_i$, $1\leq i\leq r$, strictly extend $X$ to $Y$, and $f$ is a second-order function symbol.
\end{enumerate} 
\end{definition}

\begin{remark}
Note that instead of $\varphi(\underline{X}) \cdot\big(\top+\sum_{i=1}^r Y:=\psi_i(X)\cdot f(Y)\big)$ we could have  $\sum_{i=1}^r\quantsigma Y.\, \varphi(\underline{X})\cdot\big(\top+\psi_i(X,Y)\cdot f(Y)\big)$ in grammar~(\ref{beta4}).  In the rest of this subsection we use the latter formula.
\end{remark}

We show how the generic \totp problem is expressed in \totlogiclfpth. We first describe how an NPTM run can be encoded. The idea is the same as in the previous sections, and the details are suitable for the scope of this section.
We fix an NPTM $N=(\mathcal{Q},\Sigma,\delta,q_0,q_F)$ 
% be a NPTM 
%(Nondeterministic Polynomial Time Turing Machine) 
that uses at most time $n^c-1$ on structures of size $n$. We define $\Gamma=\Sigma\cup\{\vartextvisiblespace\}=\{0,1,\vartextvisiblespace\}$, $\Gamma_\mathcal{Q}=\Gamma\times \mathcal{Q}$, and $k=\max\{c,\lceil\log(3+3|\mathcal{Q}|)\rceil\}$. W.l.o.g.\ assume that $N$ has a single tape.
We also fix a finite ordered structure $\mathcal{A}$ of size $n$.
To encode cells, time steps, and symbols in $\Gamma\cup \Gamma_\mathcal{Q}$, we use $k$-tuples of elements from $A$. The computation of $N$ is encoded using a relation $S$ of arity $3k$. If $\Vec{r}$ represents the symbol $\gamma\in\Gamma$, then $S(\Vec{c},\Vec{t},\Vec{r})$ signifies that cell $\Vec{c}$ contains symbol $\gamma$ at time step $\Vec{t}$. If $\Vec{r}$ represents the symbol-state pair $(\gamma,q)\in \Gamma_\mathcal{Q}$, then $S(\Vec{c},\Vec{t},\Vec{r})$ signifies that $\Vec{c}$ contains symbol $\gamma$, the head is at cell $\Vec{c}$, and $N$ is in state $q$ at time step $\Vec{t}$.
%
%For convenience, we use the expression $\overrightarrow{ctr}$ or $\overrightarrow{xyz}$ to denote $3k$-tuples which correspond to cell-time-symbol combinations, and 
We use the expressions $\Vec{s} \leq \Vec{u}$, $\Vec{s} < \Vec{u}$, $\Vec{s}+1$, $\Vec{s}-1$,
%$\Vec{s}{max}$, $\Vec{s}{min}$
to describe $k$-tuples $\Vec{s}=(s_1,\dots,s_k)$ and $\Vec{u}=(u_1,\dots,u_k)$ in the expected way; we use $\mathrm{min}$ to describe the smallest $k$-tuple. All of these expressions are defined in \fo.
%We also abuse notation and write $\varphi(S)$, instead of $\varphi(S/X)$, to denote that $X$ is replaced by $S$ in $\varphi$, where $S$ is a relation, $X$ is a second-order variable that occurs free in $\varphi$, and $\arity(S)=\arity(X)$.

Let $\Delta_{det}$, $\Delta_0$, and $\Delta_1$ denote the sets of deterministic, left non-deterministic, and right non-deterministic transitions of $N$, respectively. Moreover, we write $\Gamma_{det}$ and $\Gamma_{nondet}$ to denote the encodings of symbol-state combinations in $\Gamma_\mathcal{Q}$ that lead to a deterministic transition or a non-deterministic choice, respectively.
% We need relations $S_0, S_1,  S_{\vartextvisiblespace}, S_{(0,q_i)}, S_{(1,q_i)}, S_{(\vartextvisiblespace,q_i)}$, $0\leq i\leq |Q|-1$, each of arity $2k$ to encode the computation of the Turing machine. We denote by $\Vec{S}$ the sequence of all these $m:= 3+3|Q|$ relations.

\begin{definition}
    Let $S$ be a relation of arity $3k$ on a finite structure $\mathcal{A}$. We say that \emph{relation $S$ describes a partial run} $c_0c_1\cdots c_m$, equivalently a run $c_0c_1\cdots c_m$ of $N$ up to configuration $c_m$, when 
    \begin{itemize}
        \item there is some $\Vec{t} \in A^k$, such that for every $\vec{t'}\leq \vec{t}$, there are $\vec{c},\vec{r} \in A^k$, such that $S(\Vec{c},\vec{t'},\vec{r})$, and  for every $\vec{t'} > \vec{t}$ and $\vec{c},\vec{r} \in A^k$, it is not the case that $S(\Vec{c},\vec{t'},\vec{r})$;
        \item $S(-,\mathrm{min},-)$ describes the encoding of the starting configuration, $c_0$; and
        \item if $S(-,\vec{t},-)$ describes the encoding of  $c_i$, then $S(-,\vec{t}+1,-)$ either describes the encoding of  $c_{i+1}$ or is empty.
    \end{itemize}
    We say that \emph{formula $\varphi(\Vec{c},\vec{t},\vec{r})$ describes a partial run} $c_0c_1\cdots c_m$, equivalently a run $c_0c_1\cdots c_m$ of $N$ up to configuration $c_m$, when $\varphi$ defines in $\mathcal{A}$ a relation that does so.
\end{definition}

We use the standard notion of definability, where $\varphi(\Vec{x})$ defines $R$ in $\mathcal{A}$, if for every $\Vec{a}\in A^k$, $R(\Vec{a})$ iff $\mathcal{A},v[\Vec{a}/\Vec{x}]\models\varphi(\Vec{x})$. For example, let $S_0$ be a relation of arity $3k$ that describes the beginning of a run by $N$ on $\enc(\mathcal{A})$.
$S_0$ can be defined in \fo{} by $\Vec{y} = \mathrm{min} 
\land \varphi_{c_0}(\vec{x},\vec{z})$, where $\varphi_{c_0}$  encodes the starting configuration, as, for instance, in \cite{Immerman}.

% To express the problem of counting the number of branchings of $N$, we need the least fixed point of 
 The following 
formula $\mathsf{tot}(X,f)$ contains a free second-order variable and a free second-order function symbol.
Its least fixed point 
applied on $S_0$ counts 
% expresses the problem of counting 
the number of branchings of $N$ on  $\enc(\mathcal{A})$.
% $$\begin{aligned}
% &\mathsf{tot}(S_{cur},f):=S_{cur}\cdot\Big(\mathsf{exists\_branching}(S_{cur}) +\\
% &\quantsigma S_{old}.\quantsigma S_{new}.\, \big(\mathsf{exists\_branching}(S_{cur})\wedge \mathsf{\Delta_0}(S_{old},S_{new})\big)\cdot f(S_{new})+\\
% &\quantsigma S_{old}.\quantsigma S_{new}.\,\big(\mathsf{exists\_branching}(S_{cur})\wedge \mathsf{\Delta_1}(S_{old},S_{new})\big)\cdot f(S_{new})\Big).
% \end{aligned}$$
%Formula $\mathsf{tot}$ can be written in the simpler following form.
$$\begin{aligned}
&\mathsf{tot}(X,f):=%\\
&\sum_{i=0,1}\quantsigma Y. \,\mathsf{exists\_branching}(\underline{X})\cdot\Big(\top+\mathsf{branch}_i(X,Y)\cdot f(Y)\Big).
\end{aligned}$$
% Assume that
Let
$X$ 
% is 
be
interpreted as a relation $S_p$ that describes  a  run $r$ by $N$  up to a configuration. Formula $\mathsf{exists\_branching}(\underline{X})$ looks ahead in the computation and asserts that the run will reach a non-deterministic branching of $N$ from some configuration $c_{nd}$. 
Let $c_0$ and $c_1$ be the configurations that 
% .
% , which 
respectively result from the left and right non-deterministic transitions from $c_{nd}$.
%The formula then proceeds to output 1) $S_p$; and 2) $S_p$ concatenated with any extensions of the run past that branching, as $\mathsf{branch}_i(X,Y)$ updates the run from $S_p$ to $S_{new}$ up to configuration $c_{i}$, $i=0,1$.
Formula $\mathsf{exists\_branching}(\underline{X})$ then proceeds to output $S_p$, and $\mathsf{non\_det}_i(X,Y)$ extends relation $S_p$ to $S_{new}$; relation $S_{new}$ describes the extension of run $r$ that passes through $c_{nd}$ and reaches $c_{i}$, $i=0,1$.
As such, every branching is mapped to a sequence of configurations that are visited by $N$ before $N$ reaches the branching. Below we describe in detail how these formulae are built.
% is is accomplished by formula $\mathsf{tot}(S_{cur},f)$.

To start with, we introduce formulae that express the transition function of $N$. For every transition $\tau\in \Delta_{det}\cup\Delta_0\cup\Delta_1$, formula $\mathsf{update}_\tau(X,\Vec{x},\Vec{y},\Vec{z})$ expresses how a relation $S_p$ that describes a partial run, has to be updated to encode one more step of the computation determined by $\tau$. 
Let $q_1,q_2 \in \mathcal{Q}$, $b_1, b_2 \in \{0,1\}$.
%, and $D \in \{\mathrm{L},\mathrm{R}\}$.
We present $\mathsf{update}_\tau(X,\Vec{x},\Vec{y},\Vec{z})$ for transition $\tau = ((q_1,b_1),(q_2,b_2,\mathrm{R}))$; the case of transitions that move the head to the left are similar.
\begin{alignat*}{1}
% &
\mathsf{update}_\tau(X,\Vec{x},\Vec{y},\Vec{z})=
%\\
% &\exists \Vec{x'}\exists \Vec{y'}
% \big(\Vec{x'}=\Vec{x}-1\wedge \Vec{y'}=\Vec{y}-1\wedge thanks!
&X(\Vec{x}-1,\Vec{y}-1,\Vec{b}_{1q_1})\wedge 
\bigvee_{b \in \{0,1,\vartextvisiblespace\}} \big(X(\Vec{x},\Vec{y}-1,\Vec{b})\wedge \Vec{z}= \Vec{b}_{q_2}\big) \,\, \vee \\
&
% \exists \Vec{y'}
\big(
% \Vec{y'}=\Vec{y}-1 \wedge 
X(\Vec{x},\Vec{y}-1,\Vec{b}_{1q_1})\wedge \Vec{z}=\Vec{b}_2\big)\,\, \vee\\
&\exists \Vec{x'} 
% \exists \Vec{y'} \big(
% \Vec{y'}=\Vec{y}-1\wedge 
X(\Vec{x'},\Vec{y}-1,\Vec{b}_{1q_1})\wedge 
\Vec{x}\neq \Vec{x'}\wedge \Vec{x}\neq \Vec{x'}+1\wedge X(\Vec{x}, \Vec{y}-1, \Vec{z})\big)
\end{alignat*}
where $\Vec{b}$ encodes symbol $b\in\{0,1,\vartextvisiblespace\}$ and $\Vec{b}_q$ encodes the symbol-state pair $(b,q)\in \{0,1\}\times \mathcal{Q}$.

\begin{lemma}\label{lem:updatet}
If $S_p$ describes a partial run $c_0c_1\cdots c_m$ of $N$ and $N$ can transition with $\tau \in \Delta$ from $c_m$ to $c_{m+1}$, then formula
$S_p(\Vec{x},\Vec{y},\Vec{z})\vee\mathsf{update}_\tau(S_p,\Vec{x},\Vec{y},\Vec{z})$
describes the partial run $c_0c_1\cdots c_mc_{m+1}$ of $N$.
If, on the other hand, $N$ cannot transition with $\tau \in \Delta$ from $c_m$, then $S_p(\Vec{x},\Vec{y},\Vec{z})\vee\mathsf{update}_\tau(S_p,\Vec{x},\Vec{y},\Vec{z})$ describes $c_0c_1\cdots c_m$.
% If $S_p$ describes a partial run $c_0c_1\cdots c_m$ of $N$ and $N$ can transition with $\tau \in \Delta$ from $c_m$ to $c_{m+1}$, then $\mathsf{update}_\tau(S_p,\Vec{x},\Vec{y},\Vec{z})$ 
% describes a partial run $c_0c_1\cdots c_mc_{m+1}$ of $N$. 
% If, on the other hand, $N$ cannot transition with $\tau \in \Delta$ from $c_m$, then $\mathsf{update}_\tau(S_p,\Vec{x},\Vec{y},\Vec{z})$ is false for all values of $\Vec{x},\Vec{y},\Vec{z}$.
\end{lemma}
\begin{proof}
    Notice that all three disjuncts of $\mathsf{update}_\tau$ ensure that $\tau$ can be applied to the last configuration of the run described by $S_p$. Then, we observe that $\mathsf{update}_\tau$ describes exactly how the first time step for which $S_p$ does not describe a configuration is updated according to $\tau$.
\end{proof}

% We provide an example in the case of $i$ being the  deterministic transition $((q_1,0),(q_2,1,\mathrm{R}))\in\Delta_{det}$. Assume that $N$ is in state $q_1$ and the cursor is on cell $c_1$ that contains $0$. Then, according to $i$, $N$ overwrites $0$ with $1$ and moves its cursor to the next right cell $c_2$ that happens to contain $1$. Then, the following formula corresponds to this transition. The first and second rows modify the contents of cells $c_2$ and $c_1$, respectively. The last row preserves the contents of all other cells. 
% \begin{alignat*}{1}
% &\mathsf{update}_i(S,\Vec{x},\Vec{y},\Vec{z})=\\
% &\exists \Vec{x'}\exists \Vec{y'}
% \big(\Vec{x'}=\Vec{x}-1\wedge \Vec{y'}=\Vec{y}-1\wedge 
% S(\Vec{x'},\Vec{y'},\Vec{0_{q_1}})\wedge S(\Vec{x},\Vec{y'},\Vec{1})\wedge \Vec{z}= \Vec{1_{q_2}}\big) \,\, \vee \\
% &\exists \Vec{y'}
% \big(\Vec{y'}=\Vec{y}-1 \wedge S(\Vec{x},\Vec{y'},\Vec{0_{q_1}})\wedge \Vec{z}=\Vec{1}\big)\,\, \vee\\
% &\exists \Vec{x'} \exists \Vec{y'} \big(\Vec{y'}=\Vec{y}-1\wedge S(\Vec{x'},\Vec{y'},\Vec{0_{q_1}})\wedge 
% \Vec{x}\neq \Vec{x'}\wedge \Vec{x}\neq \Vec{x'}+1\wedge S(\Vec{x}, \Vec{y'}, \Vec{z})\big).
% \end{alignat*}
Define formula $\mathsf{detstep}$ to be
$\mathsf{detstep}(X,\Vec{x},\Vec{y},\Vec{z}):= X(\Vec{x},\Vec{y},\Vec{z})\vee \bigvee_{\tau\in\Delta_{det}} \mathsf{update}_\tau(X,\Vec{x},\Vec{y},\Vec{z}).$

\begin{lemma}\label{lem:detstep}
    If $S_p$ describes a partial run $c_0c_1\cdots c_m$ of $N$ and $N$ can deterministically transition from $c_m$ to $c_{m+1}$, then formula $\mathsf{detstep}(S_p,\Vec{x},\Vec{y},\Vec{z})$  
    describes the partial run $c_0c_1\cdots c_mc_{m+1}$ of $N$. 
    If, on the other hand, $N$ has no deterministic transition from $c_m$, then $\mathsf{detstep}(S_p,\Vec{x},\Vec{y},\Vec{z})$ 
    describes partial run $c_0c_1\cdots c_m$. 
\end{lemma}
\begin{proof}
    Immediate from Lemma \ref{lem:updatet}.
\end{proof}

% Let $S_{cur}$ be a relation that 
% describes a partial run 
% % encodes an ongoing computation 
% $r$ of $N$ up to some time step $t$. Suppose there is a deterministic transition of $N$ that can extend $r$. In that case, formula $\mathsf{detstep}$ 
% describes partial run $r'$ that 
% % defines a relation $S$ that encodes computation $c'$ which 
% agrees with $c$ in the first $t$ steps and includes this deterministic transition of $N$. Otherwise, $\mathsf{detstep}$ describes $c'=c$.
% $$
% % \begin{aligned}
% \mathsf{detstep}(S_{cur},\Vec{x},\Vec{y},\Vec{z}):= S_{cur}(\Vec{x},\Vec{y},\Vec{z})\vee \bigvee_{\tau\in\Delta_{det}} \mathsf{update}_\tau(S_{cur},\Vec{x},\Vec{y},\Vec{z}).
% % \end{aligned}
% $$
% \begin{lemma}\label{lem:detstep}
%     If $S_p$ describes a partial run $C_0C_1\cdots C_m$ of $N$ and $N$ can deterministically transition from $C_m$ to $C_{m+1}$, then $\mathsf{detstep}(S_p,\Vec{x},\Vec{y},\Vec{z})$ 
%     describes a partial run $C_0C_1\cdots C_mC_{m+1}$ of $N$. 
%     If, on the other hand, $N$ has no deterministic transition from $C_m$, then $\mathsf{detstep}(S_p,\Vec{x},\Vec{y},\Vec{z})$ 
%     describes partial run $C_0C_1\cdots C_m$. 
% \end{lemma}
% \begin{proof}
%     Immediate from Lemma \ref{lem:updatet}.
% \end{proof}

Using a least fixed point on $\mathsf{detstep}$, we can describe a maximal deterministic extension of a run with formula $\mathsf{detcomp}$:
$$
\mathsf{detcomp}(X,\Vec{x},\Vec{y},\Vec{z}):= 
[\lfpd_{Y,\Vec{x},\Vec{y},\Vec{z}}\,\mathsf{detstep}(Y,\Vec{x},\Vec{y},\Vec{z})
\lor X(\Vec{x},\Vec{y},\Vec{z})
].
$$

Note that $Y$ appears positive in $\mathsf{detstep}(Y,\Vec{x},\Vec{y},\Vec{z})$, so $\mathsf{detcomp}$ is well-defined. Given $S_p$ that describes a computation $c_0\dots c_m$, $\mathsf{detcomp}(S_p,\Vec{x},\Vec{y},\Vec{z})$ defines a relation that
describes 
% defines a relation 
% $S$ which 
 % that 
% encodes 
a computation which starts with $c_0\dots c_m$, continues with all possible deterministic transitions, and finally, it reaches a configuration in which $N$ terminates or can make a non-deterministic transition. In other words, this formula 
extends
% encodes 
$S_p$ 
% together 
with a maximal deterministic computation. 
% that extends it. 
%By a maximal deterministic computation, we mean that such a computation cannot be extended by a deterministic transition of $N$.

Formula $\mathsf{exists\_branching}$ updates $S_p$ as described above and it detects the existence of a new branching, i.e.\ a branching that is not in the partial run $c_0\dots c_m$ described by $S_p$, but occurs right after a maximal deterministic computation that extends $c_0\dots c_m$.
$$
\mathsf{exists\_branching}(X):=
\exists\Vec{x}\exists \Vec{y}\exists\Vec{z}\Big(\mathsf{detcomp}(X,\Vec{x},\Vec{y},\Vec{z})\wedge\bigvee_{\Vec{\gamma}\in \Gamma_{nondet}}\Vec{z}=\Vec{\gamma}\wedge
\neg X(\Vec{x},\Vec{y},\Vec{z})
\Big) 
.$$
\begin{lemma}\label{lem:exists_branching}
    If $S_p$ describes a partial run $c_0c_1\cdots c_m$ of $N$, then $\mathcal{A},V[S_p/X]\models\mathsf{exists\_branching}(X)$ 
    if and only if 
    $c_0c_1\cdots c_m$ can be extended to $c_0c_1\cdots c_{l}$, where $N(\enc(\mathcal{A}))$ has a non-deterministic choice in $c_{l}$. 
\end{lemma}
\begin{proof}
    Immediate from Lemma \ref{lem:detstep} and the definitions of formulae $\mathsf{exists\_branching}(X)$ and $\mathsf{detcomp}(X,\Vec{x},\Vec{y},\Vec{z})$.
\end{proof}
 
% $$\begin{aligned}
% &\mathsf{exists\_branching}(S_{cur}):=\\
% &\exists\Vec{ctr}\Big([\lfpd_{S,\Vec{xyz}}\mathsf{detstep}](\Vec{ctr})\wedge\bigvee_{\Vec{\gamma}\in \Gamma_{nondet}}\Vec{r}=\Vec{\gamma}\wedge\\
% &\forall \Vec{c't'r'}\big((\Vec{t'}< \Vec{t}\wedge\forall \Vec{c''r''} \neg S_{cur}(\Vec{c''t'r''}))\rightarrow [\lfpd_{S,\Vec{xyz}}\mathsf{detstep}](\Vec{c't'r'}) \rightarrow \bigvee_{\Vec{\gamma}\in \Gamma_{det}}\Vec{r'}=\Vec{\gamma}\big)\Big)
% \end{aligned}.$$

Formulae $\mathsf{branch}_i$, $i=0,1$, extend a relation $S_p$ that describes a run $r$ to a relation $S_{new}$, 
% which extends $S_{cur}$ in the following way: 
that describes a run which extends $r$ with a maximal deterministic computation,
and then with the configuration that $N$ reaches by making non-deterministic choice $i$, if such a choice is possible.
% following the one encoded  by $S_{cur}$, and also the non-deterministic choice $i$ made after that (if choice $i$ is possible).
% $$\begin{aligned}
% &\mathsf{\Delta}_i(S_{old},S_{new}):=\\
% &\forall \Vec{ctr} 
% \big(S_{old}(\Vec{ctr}) \leftrightarrow  [\lfpd_{S,\Vec{xyz}}\mathsf{detstep}](\Vec{ctr}) \,\wedge S_{new}(\Vec{ctr})\leftrightarrow(S_{old}(\Vec{ctr}) \vee \bigvee_{i\in\Delta_i}\mathsf{update}_i(S_{old}, \Vec{ctr}))\big)
% \end{aligned}$$
% or using only the second-order variable $S_{new}$:
$$\begin{aligned}
&\mathsf{branch}_i(X,Y):=\\
&\forall \Vec{x}\forall\Vec{y}\forall\Vec{z} 
\Big(Y(\Vec{x},\Vec{y},\Vec{z})\leftrightarrow  \big(\mathsf{detcomp}(X,\Vec{x},\Vec{y},\Vec{z}) \vee \bigvee_{\tau\in\Delta_i}\mathsf{update}_\tau(\mathsf{detcomp}(X,\Vec{x},\Vec{y},\Vec{z}))\big)\Big).
\end{aligned}$$

\begin{lemma}\label{lem:branch}
   If $S_p$ describes a partial run $r:=c_0\cdots c_m$ of $N$, and $\mathcal{A},V[S_p/X,S_{new}/Y]\models\mathsf{branch}_i(X,Y)$, $i=0,1$, then $S_{new}$ describes an extension $c_0\cdots c_{l-1}c_l$ of $r$, where $c_{l-1}$ is the first configuration that occurs after $c_{m-1}$ on which $N$ can make a non-deterministic choice, and $c_l$ is the configuration that $N$ transitions to, if $N$ makes choice $i$ in $c_l$. If, on the other hand, $N$ can only make a deterministic computation starting from $c_m$, then $S_{new}$  describes an extension $c_0\cdots c_l$ of $r$, where $c_l$ is a final configuration.  
\end{lemma}
\begin{proof}
    Immediate from the definition of formulae $\mathsf{detcomp}(X,\Vec{x},\Vec{y},\Vec{z})$ and $\mathsf{update}_\tau(X,\Vec{x},\Vec{y},\Vec{z})$. 
\end{proof}

% \begin{alignat*}{1}
% \mathsf{update}_i(S,\Vec{x},\Vec{y},\Vec{z})= &\exists \Vec{x'}\exists \Vec{y'}\exists \Vec{z'}\exists \Vec{z''} 
% \big(\Vec{x'}=\Vec{x}-1\wedge \Vec{y'}=\Vec{y}-1\wedge \Vec{z'}=\Vec{0_{q_1}}\wedge \Vec{z''}=\Vec{1}\wedge \\
% &S(\Vec{x'},\Vec{y'},\Vec{z'})\wedge S(\Vec{x},\Vec{y'},\Vec{z''})\wedge \Vec{z}= \Vec{1_{q_2}}\big) \,\, \vee \\
% &\exists \Vec{y'}\exists \Vec{z'}
% \big(\Vec{y'}=\Vec{y}-1 \wedge \Vec{z'}=\Vec{0_{q_1}}\wedge S(\Vec{x},\Vec{y'},\Vec{z'})\wedge \Vec{z}=\Vec{1}\big)\,\, \vee\\
% &\exists \Vec{x'} \exists \Vec{y'} \exists \Vec{z'}\big(\Vec{y'}=\Vec{y}-1\wedge \Vec{z'}= \Vec{0_{q_1}}\wedge \\ &S(\Vec{x'},\Vec{y'},\Vec{z'})\wedge 
% \Vec{x}\neq \Vec{x'}\wedge \Vec{x}\neq \Vec{x'}+1\wedge S(\Vec{x}, \Vec{y'}, \Vec{z})\big).
% \end{alignat*}

%Formula $[\lfp_f \mathsf{tot}](S_{cur})$ expresses the generic \totp{} problem

The proof of the following theorem demonstrates that by evaluating the least fixed point of formula $\mathsf{tot}$, we obtain a set of strings, that each of them corresponds to  a different branching of TM $N$.

\begin{theorem}\label{totp theorem}
    Given an NPTM $N$, $\llbracket \, [\lfp_f \mathsf{tot}](X)\,\rrbracket (\mathcal{A},v,V)=\#(\text{branchings of } N \text{ on } \enc(\mathcal{A}))$, for every $\mathcal{A},v$, and $V$, such that $V(X)$ encodes the initial configuration of $N$. 
\end{theorem}
\begin{proof}
By Lemmata~\ref{lem:exists_branching} and~\ref{lem:branch} and by the definition of $\mathsf{tot}$, $\rel[\,[\lfp_f \mathsf{tot}](X)\,](\mathcal{A},v,V)$, where $V(X)$ encodes the initial configuration, consists of strings $S_0\circl\dots \circl S_m\in(\mathcal{R}_{3k})^*$, where $S_i$ extends $S_{i-1}$ and describes a run from the initial configuration $c_0$ up to a configuration occurring exactly after a non-deterministic choice of $N$. More precisely, there is a bijection between strings in $\rel[\,[\lfp_f \mathsf{tot}](X)\,](\mathcal{A},v,V)$  and branchings of $N$; let $S_m$ describe a run of $N$ up to configuration $c$. Then, $S_0\circl\dots \circl S_m$ is mapped  to the first branching (or non-deterministic choice) that $N$ reaches after starting from $c$ and making some (or no) deterministic transitions. For every branching $b$ of $N$, there is a unique string that is contained in $\rel[\,[\lfp_f \mathsf{tot}](X)\,](\mathcal{A},v,V)$ and is mapped to $b$. Thus, $|\rel[\,[\lfp_f \mathsf{tot}](X)\,](\mathcal{A},v,V)|$ is equal to the number of branchings of $N(\enc(\mathcal{A}))$.
\end{proof}

Every \totp{} problem can be expressed in \totlogiclfpth{}, since formula $[\lfp_f \mathsf{tot}](X)$ can be easily transformed into a \totlogiclfpth formula.

\begin{proposition}\label{totp first inclusion}
$\totp \subseteq \totclasslfpu$ over finite ordered structures.
\end{proposition}
\begin{proof}
This is immediate from Theorem~\ref{totp theorem}, Remark~\ref{branchings}, and the observation that formula $\mathsf{tot}$ can be defined by grammar~(\ref{beta4}), since $\mathsf{branch}_i$, $i=0,1$, can be replaced by the following formulae that strictly extend $X$ to $Y$:
\begin{alignat*}{2}
\mathsf{branch}_i'(X,Y)&:=\forall \Vec{x}\forall\Vec{y}\forall\Vec{z} \Big(Y(\Vec{x},\Vec{y},\Vec{z})\leftrightarrow\\
&\big(X(\Vec{x},\Vec{y},\Vec{z})\vee\mathsf{detcomp}(X,\Vec{x},\Vec{y},\Vec{z}) \vee \bigvee_{\tau\in\Delta_i}\mathsf{update}_\tau(\mathsf{detcomp}(X,\Vec{x},\Vec{y},\Vec{z}))\big)\\
&\wedge\exists \Vec{x}\exists\Vec{y}\exists\Vec{z}\big(\neg X(\Vec{x},\Vec{y},\Vec{z})\wedge Y(\Vec{x},\Vec{y},\Vec{z})\big).
% \hspace{8.3cm}    %%% this pushes the qed out of the page
% &\hfill
\tag*{\qedhere}     %% let's keep this for arXiv
% \qedhere
\end{alignat*}
\end{proof}

Below we give an example of a \totp{} problem expressed in \totlogiclfpth.

\begin{figure}
    \centering
    \includegraphics[scale=1]{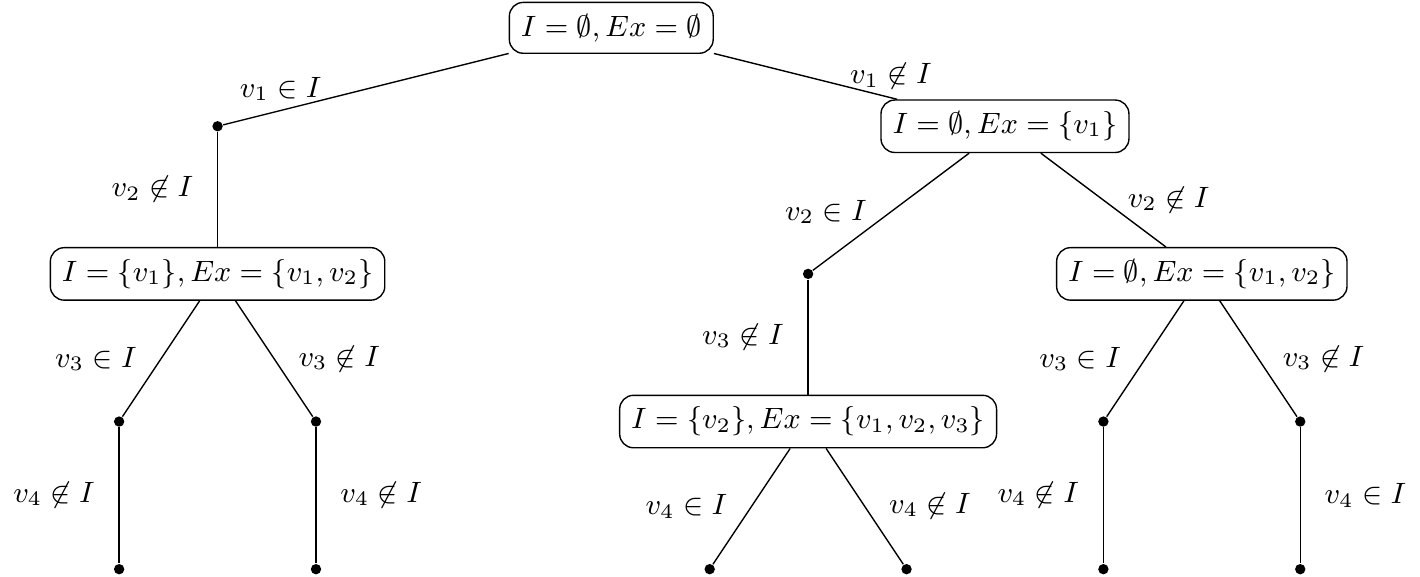}
    \caption{A strategy to compute all independent sets in $C_4$, i.e.\ the cycle of length 4. The labels on the nodes indicate the independent set and the set of vertices of $C_4$ already examined at the moment. }
    \label{fig:ind_sets}
\end{figure}

\begin{example}\label{ex:ind_sets}
To express  \textsc{\#IS} in \totlogiclfpth, we use a similar strategy as in Example~\ref{totp example 1}(b). Figure~\ref{fig:ind_sets} illustrates this strategy on input a cycle of length 4. We assume that the vertices are examined in increasing order with respect to $\leq$.
We use two unary relations $I$ and $Ex$ that represent the independent set and the set of vertices that have been already examined at the moment,  respectively.  For instance, in Figure~\ref{fig:ind_sets}, we have included the values of $I$ and $Ex$ in the cases that a new vertex can be both included and excluded from the independent set. We slightly abuse notation and use symbols  $I$ and $Ex$ to also  denote the second-order variables in the formulae given below. We define
formula $\textsf{is}(I,Ex,f)$ as follows:
$$\begin{aligned}
(Ex=\emptyset) + \quantsigma I'.\quantsigma Ex'. \textsf{can\_extend}(I,Ex) \cdot I \cdot Ex \cdot \big(\top+\textsf{include}(I, Ex, I',Ex') \cdot f(I',Ex')\big) \\+ \quantsigma I'.\quantsigma Ex'. \textsf{can\_extend}(I,Ex) \cdot I \cdot Ex \cdot \big(\top+\textsf{exclude}(I, Ex, I',Ex') \cdot f(I',Ex')\big).
\end{aligned}$$

The evaluation of $[\lfp_f \textsf{is}](I,Ex)$ produces a different output for 
every time that both including and excluding a specific vertex in $I$ are possible. 
Formula $\textsf{can\_extend}$ is true when this is the case for some vertex $v$ not examined yet, and so $I$ and $Ex$ are returned as an output.  
Formula $\textsf{include}$ defines $I'$ to be $I\cup\{v\}$, and $Ex'=\{u ~\mid~ u\leq v\}$, whereas formula $\textsf{exclude}$ defines $I'$ to be identical to $I$ and $Ex'=\{u ~\mid~ u\leq v\}$.
Note that formula $(Ex=\emptyset)$ produces an additional different output, and so $\llbracket\,[\lfp_f \textsf{is}](I,Ex)\,\rrbracket(\mathcal{A},v,V)=
\#\mathsf{IS}(\enc(\mathcal{A}))$, when $V(I)=V(Ex)=\emptyset$. Moreover, by Remark~\ref{encoding of relations}, $I$ and $Ex$ can be encoded by exactly one relation. 

All subformulae of $\textsf{is}(I,Ex,f)$ are given below. They can be easily modified so they are exactly as the definition of \totlogiclfpth requires.
\begin{itemize}
    \item $\mathsf{can\_extend}(I,Ex):=\exists x \varphi(x)$, where
    $$\begin{aligned}
        \varphi(x):= &\neg Ex(x)\wedge \forall y \big(I(y)\rightarrow \neg E(x,y) \big)\wedge
        \\
        &
        \forall {x' < x}(
        % x'<x\rightarrow 
        Ex(x')\vee \exists y \big(I(y)\wedge E(x',y)\big) \wedge \\
        &(
        % \exists y I(y) 
        I \neq \emptyset
        \vee x\neq \mathrm{max}).
    \end{aligned}$$
    Intuitively, formula $\varphi$ finds the minimum not examined vertex so far, that can be both included and excluded from the independent set. Note that a vertex $v$ can only be excluded from the independent set $I$ if it is connected with some vertex in $I$, and only be included in $I$ if $I$ is empty and $v$ is the last (maximum) vertex examined. 
    \item $\mathsf{include}(I,Ex,I',Ex'):=\forall z \big( I'(z)\leftrightarrow I(z)\vee \varphi(z)\big)\wedge \forall z \big(Ex'(z) \leftrightarrow  Ex(z)\vee 
    % \varphi(z)\vee 
    \exists 
    % y 
    % (
    y \geq  z ~
    % \wedge 
    \varphi(y)
    \big)
    $. This formula adds vertex $v$ detected by formula $\mathsf{can\_extend}$ to the independent set and marks all vertices smaller or equal to $v$ as examined. 
    \item $\mathsf{exclude}(I,Ex,I',Ex'):= \forall z \big( I'(z) \leftrightarrow I(z)\big) \wedge \forall z \big(Ex'(z) \leftrightarrow  Ex(z)\vee 
    % \varphi(z)\vee 
    \exists y \geq z ~ 
    % (y>  z \wedge 
    \varphi(y)\big)$. Formula $\mathsf{exclude}$ excludes $v$ from the independent set and marks all vertices smaller or equal to $v$ as examined. 
    \item %$(Ex=\emptyset) := \forall x \neg Ex(x)$.
    %% alt:
     For every $P$, $P = \emptyset$ stands for $\forall x \neg P(x)$.
\end{itemize}
\end{example}

To prove the inverse inclusion $\totclasslfpu\subseteq \totp$, we first need Lemmata~\ref{syntactically-defines-implication}--\ref{poly-algo-extends-formula}.

Let $\varphi(Y)=\forall \Vec{y} Y(\Vec{y})\leftrightarrow \psi(\Vec{y})$. If $\mathcal{A},v[\Vec{a}/\Vec{y}]\models\psi(\Vec{y})$ can be determined in polynomial time for every $\mathcal{A}$ and $\Vec{a}\in A^k$, then a relation $B\in\mathcal{R}_k$ such that $\mathcal{A},V[B/Y]\models\varphi(Y)$ can be found in polynomial time if such $B$ exists. Lemma~\ref{poly-algo-define-formula} demonstrates this fact.  The following lemma guarantees that if such $B$ exists, then it is unique. 

\begin{lemma}\label{syntactically-defines-implication}
Let $\mathcal{A}$ be a finite ordered structure over $\sigma$ and  $\varphi(Y)$ be a formula over $\sigma$ that syntactically defines $Y$, where $\arity(Y)=k$. Then, there is at most one $B\in\mathcal{R}_k$, such that $\mathcal{A},V[B/Y]\models\varphi(Y)$. 
\end{lemma}

\begin{lemma}\label{poly-algo-define-formula}
Let $\varphi(Y)$ be an \folfp formula over $\sigma$ that syntactically defines $Y$, where $\arity(Y)=k$. There is a poly-time algorithm that on input a finite ordered structure $\mathcal{A}$ over $\sigma$, decides whether there is $B\in\mathcal{R}_k$, such that $\mathcal{A},V[B/Y]\models\varphi(Y)$. If the answer is positive, the algorithm outputs $B$.
\end{lemma}

For a formula $\varphi(X,Y)$ as in Definition~\ref{def:extends}, if $\mathcal{A},v[\Vec{a}/\Vec{y}],V[B/X]\models \psi(X,\Vec{y})$ can be decided in polynomial time, then there is at most one relation $C\in\mathcal{R}_k$ such that $\mathcal{A},V[B/X,C/Y]\models\varphi(X,Y)$ and $C$ can be found in polynomial time. Moreover, $C$ is a superset of $B$, or in other words, $C$ extends $B$. Lemmata~\ref{poly-algo-extends-formula} and~\ref{monotone implication} state these facts.

\begin{lemma}\label{monotone implication}
Let $\mathcal{A}$ be a finite ordered structure over $\sigma$, $\varphi(X,Y)$ be a formula over $\sigma$  that (strictly) extends $X$ to $Y$, where $\arity(X)=\arity(Y)=k$, and $B\in\mathcal{R}_k$. Then there is at most one $C\in\mathcal{R}_k$ such that $\mathcal{A},V[B/X,C/Y]\models\varphi(X,Y)$. In addition, it holds that:
\begin{enumerate}[(a)]
    \item if $\varphi$ extends $X$ to $Y$ and $\mathcal{A},V[B/X,C/Y]\models\varphi(X,Y)$, then $B\subseteq C$ and
    \item if $\varphi$ strictly extends $X$ to $Y$ and $\mathcal{A},V[B/X,C/Y]\models\varphi(X,Y)$, then $B\subsetneq C$.
\end{enumerate}
\end{lemma}

\begin{lemma}\label{poly-algo-extends-formula}
Let $\varphi(X,Y)$ be an \folfp formula over $\sigma$ that (strictly) extends $X$ to $Y$, where $\arity(X)=\arity(Y)=k$. There is a poly-time algorithm that on input a finite ordered structure $\mathcal{A}$ over $\sigma$ and $B\in\mathcal{R}_k$, decides whether there is $C\in\mathcal{R}_k$, such that $\mathcal{A},V[B/X,C/Y]\models\varphi(X,Y)$. If the answer is positive, the algorithm outputs $C$.
\end{lemma}

The following statements will also be used in the proof of Proposition~\ref{totp second inclusion}. Lemma~\ref{lem:alpha-in-p}  and Corollaries~\ref{lem:time-bound-tocheck-a} and~\ref{cor:alpha-in-totp} demonstrate that  \sigmalfpu formulae can be verified and evaluated by polynomial-time TMs. 

\begin{lemma}\label{lem:alpha-in-p}
Let $\alpha$ be a \sigmalfpu formula over $\sigma$. The size of $\rel[\alpha](\mathcal{A},v,V)$ is polynomial in $|A|$, for every $\mathcal{A},v$, and $V$. Moreover, there is a deterministic poly-time TM $M$, which on input $\enc(\mathcal{A},v,V)$,  outputs all elements of $\rel[\alpha](\mathcal{A},v,V)$ in $\mathcal{O}(\mathrm{poly}(|A|))$ time.
\end{lemma}
\begin{proof}
    The lemma can be proven by induction on $\alpha$. 
    %If $\alpha=X$ then $M$ outputs $\enc(V(X))$.
    We just describe the new case of $\alpha =\quantsigma Y. \psi\cdot \alpha'$, where $\psi$ syntactically defines $Y$. In that case,  $M_\alpha$ can determine the unique $B\in \mathcal{R}_{\arity(Y)}$ such that $\mathcal{A},V[B/Y]\models\psi(Y)$ in polynomial time by Lemma~\ref{poly-algo-define-formula}. Then, the output of $M_\alpha$ is the output of  $M_{\alpha'(B)}(\mathcal{A},v,V)$, where $\alpha'(B)$ denotes formula $\alpha'(Y)$ interpreted in $\mathcal{A}$, such that $Y$ is assigned $B$.
    By Lemma~\ref{lem:bounded-length-a} the length of the output is also polynomial. 
\end{proof}

\begin{corollary}\label{lem:time-bound-tocheck-a}
Let $\alpha$ be a \sigmalfpu{} formula over $\sigma$. There is a deterministic TM $M$, such that on input 
$\enc(\mathcal{A},v,V)$, and a string $s\in(\bigcup_{i\in\mathbb{N}}\mathcal{R}_i)^*$, $M$ decides if $s \in \rel[\alpha](\mathcal{A},v,V)$ in $\mathcal{O}(\mathrm{poly}(|A|))$ time.
\end{corollary}

\begin{corollary}\label{cor:alpha-in-totp}
Let $\alpha$ be a \sigmalfpu\ formula over $\sigma$. There is an NPTM $M$, such that $tot_M(\enc(\mathcal{A},v,V))=\llbracket \alpha\rrbracket (\mathcal{A},v,V)$, for every $\mathcal{A},v$ and $V$.
\end{corollary}
\begin{proof}
Define $M$ to be the NPTM that on input $\enc(\mathcal{A},v,V)$ simulates the deterministic TM of Lemma~\ref{lem:alpha-in-p}, stores all $s\in \rel[\alpha](\mathcal{A},v,V)$ in its work tape and generates a path for every such $s$, plus an additional dummy path.
\end{proof}

Lemma~\ref{lem:time-bound-tocheck-a-2} demonstrates that the membership of any string $s$ in the intermediate interpretation of any \totlogiclfpth{} formula can be verified in polynomial time w.r.t.\ $|A|$, but exponential w.r.t.\ $|s|$. Consequently, it can be done in polynomial time w.r.t.\ $|A|$  when $|s|$ is constant.

\begin{lemma}\label{lem:time-bound-tocheck-a-2}
Let $[\lfp_f \beta](X)$ be an \totlogiclfpth{} formula over $\sigma$.
There is a deterministic TM $M_\beta$, such that on input 
$\enc(\mathcal{A},v,V)$, and a string $s\in(\bigcup_{i\in\mathbb{N}}\mathcal{R}_i)^*$, $M_\beta$ decides if $s \in \rel[\,[\lfp_f \beta](X)\,](\mathcal{A},v,V)$ in time $\mathcal{O}\big(c^{|s|}\cdot (\mathrm{poly}(|A|)+|\enc(s)|)\big)$, for some constant $c$.
\end{lemma}
\begin{proof} If $\beta\in\sigmalfpu$, 
then the lemma follows from Corollary~\ref{lem:time-bound-tocheck-a}. We define $M_\beta$ when $\beta=\alpha+\sum_{i=1}^r\quantsigma Y.\, \varphi(\underline{X})\cdot\big(\top+\psi_i(X,Y)\cdot f(Y)\big)$, where $\alpha\in\sigmalfpu$, in Algorithm~\ref{alg:totp-membership}. Let $\arity(X)=\arity(Y)=k$; let also $N_{\psi_i}$ denote the poly-time TM from Lemma~\ref{poly-algo-extends-formula} associated with $\psi_i$: on input $\enc(\mathcal{A},v,V)$, if there is $C$ such that  $\mathcal{A},V[C/Y]\models \psi_i(X,Y)$, $N_{\psi_i}$ returns $C$, and otherwise it rejects. Let $\mathrm{out}_{N_{\psi_i}}(V(X))$ denote the output of $N_{\psi_i}$'s computation on input $\enc(\mathcal{A},v,V)$.
\begin{algorithm}
\caption{$M_\beta$ when $\beta=\alpha+\sum_{i=1}^r\quantsigma Y.\, \varphi(\underline{X})\cdot\big(\top+\psi_i(X,Y)\cdot f(Y)\big)$}\label{alg:totp-membership}
\DontPrintSemicolon
\KwIn{$s, \mathcal{A},v,V$}
\lIf{$M_\alpha(s,\mathcal{A},v,V)$ accepts}{accept}
\If{$(\mathcal{A},V\models\varphi(X))$ and $(s[1] == V(X))$}{
\lIf{$|s|>1$}{$s:=s[2:]$}
\lElse{accept}
\For{$i \gets 1$ to $r$}{
\If{$N_{\psi_i}(\mathcal{A},v,V)$ does not reject}
{$B_i:=\mathrm{out}_{N_{\psi_i}}(V(X))$\;
simulate $M_\beta(s,\mathcal{A},v,V[B_i/X])$}
}}
reject
\end{algorithm}
% \verb|def| $M_\beta(s,\mathcal{A},v,V)$\\
% \verb|    if| $M_\alpha(s,\mathcal{A},v,V)$ \verb|accepts then accept|\\
%     \verb|    if| ($\mathcal{A}\models\varphi(V(X)/X)$) \verb|and| ($s(1)$ \verb|==| $V(X)$):\\
% \verb|      if| $|s|>1$ \verb|then| $s:=s(2)\circl\dots\circl s(\mathrm{last})$ \verb|else accept|\\
% \verb|      for| $i$ \verb|from| $1$ \verb|to| $r$ $\verb|do:|$\\
% \verb|        if| $N_{\psi_i}(\mathcal{A},v,V)$ \verb|does not reject|:\\
% \verb|          |$B_i:=\mathrm{out}_{N_{\psi_i}}(V(X))$\\
% \verb|          simulate| $M_\beta(s,\mathcal{A},v,V[B_i/X])$\\
% \verb|    reject|\\

Regarding the complexity of Algorithm~\ref{alg:totp-membership}, during the computation of $M_\beta$, at most $r^{|s|}$ recursive calls are made, where each call requires $\mathcal{O}(\mathrm{poly}(|A|)+|\enc(s)|)$ time; during a call, $M_\alpha$ and $N_{\psi_i}$ use poly($|A|$) time by Corollary~\ref{lem:time-bound-tocheck-a} and Lemma~\ref{poly-algo-extends-formula}, respectively, $\mathcal{A},V\models\varphi(X)$ can be checked in poly($|A|$) time, since $\varphi$ is in \folfp, and a substring of $s$ is stored which needs at most $|\enc(s)|$ time.
\end{proof}

\begin{remark}
    A more careful analysis of Algorithm \ref{alg:totp-membership} yields a linear dependency of its running time, with respect to $|s|$.
    The recursive call in line 8 does not generate a full computation for each $B_i$. Only one of these relations can appear as the first symbol of $s$, and thus with a more careful collection of the relations $B_i$ and a limited lookup, the algorithm only needs to recurse for one $i$.
    However, a linear dependency with respect to $|s|$ is not necessary for the following results, and we prefer a clear presentation for the algorithm.
\end{remark}

% \begin{proof}
% $M_\alpha$ is defined recursively on $\alpha$ and $|s|$ as in the proof of Lemma~\ref{lem:space-bound-tocheck-a}.
% We only describe the new case of $\alpha =\quantsigma Y. \psi\cdot \alpha'$, where $\psi$ defines $Y$. In that case,  $M_\alpha$ can determine the unique $B\in \mathcal{R}_{\arity(Y)}$ such that $\mathcal{A}\models\psi(B/Y)$ in polynomial time by Lemma~\ref{poly-algo-define-formula}. Then, $M_\alpha$ simulates $M_{\alpha'(B/Y)}(s,\mathcal{A},v,V)$. 
% \end{proof}

% \begin{lemma}\label{lem:alpha-in-p}
% Let $\alpha$ be a \sigmalfpu\ formula over $\sigma$. There is a deterministic poly-time TM $M$, which on input $(\mathcal{A},v,V)$,  outputs the value of $\llbracket \alpha\rrbracket (\mathcal{A},v,V)$.
% \end{lemma}
% \begin{proof}
% By induction on $\alpha$, it can be proven that the size of $\rel[\alpha](\mathcal{A},v,V)$ is polynomial in $|A|$.
% Let $\alpha$ contain second-order variables of arity at most $m$.  Using Lemmata \ref{lem:bounded-length-a} and \ref{lem:time-bound-tocheck-a}, $M$ can store every string $s \in (\bigcup_{1\leq i\leq m}\mathcal{R}_i)^*$ of length at most $|\alpha|$ and check whether $s \in \rel[\alpha](\mathcal{A},v,V)$ in polynomial time. For every stored $s$ that belongs to $\rel[\alpha](\mathcal{A},v,V)$, $M$ increases a counter by one.
% \end{proof}

We can now prove that a \totlogiclfpth{} formula $\beta$ can be evaluated by an NPTM $M$ in the sense that $M(\enc(\mathcal{A},v,V))$ generates $\llbracket \beta\rrbracket (\mathcal{A},v,V)+1$ paths.

\begin{proposition}\label{totp second inclusion}
$\totclasslfpu \subseteq \totp$ over finite ordered structures.
\end{proposition}
\begin{proof} %(of Proposition~\ref{totp second inclusion})
\begin{algorithm}
\caption{NPTM $M_\beta$ where $[\lfp_f \beta](X)\in\totlogiclfpth$}\label{alg:totp}
\DontPrintSemicolon
\KwIn{$\mathcal{A},v,V$}
\If{$\beta==\alpha$ has no function symbol}{
simulate $M_\alpha(\mathcal{A},v,V)$ defined in the proof of Corollary~\ref{cor:alpha-in-totp}\;
}
\If{$\beta==\sum_{i=1}^r\quantsigma Y.\, \varphi(\underline{X})\cdot\big(\top+\psi_i(X,Y)\cdot f(Y)\big)$}{
\lIf{$\mathcal{A},V\not\models\varphi(X)$}{stop}
\lElse{non-deterministically choose between stop and simulate $M_{rec}(\beta,\mathcal{A},v,V)$}
}
\If{$\beta==\alpha + \sum_{i=1}^r\quantsigma Y.\, \varphi(\underline{X})\cdot\big(\top+\psi_i(X,Y)\cdot f(Y)\big)$}{
\lIf{$\mathcal{A},V\not\models\varphi(X)$}{simulate $M_\alpha(\mathcal{A},v,V)$ defined in the proof of Cor.\ \ref{cor:alpha-in-totp}}
\lElse{non-deterministically choose between stop and simulate $M_{rec}(\beta,\mathcal{A},v,V)$}}
\end{algorithm}

\begin{algorithm}
\caption{NPTM $M_{rec}$}\label{alg:recursive-algorithm-totp}
\DontPrintSemicolon
\KwIn{$\gamma, \mathcal{A},v,V$}
\If{$\gamma==\sum_{i=1}^r\quantsigma Y.\, \varphi(\underline{X})\cdot\big(\top+\psi_i(X,Y)\cdot f(Y)\big)$}{
$Choices:=\emptyset$\;
\For{$i \gets 1$ to $r$} {
% $B_i:=\emptyset$\;
\If{$N_{\psi_i}(V(X))$ does not reject}{
$B_i:=\mathrm{out}_{N_{\psi_i}}(V(X))$\;
\lIf{$(\mathcal{A},V[B_i/X]\models\varphi(X))$}{$Choices:=Choices\cup\{B_i\}$}%\;
} 
% \If{$(B_i! = \emptyset)$
% and $(B_i != B_j$ for every $j<i)$  and $(\mathcal{A}\models\varphi(B_i/X))$}{$Choices:=Choices\cup\{i\}$}
% 
}
non-deterministically go to line 9  or 10  \;
stop \hspace{5.1cm}\algorithmiccomment{this path 
corresponds to $\rel[\top](\mathcal{A},v,V)$}\;
% non-deterministically choose $i\in Choices$ and simulate $M_{rec}(\beta,\mathcal{A},v,V[B_i/X])$
non-deterministically choose $B\in Choices$ and simulate $M_{rec}(\beta,\mathcal{A},v,V[B/X])$
}
\If{$\gamma==\alpha + \sum_{i=1}^r\quantsigma Y.\, \varphi(\underline{X})\cdot\big(\top+\psi_i(X,Y)\cdot f(Y)\big)$}{
% simulate $N_\alpha(\mathcal{A},v,V)$ and store $\mathrm{out}_{N_{\alpha}}(V(X))$ in the work tape\; 
$St : = \mathrm{out}_{N_{\alpha}}(V(X))$\;
$Choices:=\emptyset$\;
\For{$i \gets 1$ to $r$} {
\If{$N_{\psi_i}(V(X))$ does not reject}{
$B_i:=\mathrm{out}_{N_{\psi_i}}(V(X))$\;
\lIf{$(\mathcal{A},V[B_i/X]\models\varphi(X))$}{$Choices:=Choices\cup\{B_i\}$}%\;
} 
% $B_i:=\emptyset$\;
% \lIf{$N_{\psi_i}(V(X))$ does not reject}{$B_i:=\mathrm{out}_{N_{\psi_i}}(V(X))$}   
% \If{$(B_i! = \emptyset)$
% and $(B_i != B_j$ for every $j<i)$  and $(\mathcal{A}\models\varphi(B_i/X))$}{$Choices:=Choices\cup\{i\}$}
 }
\For{
% all stored 
$s \in St \setminus \{\varepsilon\}$}
{\For{$B\in Choices$}{
$t:=s[2:]$\;
\If{$(s[1] == V(X))$ and $(N_\beta(t,\mathcal{A},v,V[B/X])$ accepts$)$ or $(t == \varepsilon)$}{remove $s$ from $St$
% the work tape
}}
}
non-deterministically go to line 23, or 24, or 25\\
stop \hspace{5.1cm}\algorithmiccomment{this path 
corresponds to $\rel[\top](\mathcal{A},v,V)$}\;
non-deterministically choose an $s \in St$ 
% that is still stored 
and stop\;
non-deterministically choose $B\in Choices$ and simulate $M_{rec}(\gamma,\mathcal{A},v,V[B/X])$}
\end{algorithm}
Let $[\lfp_f \beta](X)$ be in $\totlogiclfpth$. 
% Algorithm~\ref{alg:totp} describes NPTM $M_\beta$, such that $tot_{M_\beta}(\enc(\mathcal{A},v,V))=\llbracket \,[\lfp_f \beta](X)\,\rrbracket (\mathcal{A},v,V)$, for every $\mathcal{A}$, $v$, and $V$. If $\beta$ contains a function symbol, then $M_\beta$ simulates $M_{rec}(\beta,\mathcal{A},v,V)$, which is defined in Algorithm~\ref{alg:recursive-algorithm-totp} together with an additional dummy path (lines 5 and 8 of Algorithm~\ref{alg:totp}). 
%Let $k$ denote $\arity(X)=\arity(Y)$ and $m$ denote the maximum arity of any second-order variable that appears in $\beta$. 
Let $N_{\psi_i}$ denote the poly-time TM from Lemma~\ref{poly-algo-extends-formula} associated with $\psi_i$, and $\mathrm{out}_{N_{\psi_i}}(V(X))$ denote the output of $N_{\psi_i}$'s computation on input $\enc(\mathcal{A},v,V)$.
Let $N_\gamma$ be the deterministic poly-time TM from Lemma~\ref{lem:alpha-in-p} that is associated with each $\gamma$, and $\mathrm{out}_{N_{\gamma}}(V(X))$ denote the set that $N_{\gamma}$ returns on input $\enc(\mathcal{A},v,V)$. Let also $N_\beta$ be the TM associated with $[\lfp_f \beta](X)$ from Lemma~\ref{lem:time-bound-tocheck-a-2}. 
%We also assume that the empty string is encoded by some tuple of elements of $A$.

Algorithm~\ref{alg:totp} describes NPTM $M_\beta$, such that $tot_{M_\beta}(\enc(\mathcal{A},v,V))=\llbracket \,[\lfp_f \beta](X)\,\rrbracket (\mathcal{A},v,V)$, for every $\mathcal{A}$, $v$, and $V$. 
If $\beta$ contains a function symbol, then $M_\beta$ 
first verifies that $\varphi(X)$ is satisfied, and if not, only the first summand, $\alpha$ needs to be considered.
Otherwise, $M_\beta$ 
simulates $M_{rec}(\beta,\mathcal{A},v,V)$, which is defined in Algorithm~\ref{alg:recursive-algorithm-totp} together with an additional dummy path (lines 5 and 8 of Algorithm~\ref{alg:totp}). 
It is important to note that whenever 
$M_{rec}(\beta,\mathcal{A},v,V)$ is called, we are guaranteed that $\mathcal{A},V\models \varphi(X)$.
$M_\beta$ is similar to the one defined in the proof of Proposition~\ref{spspace second inclusion}. 
However, 
we must also ensure that the machine does not generate redundant computation paths.
Note that 
the processing of lines 4, 5, and 6, or 11--21 ensures that 
each path that is generated by the non-deterministic choices of lines 8 and 9, or 23, 24, and 25, represents different strings from 
$\rel[\,[\lfp_f \beta](X)\,](\mathcal{A},v,V)$.
% If two different paths correspond to the same string, only one of them must be generated. 
% $M_\beta$ guarantees that this is the case by making a deterministic polynomial time computation before the paths are generated and eliminating one of them.

Regarding the time complexity used by $M_{rec}$, the body of the for-loops in lines 3, 13, and 17 is executed a constant number of times, whereas the body of the for-loop in line 18 is executed a polynomial number of times by Lemma~\ref{lem:alpha-in-p}. By Lemmata~\ref{lem:time-bound-tocheck-a-2} and~\ref{lem:bounded-length-a}, the simulation of $N_\beta(s[2:]),\mathcal{A},v,V[s[1]/X])$ needs at most $\mathcal{O}\big(c^{|s|}\cdot(\mathrm{poly}(|A|)+|\enc(s)|)\big)=\mathcal{O}(c^{|\alpha|}\cdot\mathrm{poly}(|A|))$ time, which is polynomial in $|A|$.
Finally, the number of recursive calls made during the computation of a path $p$ of $M_{rec}(\mathcal{A},v,V)$ is polynomially bounded: let $\enc(\mathcal{A},v,V[B/X])$ be the input to a recursive call made during the computation of $p$. Then, the next recursive call will be on input $\enc(\mathcal{A},v,V[B'/X])$, where $B'$ is the unique relation such that $\mathcal{A},V[B/X,B'/Y]\models\psi_i(X,Y)$, for some $1\leq i\leq r$. Since $\psi_i(X,Y)$ strictly extends $X$ to $Y$, $B\subsetneq B'$. Moreover, $B\in\mathcal{R}_k$, and so it needs at most $|A|^k$ recursive steps to be extended to some $B^*$ that cannot be strictly extended by any $\psi_i$, and so path $p$ comes to an end. \qedhere
\end{proof}

\begin{theorem}\label{totp main theorem}
$\totp = \totclasslfpu$ over finite ordered structures.
\end{theorem}
\begin{proof}
    The theorem  is immediate from Propositions~\ref{totp first inclusion} and~\ref{totp second inclusion}.
\end{proof}

\subsection{\totlogiclfpf captures \totp over finite ordered structures}\label{totp subsection}

In this subsection, we use definitions and notation from Subsection~\ref{subsec:totp-first-logic}.

\begin{notation}
$\underline{Y}:=\varphi(X)\cdot f(Y)$ denotes $\quantsigma Y. \varphi(X,Y)\cdot Y\cdot f(Y)$, where $\varphi$ (strictly) extends $X$ to $Y$.
\end{notation}

\totlogiclfpf is a fragment of \sigmasofo with recursion.

\begin{definition} \label{totp definition-two}
\begin{enumerate}[(a)]
    \item The \sigmaforu formulae over $\sigma$ are 
    the $x$-free \sigmasofo formulae with the restriction that 
    the second-order sum operator only appears as
    $Y:=\varphi\cdot\alpha$,  
    $\varphi\in\fo$.
    \item 
    % We define the logic 
    \totlogiclfpf over $\sigma$ 
    is
    % to be 
    the set of  formulae $[\lfp_f\, \beta](X)$, where $\beta$ is defined by: 
    % the following grammar:
\begin{equation}
\begin{aligned}
\beta::=~ \alpha ~\mid~ \underline{Y}:=\psi(X)\cdot f(Y)~\mid~ \alpha+\beta ~\mid~ \varphi\cdot\beta ~\mid~ \beta+\beta+\top ~\mid~ \varphi\cdot\beta +\neg \varphi \cdot \beta 
\end{aligned}
\label{beta4-two}
\end{equation}
    \noindent where $X,Y$ are  second-order variables, $\alpha$ is a \sigmaforu formula over $\sigma$, $\varphi,\psi$ are $\fo$ formulae over $\sigma$, $\psi$ strictly extends $X$ to $Y$, and $f$ is a second-order function symbol.
\end{enumerate} 
\end{definition}

Below we define the \totlogiclfpf formula $\mathsf{total}(X,f)$, the least fixed point of which applied on $S_0$ is equal to the number of branchings of $N$ on input $\enc(\mathcal{A})$:
$$\begin{aligned}
\mathsf{branch}(X)\big(\sum_{i=0,1}\underline{Y}:=\mathsf{ndet}_i(X)\cdot
f(Y)+\top\big)+
\neg \mathsf{branch}(X) \big(\mathsf{nfinal}(X) \cdot \underline{Y}:=\mathsf{det}(X)\cdot
f(Y)\big).
\end{aligned}$$
% $$\begin{aligned}
% &\mathsf{branch}(X)\big(\underline{Y}:=\mathsf{ndet}_0(X)\cdot
% f(Y)+\underline{Y}:=\mathsf{ndet}_1(X)\cdot
% f(Y)+\top\big)+\\
% &\neg \mathsf{branch}(X) \big(\mathsf{nfinal}(X) \cdot \underline{Y}:=\mathsf{det}(X)\cdot
% f(Y)\big).
% \end{aligned}$$

Let $\mathsf{current}(X,\Vec{y}):= \forall \Vec{y}' (\Vec{y}'\leq \Vec{y}\rightarrow \exists \Vec{x} \exists\Vec{z} X(\Vec{x},\Vec{y}',\Vec{z}))\wedge \forall \Vec{y}' (\Vec{y}'> \Vec{y}\rightarrow \forall \Vec{x} \forall \Vec{z} \neg X(\Vec{x},\Vec{y}',\Vec{z}))$ be the formula that when $X$ is interpreted as a relation $S$ encoding a run $c_0\dots c_m$ of $N$, and $\Vec{y}$ is interpreted as a time step $\Vec{t}$, it expresses that $\Vec{t}$ is the current time step, i.e.\ the one corresponding to configuration $c_m$. Then, $\mathsf{branch}$ and $\mathsf{nfinal}$ are defined as follows.

$$\mathsf{branch}(X):= \exists \Vec{x} \exists \Vec{y} \exists\Vec{z} \big(X(\Vec{x},\Vec{y},\Vec{z})\wedge \mathsf{current}(X,\Vec{y})\wedge \bigvee_{\Vec{\gamma}\in\Gamma_{nondet}} \Vec{z}=\Vec{\gamma}\big),$$

$$\mathsf{nfinal}(X):=\exists \Vec{x} \exists \Vec{y} \exists\Vec{z} \big(X(\Vec{x},\Vec{y},\Vec{z})\wedge \mathsf{current}(X,\Vec{y})\wedge \bigvee_{\Vec{\gamma}\in\Gamma_{det}\cup\Gamma_{nondet}} \Vec{z}=\Vec{\gamma}\big).$$

\begin{lemma}\label{lem:exists_branching}
If $S_p$ describes a partial run $c_0\cdots c_m$ of $N$, then:
\begin{itemize}
    \item $\mathcal{A},V[S_p/X]\models\mathsf{branch}(X)$ 
if and only if $N(\enc(\mathcal{A}))$  has a non-deterministic choice in $c_m$,
\item $\mathcal{A},V[S_p/X]\models\mathsf{nfinal}(X)$ 
if and only if $N(\enc(\mathcal{A}))$ makes at least one  transition in $c_m$.
\end{itemize}
\end{lemma}
\begin{proof}
    Immediate from the definitions of $\mathsf{branch}(X)$ and $\mathsf{nfinal}(X)$.
\end{proof}

Formulae $\mathsf{ndet}_i(X)$, $i=0,1$, and $\mathsf{det}(X)$ are defined below.

$$\begin{aligned}
    \mathsf{ndet}_i(X,Y):=&\forall \Vec{x} \forall \Vec{y}\forall \Vec{z} \big(Y(\Vec{x},\Vec{y},\Vec{z})\leftrightarrow X(\Vec{x},\Vec{y},\Vec{z}) \vee \bigvee_{\tau\in \Delta_i} \mathsf{update}_\tau(X,\Vec{x},\Vec{y},\Vec{z})\big)\wedge\\ &\exists\Vec{x} \exists\Vec{y}\exists \Vec{z} \big(\neg X(\Vec{x},\Vec{y},\Vec{z})\wedge Y(\Vec{x},\Vec{y},\Vec{z}) \big),~i=0,1,
\end{aligned}$$

$$\begin{aligned}
    \mathsf{det}(X,Y):=&\forall \Vec{x} \forall \Vec{y}\forall \Vec{z} \big(Y(\Vec{x},\Vec{y},\Vec{z})\leftrightarrow X(\Vec{x},\Vec{y},\Vec{z}) \vee \bigvee_{\tau\in \Delta_{det}} \mathsf{update}_\tau(X,\Vec{x},\Vec{y},\Vec{z})\big)\wedge\\ &\exists\Vec{x} \exists\Vec{y}\exists \Vec{z} \big(\neg X(\Vec{x},\Vec{y},\Vec{z})\wedge Y(\Vec{x},\Vec{y},\Vec{z}) \big).
\end{aligned}$$

\begin{lemma}\label{lem:branch}
   If $S_p$ describes a partial run $c_0\cdots c_m$ of $N$, and 
   \begin{itemize}
       \item $\mathcal{A},V[S_p/X,S_{new}/Y]\models\mathsf{ndet}_i(X,Y)$, $i=0,1$, then $S_{new}$ describes the run $c_0\cdots c_m c_{m+1}$, where $c_{m+1}$ is the configuration that $N$ reaches after making non-deterministic choice $i$ in $c_m$,
       \item $\mathcal{A},V[S_p/X,S_{new}/Y]\models\mathsf{det}(X,Y)$, $i=0,1$, then $S_{new}$ describes the run $c_0\cdots c_m c_{m+1}$, where $c_{m+1}$ is the configuration that $N$ reaches after making a deterministic transition in $c_m$.
   \end{itemize}  
\end{lemma}
\begin{proof}
    Immediate from Lemma~\ref{lem:updatet} and the definitions of $\mathsf{ndet}_i(X,Y)$ and  $\mathsf{det}(X,Y)$.
\end{proof}

Let $X$ be interpreted as a relation $S_p$ that describes a partial run $c_0\dots c_m$ of $N$. Formula $\mathsf{branch}$ checks whether the current configuration $c_m$ creates a branching.
%and if this is true, a new string is returned as an output, because of $\mathsf{branch}(X)\cdot\top$. 
Formulae  $\mathsf{ndet}_i$, $i=0,1$, and $\mathsf{det}$ extend $S_p$ to a relation $S_{new}$, that describes the run $c_0\dots c_m c_{m+1}$, where $c_{m+1}$ is the configuration that $N$ reaches from $c_m$ by making non-deterministic choice $i$ 
%in $\Delta_i$,  
or a deterministic transition,
%$\Delta_{det}$, 
respectively. The evaluation continues recursively on $S_{new}$. 
%If $c_m$ leads to a deterministic transition, the formula proceeds analogously, but does not output a new string. 
Finally, if $c_m$ is a configuration where $N$ halts, $\mathsf{nfinal}$ becomes false and recursion stops. Moreover, $\mathsf{ndet}_i(X,Y)$, $i=0,1$, and $\mathsf{det}(X,Y)$ are \fo formulae that strictly extend $X$ to $Y$.
As a result, 
%$\rel[\, [\lfp_f\,\mathsf{tot}](X)\,](\mathcal{A},v,V)$, where $V(X)=S_0$, is a set of strings, that each of them corresponds to a different branching of $N$ on $\enc(\mathcal{A})$. In specific, 
there is a bijection between the strings in $\rel[\, [\lfp_f\,\mathsf{tot}](X)\,](\mathcal{A},v,V)$ and branchings of $N(\enc(\mathcal{A}))$. Assume that $c_m$ is a 
configuration that is not the initial configuration $c_0$ and leads to a non-deterministic choice. Then, $c_m$ can be mapped to a string $S_1\circl \dots \circl S_i\in(\mathcal{R}_{3k})^*$ in $\rel[\, [\lfp_f\,\mathsf{tot}](X)\,](\mathcal{A},v,V)$, where $S_j$ extends $S_{j-1}$, for every $2\leq j\leq i$, and $S_i$ describes $c_0\dots c_m$. If $c_0$ leads to a non-deterministic choice, it is mapped to string $\varepsilon$.
%Thus, the following proposition holds and since $[\lfp_f\,\mathsf{tot}](X)\in\totlogiclfpf$, the proposition implies that $\totp\subseteq\totclasslfpf$ over finite ordered structures.

\begin{proposition}\label{totp first inclusion-two}
Given an NPTM $N$, $\llbracket\, [\lfp_f\,\mathsf{total}](X)\,\rrbracket(\mathcal{A},v,V)=\#($branchings of  $N(\enc(\mathcal{A}))$,
%\text{ on } \enc(\mathcal{A}))$
where $V(X)$ encodes the initial configuration of $N$.
\end{proposition}

\begin{example}\label{ex:ind_sets-two}
 \textsc{\#IS} on $\enc(\mathcal{A})$ is equal to  $\llbracket\,[\lfp_f \textsf{ind\_sets}](I,Ex)\, \rrbracket (\mathcal{A},v,V)$ with $V(I)=V(Ex)=\emptyset$, where 
 $\textsf{ind\_sets}(I,Ex,f)$ is the following \totlogiclfpf formula: 
$$\begin{aligned}
(I=\emptyset\wedge Ex=\emptyset)\cdot I\cdot Ex + 
\textsf{can\_extend}(I,Ex)\cdot \big( & \underline{I'},\underline{Ex'}:=\textsf{include}(I, Ex)\cdot f(I',Ex')+\\
& \underline{I'},\underline{Ex'}:=\textsf{exclude}(I, Ex)\cdot f(I',Ex')+\top\big).
\end{aligned}$$
All subformulae of \textsf{ind\_sets} have been described in Example~\ref{ex:ind_sets}.
\end{example}

To prove $\totclasslfpf\subseteq \totp$, we use  Lemmata~\ref{syntactically-defines-implication}--\ref{poly-algo-extends-formula} and similar results to Lemma~\ref{lem:alpha-in-p}  and Corollaries~\ref{lem:time-bound-tocheck-a} and~\ref{cor:alpha-in-totp}. 

\begin{lemma}\label{lem:alpha-in-p-two}
Let $\alpha$ be a \sigmaforu formula over $\sigma$. The size of $\rel[\alpha](\mathcal{A},v,V)$ is polynomial in $|A|$, for every $\mathcal{A},v$, and $V$. Moreover, there is a deterministic poly-time TM $M$, which on input $\enc(\mathcal{A},v,V)$,  outputs all elements of $\rel[\alpha](\mathcal{A},v,V)$ in $\mathcal{O}(\mathrm{poly}(|A|))$ time.
\end{lemma}

\begin{corollary}\label{lem:time-bound-tocheck-a-two}
Let $\alpha$ be a \sigmaforu formula over $\sigma$. There is a deterministic TM $M$, such that on input 
$\enc(\mathcal{A},v,V)$, and a string $s\in(\bigcup_{i\in\mathbb{N}}\mathcal{R}_i)^*$, $M$ decides if $s \in \rel[\alpha](\mathcal{A},v,V)$ in $\mathcal{O}(\mathrm{poly}(|A|))$ time.
\end{corollary}

\begin{corollary}\label{cor:alpha-in-totp-two}
Let $\alpha$ be a \sigmaforu formula over $\sigma$. There is an NPTM $M$, such that $tot_M(\enc(\mathcal{A},v,V))=\llbracket\alpha\rrbracket(\mathcal{A},v,V)$, for every $\mathcal{A},v$ and $V$.
\end{corollary}

Lemma~\ref{lem:time-bound-tocheck-a-2-two} implies that the membership of $s$ in the intermediate interpretation of any \totlogiclfpf formula can be verified in polynomial time w.r.t.\ $|A|$, for every $s$ of polynomial size.

\begin{lemma}\label{lem:time-bound-tocheck-a-2-two}
Let $[\lfp_f \beta](X)$ be a \totlogiclfpf formula over $\sigma$.
There is a deterministic TM $M_\beta$, such that on input 
$\enc(\mathcal{A},v,V)$, and a string $s\in(\bigcup_{i\in\mathbb{N}}\mathcal{R}_i)^*$, $M_\beta$ decides if $s \in \rel[\,[\lfp_f \beta](X)\,](\mathcal{A},v,V)$ in $\mathcal{O}\big(|s|\cdot \mathrm{poly}(|A|),|\enc(s)|\big)$ time.
\end{lemma}
\begin{proof} To prove the lemma we use the first part of the following claim. This claim will also be of use in the proof of Proposition~\ref{totp second inclusion-two}.\\
\textit{Claim.} For any formula $[\lfp_f \beta](X)\in \totlogiclfpf$, $\beta$ is of the form $\alpha(X)+\sum_{i=1}^r \varphi_i(X) \cdot \underline{Y}:=\psi_i(X)\cdot f(Y)$, $r\geq 0$, where $\alpha\in\sigmaforu$, and  $\varphi_i,\psi_i\in\fo$, $1\leq i\leq r$. 
Moreover, for every $\mathcal{A},v,V$, if $\mathcal{A},V[B/X]\models\varphi_i(X)$ and $\mathcal{A},V[B/X]\models\varphi_j(X)$, for some $i\neq j$, then $\varepsilon\in\rel[\alpha](\mathcal{A},v,V[B/X])$.\\
\textit{Proof of Claim.} By straightforward induction on the structure of $\beta$. For example, if $\beta=\varphi\cdot\beta_1+\neg\varphi\cdot\beta_2$, then by the inductive hypothesis, $\beta_j=\alpha_j+\sum_{i=1}^{r_j} \varphi_{ji}(X) \cdot \underline{Y}:=\psi_{ji}(X)\cdot f(Y)$, $j=1,2$. So, $\beta$ can be transformed into the following formula: 
$$\varphi \cdot\alpha_1+\sum_{i=1}^{r_1}\varphi \cdot\varphi_{1i}(X) \cdot \underline{Y}:=\psi_{1i}(X)\cdot f(Y)+\neg\varphi\cdot\alpha_2+\sum_{i=1}^{r_2} \neg\varphi\cdot\varphi_{2i}(X) \cdot \underline{Y}:=\psi_{2i}(X)\cdot f(Y)$$
which is equivalent to $\alpha+\sum_{i=1}^{r}\chi_i \cdot \underline{Y}:=t_{i}(X)\cdot f(Y)$, where $\alpha=\varphi \cdot\alpha_1+\neg\varphi\cdot\alpha_2$, $r=r_1+r_2$, $\chi_i=\varphi \cdot\varphi_{1i}$, $t_i=\psi_{1i}$, $1\leq i\leq r_1$, $\chi_i=\neg\varphi \cdot\varphi_{2(i-r_1)}$, $t_i=\psi_{2i}$,  $r_1+1\leq i\leq r$. If $\mathcal{A},V[B/X]\models\chi_i(X)$ and $\mathcal{A},V[B/X]\models\chi_j(X)$, for some $i\neq j$, then either $\chi_i=\varphi \cdot\varphi_{1i}$ and $\chi_j=\varphi \cdot\varphi_{1j}$, or $\chi_i=\neg\varphi \cdot\varphi_{2(i-r_1)}$ and $\chi_j=\neg\varphi \cdot\varphi_{2(j-r_1)}$, since $\varphi,\neg \varphi$ cannot both be satisfied. W.l.o.g.\ assume the former case. By the inductive hypothesis, $\varepsilon\in \rel[\alpha_1](\mathcal{A},v,V[B/X])$, and since $\mathcal{A},V[B/X]\models\varphi(X)$, we have that $\varepsilon\in \rel[\varphi\cdot\alpha_1](\mathcal{A},v,V[B/X])\subseteq \rel[\alpha](\mathcal{A},v,V[B/X])$.\\

If $\beta$ is some $\alpha\in\sigmaforu$, 
then the lemma follows from Corollary~\ref{lem:time-bound-tocheck-a-two}. In this case, let $M_\alpha$ denote the deterministic poly-time TM from Corollary~\ref{lem:time-bound-tocheck-a-two}. If $\beta$ is $\alpha+\sum_{i=1}^r \varphi_i \cdot \underline{Y}:=\psi_i(X)\cdot f(Y)$, $r\geq 1$, then $M_\beta$ is defined in Algorithm~\ref{alg:totp-membership-two}. Let $\arity(X)=\arity(Y)=k$; let also $N_{\psi_i}$ denote the poly-time TM from Lemma~\ref{poly-algo-extends-formula} associated with $\psi_i$: on input $\enc(\mathcal{A},v,V)$, if there is $C$ such that  $\mathcal{A},V[C/Y]\models \psi_i(X,Y)$, $N_{\psi_i}$ returns $C$, and otherwise it rejects. Let $\mathrm{out}_{N_{\psi_i}}(V(X))$ denote the output of $N_{\psi_i}$'s computation on input $\enc(\mathcal{A},v,V)$.
\begin{algorithm}
\caption{$M_\beta$ when $\beta=\alpha+\sum_{i=1}^r \varphi_i \cdot \underline{Y}:=\psi_i(X)\cdot f(Y),~r\geq 1$}\label{alg:totp-membership-two}
\DontPrintSemicolon
\KwIn{$s, \mathcal{A},v,V$}
{simulate $M_\alpha(s,\mathcal{A},v,V)$ from Corollary~\ref{lem:time-bound-tocheck-a-two}}\;
\lIf{$M_\alpha(s,\mathcal{A},v,V)$ accepts}{accept}
\For{$i \gets 1$ to $r$}{
$C:= \mathrm{out}_{N_{\psi_i}}(V(X))$\;
\If{$(\mathcal{A},V\models\varphi_i(X))$ and $(s[1] == C)$}{
%\If{ $(\mathcal{A},V[C_i/X]\models\psi_i(X))$ or $(\rel[\alpha](\mathcal{A},v,V[C_i/X]\neq\emptyset)$}
{simulate $M_\beta(s[2:],\mathcal{A},v,V[C/X])$}}{
}}
reject
\end{algorithm}

Algorithm~\ref{alg:totp-membership-two} accepts in line 1 if $s\in\rel[\alpha](\mathcal{A},v,V)$. Otherwise, it checks whether $s$ starts with some $C$ such that $\mathcal{A},V[C/Y]\models \psi_i(X,Y)$  and also $\mathcal{A},V\models\varphi_i(X)$, for the input second-order assignment $V$. Then, it removes the starting element of $s$ and recurses on $s[2:]$ and $\mathcal{A},v,V[C/X]$. If at some point the remaining part of $s$ does not belong to $\rel[\alpha](\mathcal{A},v,V)$ for the current input $V$ and does not start with some relation generated by the second summand $\sum_{i=1}^r \varphi_i \cdot \underline{Y}:=\psi_i(X)\cdot f(Y)$, then Algorithm~\ref{alg:totp-membership-two} rejects.

Regarding the complexity of Algorithm~\ref{alg:totp-membership-two}, $M_\alpha$ and $N_{\psi_i}$ are deterministic poly-time TMs. The body of the for-loop in line 3 is repeated at most $r$ times, i.e.\ a constant number of times. The if condition can be checked in polynomial time w.r.t.\ $|A|$ and at most one recursive call is made in line 6 the first time the if condition becomes true. Since the length of the input to the next recursive call is reduced by one, at most $|s|$ recursive calls are made in total. Of course, additional $|\enc(s)|$ space and time is required to store and proccess  string $s$.
\end{proof}

We can now prove that $\totclasslfpf\subseteq \totp$ over finite ordered structures.

\begin{proposition}\label{totp second inclusion-two}
$\totclasslfpf\subseteq \totp$ over finite ordered structures.
\end{proposition}
\begin{proof} 
To prove the proposition, we use the claim that was stated in the proof of Lemma~\ref{lem:time-bound-tocheck-a-2-two}. We first restate the claim.\\
\textit{Claim.} For any formula $[\lfp_f \beta](X)\in \totlogiclfpf$, $\beta(X)$ is of the form $\alpha(X)+\sum_{i=1}^r \varphi_i(X) \cdot \underline{Y}:=\psi_i(X)\cdot f(Y)$, $r\geq 0$, where $\alpha\in\sigmaforu$, and  $\varphi_i,\psi_i\in\fo$, $1\leq i\leq r$. Moreover, for every $\mathcal{A},v,V$, if $\mathcal{A},V[B/X]\models\varphi_i(X)$ and $\mathcal{A},V[B/X]\models\varphi_j(X)$, for some $i\neq j$, then $\varepsilon\in\rel[\alpha](\mathcal{A},v,V[B/X])$.\\

%\begin{algorithmic}[1]
\begin{algorithm}
\caption{NPTM $M_\beta$ where $[\lfp_f \beta](X)\in\totlogiclfpf$}\label{alg:totp-two}
\DontPrintSemicolon
\KwIn{$\mathcal{A},v,V$}
\If{$\beta==\alpha$ has no function symbol}{
simulate $M_\alpha(\mathcal{A},v,V)$ from Corollary~\ref{cor:alpha-in-totp-two}
}
\If{$\beta==\alpha + \sum_{i=1}^r\varphi_i\cdot\underline{Y}:=\psi_i(X)\cdot f(Y),~r\geq 1$}{
{$Choices\gets\emptyset$}\;
\For{$i \gets 1$ to $r$}{
\If{$(\mathcal{A},V\models\varphi_i(X))$ and $(N_{\psi_i}(\mathcal{A},v,V)$ does not reject$)$}{
{$Choices:=Choices\cup\{\mathrm{out}_{N_{\psi_i}}(V(X))\}$}\;
}
}
{$Choices\gets $ \Valid{$Choices$}}\;
\lIf{$Choices==\emptyset$}{simulate $M_\alpha(\mathcal{A},v,V)$ from Corollary~\ref{cor:alpha-in-totp-two}}
\lElse{non-deterministically choose between stop and  \Recursion{$\beta,\mathcal{A},v,V$}
}}
\end{algorithm}
%\end{algorithmic}

%begin{algorithmic}[1]
\begin{algorithm}
\caption{Procedures called by NPTM $M_\beta$ of Algorithm~\ref{alg:totp-two}}\label{alg:recursive-algorithm-totp-two}
\DontPrintSemicolon
 \DontPrintSemicolon
  \SetKwFunction{GenPath}{Gen\_Path}
  \SetKwProg{Fn}{procedure}{:}{}
  \Fn{\GenPath{$R$}}{
  \lIf{$\mathrm{out}_{N_{\alpha}}(R)~!= \emptyset$}{return true}
  \Else{
  \For{$i\gets 1$ to $r$}
  {\If{$(\mathcal{A},V[R/X]\models\varphi_i(X))$ and $(N_{\psi_i}(\mathcal{A},v,V[R/X])$ does not reject$)$}{return \GenPath{$\mathrm{out}_{N_{\psi_i}}(R)$}}}
  {return false}}
  }
  \;
  \SetKwFunction{Valid}{Valid}
  \SetKwProg{Fn}{procedure}{:}{}
  \Fn{\Valid{$S$}}{
      \For{$C \in S$}{
       \If{not \GenPath{$C$}}{
         remove $C$ from $S$}} 
    return $S$}
  \; 
  \SetKwFunction{Recursion}{Recursion}
  \SetKwProg{Fn}{procedure}{:}{}
  \Fn{\Recursion{$\beta,\mathcal{A},v,V$}}{
  $St \gets \mathrm{out}_{N_{\alpha}}(V(X))$\;
  $Choices\gets\emptyset$\;
  \For{$i \gets 1$ to $r$} {
  \If{$(\mathcal{A},V\models\varphi_i(X))$ and $(N_{\psi_i}(\mathcal{A},v,V)$ does not reject$)$}{$Choices\gets Choices\cup\{\mathrm{out}_{N_{\psi_i}}(V(X))\}$}}
  {$Choices\gets$ \Valid{$Choices$}}\;
  \For{$s\in St$}{
  \For{$B\in Choices$}
  {$t\gets s[2:]$\;
  \If{$(s[1]==B)$ and $(N_\beta(t,\mathcal{A},v,V[B/X])$ accepts$)$}{remove $s$ from $St$}}}
  {non-deterministically go to line 28  or 29}\;
  {non-deterministically choose $s\in St$ and stop}\;
  {non-deterministically choose $B\in Choices$ and \Recursion{$\beta,\mathcal{A},v,V[B/X])$}
  }
  }
   % \;
   % \SetKwFunction{Machine}{$M_{\beta}$}
   % \SetKwProg{Fn}{procedure}{:}{}
   % \Fn{\Machine{$\mathcal{A},v,V$}}{
   % \If{$\beta==\alpha$ has no function symbol}{
   % simulate $M_\alpha(\mathcal{A},v,V)$ defined in the proof of Corollary~\ref{cor:alpha-in-totp}
   % }
   % \If{$\beta==\alpha + \sum_{i=1}^r\varphi_i\cdot\underline{Y}:=\psi_i(X)\cdot f(Y),~r\geq 1$}{
   % {$Choices=\emptyset$}\;
   % \For{$i \gets 1$ to $r$}{
   % \If{$(\mathcal{A},V\models\varphi_i(X))$ and $(N_{\psi_i}(\mathcal{A},v,V)$ does not reject$)$}{
   % {$Choices:=Choices\cup\{\mathrm{out}_{N_{\psi_i}}(V(X))\}$}\;
   %  }
   %  }
   % {$Choices$ \gets \Valid{$Choices$}}\;
   % \lIf{$Choices=\emptyset$}{simulate $M_\alpha(\mathcal{A},v,V)$ defined in the proof of Cor.~\ref{cor:alpha-in-totp}}
   % \lElse{non-deterministically choose between stop and \Recursion{$\beta,\mathcal{A},v,V$}
   % }}}
\end{algorithm}
%\end{algorithmic}

Let $[\lfp_f\,\beta](X)$ be in $\totlogiclfpf$.
Let $N_{\psi_i}$ denote the poly-time TM from Lemma~\ref{poly-algo-extends-formula} associated with $\psi_i$, and $\mathrm{out}_{N_{\psi_i}}(B)$ denote the output of $N_{\psi_i}$'s computation on input $\enc(\mathcal{A},v,V[B/X])$. For $\alpha\in\sigmaforu$, let
$N_\alpha$ denote the deterministic poly-time TM from Lemma~\ref{lem:alpha-in-p-two} that is associated with $\alpha$, and $\mathrm{out}_{N_{\alpha}}(B)$ denote the set that $N_{\alpha}$ returns on input $\enc(\mathcal{A},v,V[B/X])$. For $[\lfp_f\,\beta](X)\in\totlogiclfpf$, let $N_\beta$ denote the TM associated with $[\lfp_f\,\beta](X)$ from Lemma~\ref{lem:time-bound-tocheck-a-2-two}.

Algorithm~\ref{alg:totp-two} describes NPTM $M_\beta$, such that $tot_{M_\beta}(\enc(\mathcal{A},v,V))=\llbracket\,[\lfp_f\,\beta](X)\,\rrbracket(\mathcal{A},v,V)$, for every $\mathcal{A}$, $v$, and $V$. 
If $\beta$ contains a function symbol, then $M_\beta$ 
first checks whether recursion occurs, and if not, only the first summand, that is $\alpha$, is considered (lines 4--9).
Otherwise, $M_\beta$ 
calls \Recursion{$\beta,\mathcal{A},v,V$} and also generates  an additional dummy path (line 10). 
$M_\beta$ is similar to the one defined in the proof of Proposition~\ref{totp second inclusion}: since we are interested in the total number of paths of $M_\beta$, $M_\beta$ must ensure that it does not generate redundant computation paths; for instance, creating a path and then rejecting is not appropriate now. 
%Note that the processing of lines 4, 5, and 6, or 11--21 ensures that each path that is generated by the non-deterministic choices of lines 8 and 9, or 23, 24, and 25, represents different strings from $\rel[\,[\lfp_f\,\beta](X)\,](\mathcal{A},v,V)$.

Algorithm~\ref{alg:recursive-algorithm-totp-two} describes procedures \GenPath{$R$}, \Valid{$S$}, and \Recursion{$\beta,\mathcal{A},v,V$} that are called by $M_\beta$. In the case of $\beta=\alpha + \sum_{i=1}^r\varphi_i\cdot\underline{Y}:=\psi_i(X)\cdot f(Y)$, $r\geq 1$, procedure \Recursion{$\beta,\mathcal{A},v,V$} non-deterministically chooses to generate a path  that corresponds to either  the first or the second summand of $\beta$ (lines 27--29). Before generating a path that corresponds  to $s\in \rel[\alpha](\mathcal{A},v,V)$ (lines 16 and 28), it verifies that $s$ is not also generated by the second summand $\sum_{i=1}^r\varphi_i\cdot\underline{Y}:=\psi_i(X)\cdot f(Y)$ (lines 17--26).  Procedure \Valid{$Choices$}---which is called by $M_\beta$ in line 8 and by \Recursion{$\cdot$} in line 21---is necessary, so that a path corresponding to a sequence of recursive calls starting from a relation $C\in Choices$ is generated only if this sequence produces a string. In order for a string to be produced, the sequence of recursive calls has to end up at $\rel[\alpha](\mathcal{A},v,V)\neq \emptyset$ after a number of steps (which is checked by \GenPath{$C$} in line 2). If $\rel[\alpha](\mathcal{A},v,V[C/X])= \emptyset$, then there is at most one $1\leq i\leq r$, such that $\mathcal{A},V[C/X]\models\varphi_i(X)$ by the claim. If such an $i$ exists, then \GenPath{$C$} recursively checks whether $\mathrm{out}_{N_{\psi_i}}(C)$ gives rise to a sequence of recursive calls that produces a string. Otherwise, it deletes $C$ from the non-deterministic choices of \Recursion{$\beta,\mathcal{A},v,V$} (line 12).

Regarding the time complexity used by $M_\beta$, the body of the for-loop in line 22 is executed  a polynomial number of times by Lemma~\ref{lem:alpha-in-p-two}, whereas the body of all other for-loops is executed a constant number of times. By Lemmata~\ref{lem:time-bound-tocheck-a-2-two} and~\ref{lem:bounded-length-a}, the simulation of $N_\beta(s[2:],\mathcal{A},v,V[s[1]/X])$ in line 25, needs at most $\mathcal{O}\big(|s|\cdot\mathrm{poly}(|A|)+|\enc(s)|\big)=\mathcal{O}(|\alpha|\cdot\mathrm{poly}(|A|))$ time, which is polynomial in $|A|$.
Finally, the number of recursive calls to \Recursion{$\cdot$} during the computation of a path $p$ of $M_\beta(\mathcal{A},v,V)$ is polynomially bounded: let $(\beta,\mathcal{A},v,V[B/X])$ be the input to such a call made during the computation of $p$. Then, the next  call to \Recursion{$\cdot$} will be on input $(\beta,\mathcal{A},v,V[B'/X])$, where $B'$ is the unique relation such that $\mathcal{A},V[B/X,B'/Y]\models\psi_i(X,Y)$, for some $1\leq i\leq r$. Since $\psi_i(X,Y)$ strictly extends $X$ to $Y$, $B\subsetneq B'$. Moreover, $B\in\mathcal{R}_k$, and so it needs at most $|A|^k$ recursive steps to be extended to some $B^*$ that cannot be strictly extended by any $\psi_i$, and so path $p$ comes to an end. For the same reason, procedure \GenPath{$C$} makes a polynomial number of calls to itself for any $C$.\qedhere
\end{proof}

\begin{restatable}{theorem}{totpmaintheorem}\label{totp main theorem-two}
$\totclasslfpf =\totp$ over finite ordered structures.
\end{restatable}
\begin{proof}
    $\totp\subseteq\totclasslfpf$ follows from Proposition~\ref{totp first inclusion-two} and the fact that $[\lfp_f \textsf{total}](X)$ is in \totlogiclfpf. $\totclasslfpf\subseteq\totp$ was stated in Proposition~\ref{totp second inclusion-two}.
\end{proof}

The main theorems of Subsections~\ref{subsec:totp-first-logic} and~\ref{totp subsection} imply that the  logics \totlogiclfpth and \totlogiclfpf express exactly the same problems over finite ordered structures. So the use of \folfp instead of \fo does not make \totlogiclfpf more expressive.

\begin{corollary}
    $\totclasslfpu =\totclasslfpf$ over finite ordered structures.
\end{corollary}
\begin{proof}
    The corollary is immediate from Theorems~\ref{totp main theorem} and~\ref{totp main theorem-two}. 
\end{proof}

\section{Conclusions and open questions}\label{conclusions-section}

Inspired by the two-step semantics developed in the context of weighted logics, we introduced two-step semantics that enriches the existing framework of quantitative logics, i.e.\ logics for expressing counting problems. 
% In addition, w
We introduced least fixed formulae that use recursion on second-order function symbols
and 
% . 
% In such a manner, we 
provided logical characterizations of \spanl and \totp, answering an open question of~\cite{Arenas}. Furthermore, we determined logics that capture \spanpspace and \fpspace.
%
%Some of the quantitative logics we introduced in this work naturally capture classes of functions that count different valid outputs of space-restricted transducers. 
% in a natural way. 
Compared to the other classes, the logic that captures \totp over finite ordered structures, was defined in a more complicated way that is related to the properties of \totp problems: recursion of the logic expresses self-reducibility and the restricted form of the recursion captures the  easy-decision property. It would be interesting to investigate whether \totp is
% could be 
captured by a simpler, 
% and 
more elegant logic. 

The intermediate semantics can express sets of computation paths of TMs, different valid outputs of transducers, or solutions to computational problems. In specific, in the  case of \spanl and \spanpspace, union and concatenation of sets is more suitable than addition and multiplication of \qso; when the union (resp.\ concatenation) of two sets of strings is computed, identical outputs will contribute  one string to the resulting set. In general, using the intermediate semantics, it becomes possible to keep track of paths, outputs, and solutions, apply operations on them, and then count them. Another difference between our logics and  quantitative logics from~\cite{Arenas}, is that in~\cite{Arenas}, only first-order function symbols were considered and  interpreted as functions $h:A^k\rightarrow\mathbb{N}$. Then, the respective lattice $(\mathcal{F},\leq_F)$ is not complete and the least fixed point was defined by considering the supports of functions in $\mathcal{F}$~\cite[Section 6]{Arenas}.  By defining here, functions that their values are sets of strings, the lattice $(\mathcal{F},\leq_F)$  becomes complete,
%(see Section~\ref{relational semantics}),
and the definition of the least fixed point is straightforward.

% \textcolor{red}{rewrite this paragraph} The two-step semantics that we propose here presents its own interest for other reasons as well. First, it can be generalized so that the intermediate semantics maps formulae to, not necessarily sets of strings, but elements of any structure $\mathcal{S}$ that is equipped with two operations $\cup$ and $\circl$. Alternatively, it can be restricted. For example, by defining the concrete semantics to map any non-empty set to $1$ and the empty set to $0$, our results give 
The two-step semantics that we propose in this work is noteworthy for 
% additional 
reasons beyond its primary objective. 
% Firstly, I
It can be generalized to map formulae to elements of any structure $\mathcal{S}$ equipped with operations $\cup$ and $\circl$, instead of solely sets of strings. 
%Such generalization can enrich 
Conversely, it can also be specialized. For instance, by specifying the concrete semantics such that any non-empty set maps to $1$ and the empty set to $0$, our results yield
least-fixed-point logical characterizations of  \NL and \PSPACE, the decision variants of \spanl{} and \fpspace, respectively. It is known that these two classes are captured by \fo and \SO, equipped with the transitive closure operator, namely \fotc and \sotc, respectively~\cite{Immerman}. The logics defined here combine the least fixed point with quite natural syntactic definitions, 
without 
% having to introduce 
resorting to 
different fixed-point operators for each logic.
% whereas one can argue that the transitive closure operator is an artificial syntactic restriction of the least fixed point.

%It is worth investigating whether 
We believe that the logical characterization of \spanl 
%sheds more light on the inclusion of \spanl in \FPRAS.
% can enlighten the approximability of 
% can shed light on 
can yield 
more direct ways to approximate
its problems.
Logical formulae in \spanlogic bear some resemblance to regular grammars (or, equivalently, to NFAs), since the syntax of the logic, at each recursive call, concatenates a string of fixed length from the left with $f(\Vec{x})$. An interesting question is  whether one can adjust the fpras for \snfa{} and apply it directly to the syntax of \spanlogic, giving an fpras metatheorem for the logic. Moreover, it is only natural to investigate the class that results from allowing arbitrary concatenations of recursive calls, and to expect a natural connection to context-free languages. Note that the problem of counting the strings of a specific length accepted by a context-free grammar admits a quasi-polynomial randomized approximation algorithm~\cite{GoreJKSM97} and it is open whether it has an fpras. 

Another interesting question remains the logical characterization of a subclass of \sP for which computing the permanent of a matrix is complete under parsimonious reductions. The \textsc{Permanent} problem is equivalent to counting perfect matchings in a bipartite graph, and it has an fpras~\cite{JSV04}. It was the first problem shown in~\cite{Valiant79} to  be \sP-complete under Turing reductions, i.e.\ reductions that use oracle calls. Therefore, such a result would provide a new subclass of \FPRAS and at the same time, it would refine the complexity of the well-studied \textsc{Permanent} problem.

\bibliography{lipics-v2021-sample-article}

\end{document}